\documentclass{aa}  
\usepackage{graphicx}
\usepackage{natbib}
\usepackage{twoopt}
\usepackage[varg]{txfonts} 
\usepackage[percent]{overpic}
\usepackage{amsmath}
\usepackage{multicol}
\usepackage{dcolumn}
\usepackage{footnote}
\usepackage{url}
\makesavenoteenv{tabular}
\makesavenoteenv{table}
\bibpunct{(}{)}{;}{a}{}{,}             
\begin{document}
\title{Using rotation measure grids to detect cosmological magnetic fields -- a Bayesian approach}

   \subtitle{}
\author{  V. Vacca\inst{1}
          \and
          N. Oppermann\inst{2}
          \and
          T. En{\ss}lin\inst{1,3}
          \and
          J. Jasche\inst{4,5} 
          \and
          M. Selig\inst{1,3,6}
          \and
          M. Greiner\inst{1,3} 
          \and
          H. Junklewitz\inst{7} 
          \and
          M. Reinecke\inst{1}
          \and
          M. Br{\"u}ggen\inst{8}
          \and
          E. Carretti\inst{9}
          \and
          L. Feretti\inst{10}
          \and
          C. Ferrari\inst{11}
          \and
          C.\ A. Hales\inst{12,13}
           \and
          C. Horellou\inst{14}
          \and
          S. Ideguchi\inst{15}
          \and
          M. Johnston-Hollitt\inst{16}
          \and
          R.\ F. Pizzo\inst{17}
          \and
          H. R{\"o}ttgering\inst{18}
          \and
          T. W. Shimwell\inst{18}
           \and
          K. Takahashi\inst{19}
          }
\institute{Max Planck Institute for Astrophysics, Karl-Schwarzschild-Str. 1,  85748 Garching, Germany
              \and   
              Canadian Institute for Theoretical Astrophysics, University of Toronto, 60 St. George Street, Toronto ON, M5S 3H8, Canada
              \and
              Ludwig-Maximilians - Universit{\"a}t M{\"u}nchen, Geschwister-Scholl-Platz 1, 80539, München, Germany
              \and
               CNRS, UMR 7095, Institut d'Astrophysique de Paris, 98 bis, boulevard Arago, F-75014 Paris, France
              \and
              Excellence Cluster Universe, Boltzmannstr. 2, D-85748 Garching, Germany
              \and
              IBM Deutschland Research and Development GmbH Sch\"onaicher Stra{\ss}e 220, D-71032 B\"oblingen
              \and
              Argelander-Institut f\"ur Astronomie, Auf dem H\"ugel 71, 52121 Bonn, Germany
              \and
              Hamburger Sternwarte, University of Hamburg, Gojenbergsweg 112, D-21029 Hamburg, Germany
              \and
              INAF-Osservatorio Astronomico di Cagliari, Via della Scienza 5, 09047 Selargius (CA), Italy
              \and
              INAF-Istituto di Radioastronomia, via Gobetti 101, 40129 Bologna, Italy
              \and
              Laboratoire Lagrange, Universit\'e  C\^ote d'Azur, Observatoire de la C\^ote d'Azur, CNRS, Blvd de l'Observatoire, CS 34229, 06304 Nice cedex 4, France 
              \and
              National Radio Astronomy Observatory, PO Box 0, Socorro, NM 87801, USA
              \and
              Jansky Fellow of the National Radio Astronomy Observatory
              \and
              Department of Earth and Space Sciences, Chalmers University of Technology, OSO, SE-439 92 Onsala, Sweden
              \and
              Department of Physics, UNIST, Ulsan 689-798, Korea
              \and
              School of Chemical \& Physical Sciences, Victoria University of Wellington, PO Box 600, Wellington 6014, New Zealand
               \and
              ASTRON, Postbus 2, 7990 AA, Dwingeloo, The Netherlands
              \and
              Leiden Observatory, Leiden University, PO Box 9513, NL-2300 RA, Leiden, the Netherlands
              \and
              Kumamoto University, 2-39-1, Kurokami, Kumamoto 860-8555, Japan
              }

   \date{Received MM DD, YY; accepted MM DD, YY}
\abstract{
 
  Determining magnetic field properties in different environments of the cosmic large-scale structure as well as their evolution over redshift is a fundamental step toward uncovering the origin of cosmic magnetic fields. 
      Radio observations permit the study of extragalactic magnetic fields via
      measurements of the Faraday depth of extragalactic radio sources. Our
      aim is to investigate how much different extragalactic environments
      contribute to the Faraday depth variance of these sources. 
    We develop a Bayesian algorithm to distinguish statistically
    Faraday depth variance contributions intrinsic to the source from
    those due to the medium between the source and the observer. In our algorithm 
    the Galactic foreground and the measurement noise are taken
    into account as the uncertainty correlations of the
    galactic model. Additionally, our algorithm allows for the investigation of 
    possible redshift evolution of the extragalactic contribution. 
    This work presents the derivation of the algorithm and tests performed
    on mock observations. With cosmic magnetism being one of the key science projects of the new generation of radio interferometers we have made predictions for the algorithm's performance on data from the next generation of radio interferometers. 
    Applications to real data are left for future work. 

    }

   \maketitle

   \section{Introduction}

The origin and evolution of cosmic magnetism are at present poorly
understood. Answering the many open questions surrounding the physics
of astrophysical magnetic fields is a difficult task
since magnetic fields can be significantly affected by structure and galaxy
formation and evolution processes. Their strength can be amplified for example
in galaxy clusters by mergers, and in galaxies by large-scale
dynamos, invoking differential rotation and turbulence.  Insights into
the origin and properties of magnetic fields in the Universe could be
provided by probing them on even larger scales. Along
filaments and voids of the cosmic web, turbulent intracluster gas motions
have not yet enhanced the magnetic field; its
strength thus still depends on the seed field intensity, in contrast to
galaxy clusters, where it probably mostly reflects the present level
of turbulence (see e.g., \citealt{DDLM2009,XLCLN2010,XLCLN2011}).
Intervening magneto-ionic media cause a difference in the phase velocity 
between the left-handed and right-handed circular
polarization components of the linearly polarized
synchrotron radiation emitted by a background radio source (e.g. \citealt{CT2002,GF2004}). This effect
translates into a rotation of the intrinsic polarization angle,
$\psi_0$,
\begin{equation}
\psi(\lambda^2)=\psi_0+\phi\lambda^2.
\end{equation}
Following \cite{B1966}, the observed polarization angle, $\psi$, depends
on the observation wavelength, $\lambda$, through the Faraday depth,
$\phi$,
\begin{equation}
\phi=a_0\int_0^{z_{\rm s}}B_l(z)~n_{\rm e}(z)~\frac{\mathrm{d}l}{\mathrm{d}z}~\mathrm{d}z,
\label{definition}
\end{equation}
where $a_0$ depends only on fundamental constants, 
$n_{\rm e}$ is the electron density, $B_l$ the component of the
magnetic field along the line of sight, and $z_{\rm s}$ the redshift
of the source.  When the rotation is completely due to a foreground
screen, the Faraday depth has the same value as the rotation measure
(RM), defined by
\begin{equation}
{\rm RM}\equiv\frac{\partial \psi}{\partial \lambda^2}.
\end{equation}
The Faraday depth is assumed to be positive when the line of sight average component of the magnetic field
points toward the observer, otherwise it is negative for a field with an average
component pointing away from the observer.
The amount of Faraday depth
measured by radio observations along a given line of sight is the sum
of the contributions from the Milky Way, the emitting radio source
and any other sources and large-scale structures in between hosting a magnetized plasma. 
The investigation of these contribution and of their possible dependence
on redshift is essential in order to discriminate among the different
scenarios of magnetic field formation and evolution and therefore 
crucial for the understanding of cosmic magnetism. Sensitive observations, a good
knowledge of the Galactic Faraday foreground screen, and a statistical approach able
to properly combine all the observational information are necessary.
An all-sky map of the Galactic Faraday rotation foreground and an
estimate of the overall extragalactic contribution has been derived by
\cite{OJR2012,OJG2014} in the framework of \emph{Information Field
  Theory} \citep{EFK2009}, by assuming a correlated Galactic
foreground and a completely uncorrelated extragalactic term.
In this paper, we propose a
new, fully Bayesian, approach aiming at further disentangling the
contribution intrinsic to emitting sources from the contribution due to the
intergalactic environment between the source and the observer, and
at investigating the dependence of these contributions on redshift.

The first direct proof of the existence
of magnetic fields in large-scale extragalactic environments, i.e.,
galaxy clusters, dates back to the 1970s with the discovery of
extended, diffuse, central synchrotron sources called radio halos (see
e.g. \citealt{FGGM2012} for a review). Later, indirect evidence of the existence
of intracluster magnetic fields has been given by several statistical studies
on the effect of the Faraday rotation on the radio signal from background galaxies
or galaxies embedded in galaxy clusters \citep{LD1982,VMB1986,CKB2001,JH2003,C2004,JHE2004}.
On scales up to a few Mpc from the nearest galaxy cluster, possibly
along filaments, only a few diffuse synchrotron sources have been reported
\citep{HSWD1993,BEM2002,KKSP2007,GVG2013,GBB2015}. Magnetic fields with strengths of the order of
$10^{-15}$\,G in voids might be indicated by $\gamma$-ray observations (see
\citealt{NV2010,TGF2010,TMII2012,TMIIT2013}, but see \citealt{BPPCS2014,BPPC2014} for
alternative possibilities). Nevertheless, up to now, a robust confirmed detection
of magnetic fields on scales much larger than clusters is not available.
\cite{SNDB2010} and \cite{AGR2014} investigated the
possibility to statistically measure Faraday rotation from
intergalactic magnetic fields with present observations, showing that
only the Square Kilometre Array (SKA) and its pathfinders are likely
to succeed in this respect.
By comparing the observations with single-scale magnetic field simulations,
\cite{PTU2015} infer an upper limit of 1.2\,nG for extragalactic large-scale
magnetic fields, while the \cite{PC2015} derived a more stringent upper limit for
primordial large-scale magnetic fields of $B<0.67$\,nG from the analysis of the
CMB power spectra and the effect on the ionization history (but see also \citealt{TIOH2005}; \citealt{ITOHS2006}). 

A number of authors
examined a possible dependence of extragalactic Faraday depths on the redshift of the observed radio source, but no firm
conclusion has yet been drawn.
\cite{KP1982} found an increased variance of the Faraday depth in
conjunction with higher redshifts, as also found in some later studies
(e.g. \citealt{WPK1984,KBM2008}). However, \cite{OW1995} did not find
any evidence of an increase of the variance of the Faraday depth
as a function of the redshift, as suggested also by the
recent work by \cite{HRG2012} and \cite{PTU2015}.

The rest of this paper is organized as follows.  In \S\,\ref{approach}
we describe the theory behind our method. In \S\,\ref{allresults}
the tests performed are presented, with predictions for the
new generation of radio interferometers.
Moreover, we outline a
generalization of the algorithm in order to discriminate the
contribution from different large-scale structures
along the line of sight. Finally, in \S\,\ref{conclusions} we draw our
conclusions.  The application of the algorithm to real data
is left for future work, as explained in the text.  In the
following we adopt a $\Lambda$CDM cosmology with
$H_0=67.3$\,km\,s$^{-1}$Mpc$^{-1}$, $\Omega_{\rm m}$=0.315,
$\Omega_{\Lambda}$=0.685, and $\Omega_{\rm c}$=0.0 \citep{PC2014}.

\begin{figure}[h!]
  \centering
  \includegraphics[width=9cm, angle=0]{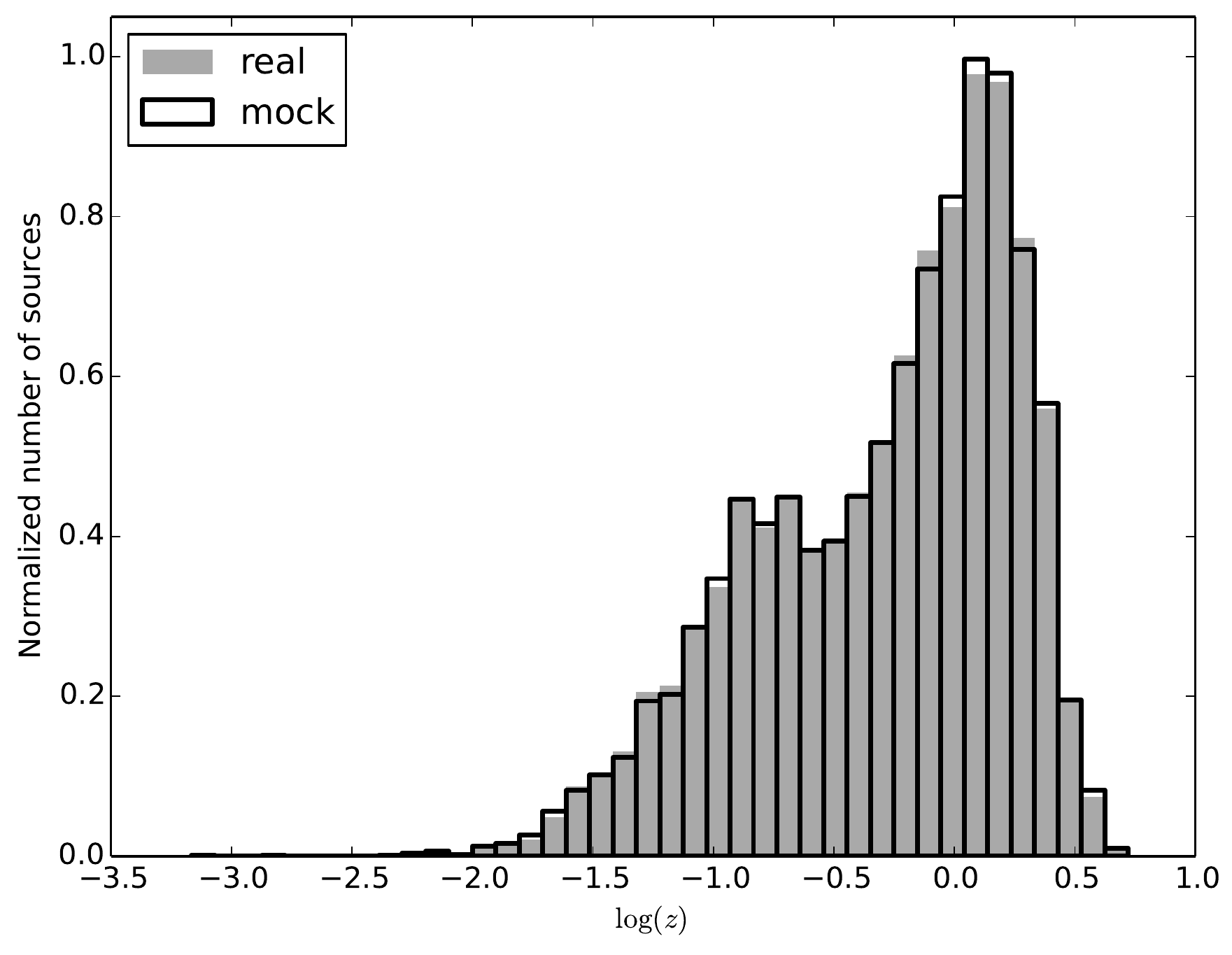}\\
  \includegraphics[width=9cm, angle=0]{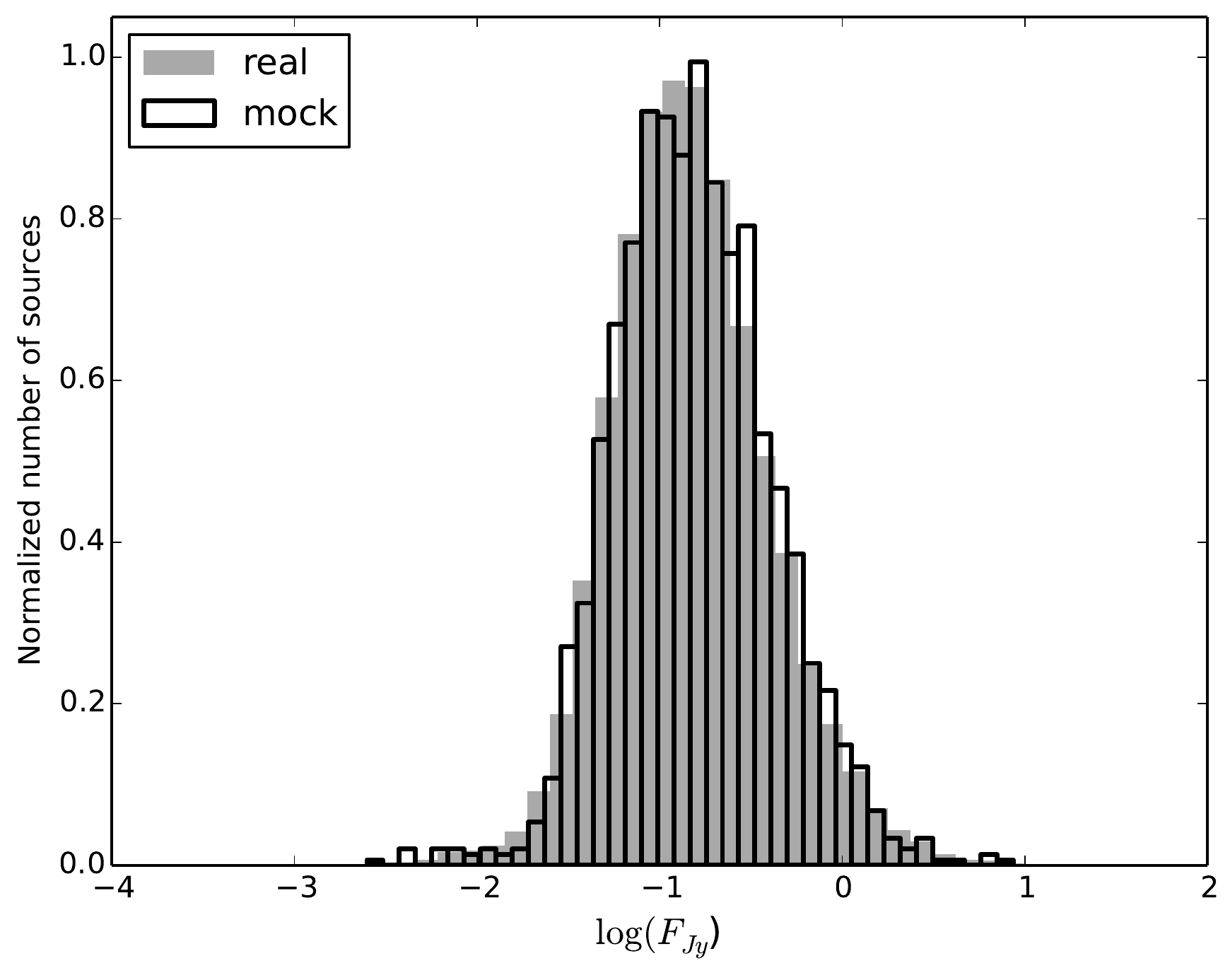}\\
  \includegraphics[width=9cm, angle=0]{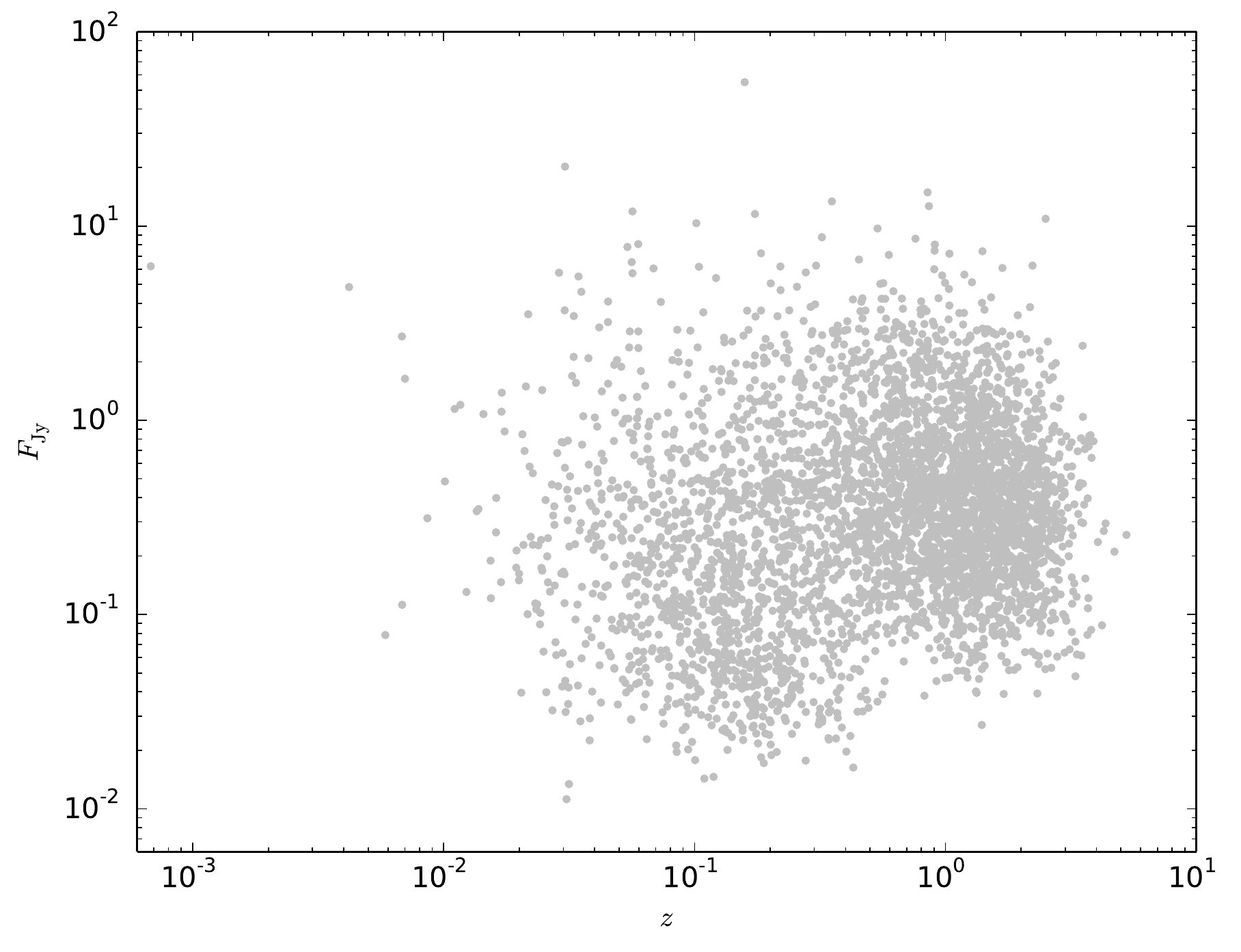}

    \caption{Distribution of the real and mock sample of redshifts $z$
      (\emph{top}), and flux densities $F_{\rm Jy}$ at 1.4\,GHz (\emph{middle}).
      In the \emph{bottom} panel, flux density versus redshift for the sources with both measurements available (\citealt{TSS2009}; \citealt{HRG2012}). }
  \label{reddistr}
\end{figure}

\section{Theoretical framework}
\label{approach}

The probability for the extragalactic contribution, $\phi_{\rm e}$, along the line of sight,
$i$, to take on a value within the infinitesimal interval between $\phi_{{\rm    e},i}$ and
$\phi_{{\rm    e},i}+\mathrm{d}\phi_{{\rm    e},i}$ given the data, $d$, and a model
parameterization, $\Theta$,  is given by the probability density distribution.
In turbulent environments (see, e.g., \citealt{LD1982}; \citealt{T1991}; \citealt{FDGT1995}; \citealt{F1996},
for galaxy clusters), this probability density
distribution can be represented with a Gaussian.
The mean and variance,
\begin{eqnarray}
\langle\phi_{{\rm e},i}\rangle&=&\langle \phi_{{\rm e}, i}\rangle_{(\phi_{{\rm e}, i}|d,\Theta)}\\
\langle\phi_{{\rm    e},i}^{2}\rangle&=&\langle (\phi_{{\rm e}, i}-\langle \phi_{{\rm e}, i}\rangle_{(\phi_{{\rm e}, i}|d)})(\phi_{{\rm e}, i}-\langle \phi_{{\rm e}, i}\rangle_{(\phi_{{\rm e}, i}|d)})^{\dag}\rangle_{(\phi_{{\rm e}, i}|d, \Theta)},\nonumber%
\end{eqnarray}
 of this distribution  are
\begin{eqnarray}
\langle\phi_{{\rm e},i}\rangle&=&0\\
\langle\phi_{{\rm
    e},i}^{2}\rangle&=&a_0^2\int_{0}^{z_{i}}\int_{0}^{z_{i}}\frac{\mathrm{d}l}{\mathrm{d}z}\frac{\mathrm{d}l^{\prime}}{\mathrm{d}z^{\prime}}\mathrm{d}z
\mathrm{d}z^{\prime}\langle n(z) n(z^{\prime}) B_{l}(z) B_{l}(z^{\prime}).\nonumber
\label{overall}
\end{eqnarray}
The zero mean results from the fact that we do not have any reason to suppose either a positive or a negative mean Faraday depth value. 
Frozen-in magnetic fields are expected to have strengths depending on
redshift $z$.  Because of the expansion of the Universe, lengths are
stretched
\begin{equation}
  \frac{l_{\rm 0}}{l}=(1+z).
  \label{length}
\end{equation}
In an isotropically expanding fluid with no significant
fluctuations, the mean electron density evolves as
\begin{equation}
  \frac{\langle n_{\rm e}\rangle}{\langle n_{\rm 0} \rangle}=(1+z)^3,
  \label{density}
\end{equation}
and, if the magnetic flux density is conserved, the magnetic field strength as
\begin{equation}
  \frac{\langle B\rangle}{\langle B_{\rm 0}\rangle}=\left(\frac{\langle n_{\rm e}\rangle}{\langle n_{\rm 0}\rangle}\right)^{\frac{2}{3}}=(1+z)^2,
  \label{mf}
\end{equation}
where $l_0$, $\langle n_0\rangle$, and $\langle B_0\rangle$ are the present-day values, and
$l$, $\langle n_{\rm e}\rangle$, and $\langle B\rangle$ are the values at the time
when the signal was emitted by the source.  
If we consider these assumptions and we define a length scale
$\Lambda_{l}:=\int \mathrm{d}l^{\prime}\langle B(l)B(l^{\prime})\rangle/\langle B(l)^2\rangle$, 
we obtain 
\begin{equation}
\langle\phi_{{\rm e}, i}^2\rangle\approx a_0^2\int_0^{z_{i}}\langle n_{\rm 0}^2\rangle\langle B_{l{\rm 0} }^2\rangle\Lambda_{l {\rm 0} }(1+z)^4\frac{c}{H(z)}\mathrm{d}z.
\label{diffenv}
\end{equation}
For the derivation of this expression see Appendix\,\ref{derivation}.
Here, we assumed an unstructured Universe and used the definition of proper displacement along a
light-ray, $\frac{\mathrm{d}l}{dz}=\frac{c}{(1+z)H(z)}$. 
Moreover, we assumed that, within a
correlation length $\Lambda_{l {\rm 0}}$, the redshift can be
approximated to be constant. Magnetic field strength and structure as
well as the electron density have different values in different
environments, $j$. In the following, we assume them to depend only on
the environment and not on the location within an environment. 
This simplification renders the problem feasible.

In this paper, as a first step, we will restrict our analysis to a two components (scenario 2C),
the emitting radio source itself, whose contribution is $\sigma_{{\rm
    int},i}^2(z_i, \Theta)$, and the medium between the source and the
Galaxy, whose contribution is $\sigma_{{\rm env}, i}^2(z_i, \Theta)$, such that
\begin{equation}
\langle\phi_{{\rm e},i}^{2}\rangle=\sigma_{{\rm e},i}^2(z_i, \Theta)=\sigma_{{\rm int},i}^2(z_i, \Theta)+\sigma_{{\rm env},i}^2(z_i, \Theta).
\label{split}
\end{equation}
We denote with $\Theta=\{\sigma_{{\rm int,0}},\sigma_{{\rm env,0}},\chi_{\rm lum}, \chi_{\rm red}\}$ an $N$-dimensional vector,
where $N$ is the number of parameters
used in the representation of the Faraday depth variance.
These are the parameters we want to infer.
We choose the following parameterization for the intrinsic contribution to the variance in Faraday depth, 
\begin{equation}
\sigma_{{\rm int},i}^2(z_i, \Theta)=\left( \frac{L_i}{L_0}\right)^{\chi_{\rm lum}}\frac{\sigma_{{\rm int,0}}^2}{(1+z_i)^{4}}
\label{intr}
\end{equation} 
where $L_i$ is the luminosity\footnote{This luminosity refers to the mean-frequency of the frequency band used for the computation of the Faraday depth values.} of the source $i$, $L_0=10^{27}\,{\rm W/Hz}$, and $\chi_{\rm lum}$ absorbs possible dependencies on the luminosity of the source since faint sources may not be detected, and for the environmental contribution
\begin{equation}
\sigma_{{\rm env},i}^2(z_i, \Theta)=\frac{D_i(z_i, \chi_{\rm red})}{D_0}\sigma_{{\rm env,0}}^2,
\label{env}
\end{equation}
where $D_0=1$\,Gpc, and $D_i(z,\chi_{\rm red})$ is defined as
\begin{equation}
D_i(z_i, \chi_{\rm red})=\int_0^{\rm z_{i}}\frac{c}{H(z)}(1+z)^{4+\chi_{\rm red}}\mathrm{d}z,
\label{coeff}
\end{equation}
to capture the redshift scaling of Eq.\,(\ref{diffenv}) as well as auxiliary modifications by the clumpiness of the Universe via $\chi_{\rm red}$.
Since the length of the path covered by the signal in the source is unknown,
we factored it in $\sigma_{{\rm int,0}}^2$ in
Eq.\,(\ref{intr}).  We note that $\sigma_{{\rm int,0}}$ and $\sigma_{{\rm env,0}}$
have been assumed to be independent of the redshift.  In
Eq.\,(\ref{intr}) the only redshift dependence is absorbed by the
factor $(1+z_i)^{-4}$ that takes into account the effect of redshift
on Faraday rotation (squared), while in Eq.\,(\ref{env}) it is absorbed by 
$D_i(z_i, \chi_{\rm red})$ which takes into account Eq.\,(\ref{coeff}).

This parameterization describes the simplest scenario. For more complex scenarios that include three components, we refer to Appendix\,\ref{alternative_app}, where we introduce an additional constant and latitude dependent term. 
Moreover, galaxies along the line of sight between a source and the observer
can be responsible for high Faraday depth values (e.g. \citealt{KP1982,WPK1984,KBM2008,BML2012}),
indicating magnetic field strengths in these intervening galaxies
of (1.8$\pm$0.4)\,$\mu$G (\citealt{FOCG2014}).
The rotation
of the polarization angle due to these sources adds to that associated
with the large-scale structure and, therefore, should be taken into
account in a proper modeling. Since
the aim of this paper is to give a proof of concept, we leave this for future work.

\subsection{Bayesian inference}
\label{Bayesian inference}
To constrain the vector $\Theta=\{\sigma_{{\rm int,0}},\sigma_{{\rm env,0}},\chi_{\rm lum}, \chi_{\rm red}\}$
on the basis of these data, $d$, we propose a Bayesian approach.
The \emph{posterior} probability distribution, $P(s|d)$, on a signal, $s$, after a
dataset, $d$, is acquired, can be expressed with Bayes' theorem,
\begin{equation}
P(s|d)=\frac{P(d|s)P(s)}{P(d)}.
\label{ourposterior}
\end{equation}
The \emph{prior} probability distribution, $P(s)$, is modified by the data, $d$, through
the \emph{likelihood}, $P(d|s)$. The \emph{evidence}, $P(d)$, is a
normalization factor, obtained by marginalizing the joint probability,
$P(d,s)=P(d|s)P(s)$, over all possible configurations of the signal,
$s$.

In this context, the data, $d$, can be represented as a vector with
elements $d_{i}$, with $i=1,...,N_{\rm los}$, where $N_{\rm los}$ is
the total number of lines of sight. Each measurement, $d_{i}$, is the
Faraday depth evaluated in the direction of the source, $i$, and is the
result of the sum of a Galactic and an extragalactic contribution,
$\phi_{{\rm g},i}$ and $\phi_{{\rm e},i}$, and the noise, $n_{i}$, of
the measurement process,such that
\begin{equation}
d_{i}=\phi_{\rm g, i }+\phi_{\rm e, i}+ n_{i}.
\end{equation}
From the observed data, the Galactic foreground should be removed as
well as possible to reveal the extragalactic contribution. However,
any estimation of the Galactic foreground is based on the same data
and is facing the separation problem for Galactic and extragalactic
contributions. The only available discriminant (so far) is the large
angular correlation the Galactic contribution shows. This allow for
a statistical separation and Galactic model construction. Such a model
will inevitably have uncertainties and correlations among such
uncertainties, which have to be properly taken into account in a
statistical search for extragalactic Faraday signals.  To this end,
\cite{OJR2012,OJG2014} developed a fully Bayesian approach to
reconstruct the Galactic Faraday depth foreground and estimate the
extragalactic contribution as well as the involved uncertainties by
using the Faraday depth catalogs available to date.  Their posterior for the
extragalactic contribution can be used in order to further disentangle
the intrinsic and environmental contributions.  
\cite{OJR2012,OJG2014}'s analysis relies on the assumption that
for each source the prior knowledge can be described by a Gaussian
probability density distribution,
\begin{equation}
P(\phi_{\rm e})=\prod_{i=1}^{\rm N_{\mathrm{los}}}\mathcal{G}(\phi_{{\rm e}, i}, \sigma_{\rm e}^2),
\label{priorniels}
\end{equation}
with a standard deviation, $\sigma_{\rm e}\approx 7$\,rad\,m$^{-2}$,
irrespective of the line of sight\footnote{The notation
  $\mathcal{G}(x,X)$ indicates a one-dimensional Gaussian distribution
  for a variable $x$ with zero mean and variance $X$.}.  Here, on the
other hand, we want to test whether the variance is different for each
line of sight, depending on the redshift of the source, according to
Eq.\,(\ref{intr}) and Eq.\,(\ref{env}). We note that in our approach
the angular separation of sources is assumed to be large enough that
the magnetic fields probed by different lines of sight can be modeled
as uncorrelated.  All angular correlations of Faraday depth on
scales down to the effective resolution 
of the catalog ($\approx$0.5$^{\circ}$), 
are absorbed by the
Galactic component in this model. A more complete modeling would
require to take into account possible correlations in the
extragalactic component. Thus, our assumptions imply that the values
derived with the proposed
algorithm for the contributions of different environments to the Faraday depth 
dispersion are a lower limit.

We cannot follow the prescription described in Appendix D.2.3 of
\cite{OJG2014} because our prior assumptions are too different from
theirs. Instead, we resort to Gibbs sampling \citep{GG1984,WLL2004}. 
This approach relies on the fact that sampling from the conditional probability densities,
\begin{equation}
\phi_{\rm e}\hookleftarrow P(\phi_{\rm e}|\Theta,d),
\label{cpphie}
\end{equation}
and,
\begin{equation}
\Theta \hookleftarrow P(\Theta|\phi_{\rm e},d),
\label{cpchi}
\end{equation}
in a two-step iterative process is equivalent to drawing samples from
the joint probability density
\begin{equation}
\phi_{\rm e}, \Theta \hookleftarrow P(\Theta,\phi_{\rm e}|d),
\end{equation}
if the process is ergodic.

For the parameters $\sigma_{{\rm int},0}$ and $\sigma_{{\rm env},0}$ we choose a prior,
\begin{equation}
P(\sigma_{{\rm int},0},\sigma_{{\rm env},0})\propto {\rm const}. 
\end{equation}
Results with other priors are
discussed in Appendix\,\ref{priorappendix}.  Conversely, we do not expect the parameters
$\chi_{\rm lum}$ and $\chi_{\rm red}$ to differ greatly from $0$,
since we have already accounted for all obvious redshift effects.
This requirement is satisfied if we use the following Gaussian priors
\begin{equation}
  P(\chi_{{\rm lum}},\chi_{{\rm red}})=\mathcal{G}(\chi_{{\rm lum}},1)\mathcal{G}(\chi_{{\rm red}},1),
  \label{gaussprior}
\end{equation}
and in combination
\begin{equation}
P(\Theta)=P(\sigma_{{\rm int},0},\sigma_{{\rm env},0})P(\chi_{{\rm lum}},\chi_{{\rm red}}).
\end{equation}

\subsection{Description of the algorithm}
Here, we describe the Gibbs sampling procedure mentioned in the
previous section. 
We run the algorithm starting from values of the parameters $\Theta$
randomly drawn from their prior.  This $\Theta$ vector is used to
compute a variance for the prior of the extragalactic contribution,
\begin{equation}
P(\phi_{\rm e}|\Theta)=\prod_{i=1}^{\rm N_{\mathrm{los}}}\mathcal{G}(\phi_{{\rm e}, i}, \sigma_{{\rm e},i}^2(z_i,\Theta)),
\end{equation}
where $\sigma_{{\rm e},i}^2(z_i,\Theta)$ is evaluated according to
Eq.\,(\ref{split}). A sample for the extragalactic contribution is
drawn from the posterior
\begin{equation}
P(\phi_{\rm e}|\Theta,d)=\frac{P(d|\phi_{\rm e},\Theta)P(\phi_{\rm e}|\Theta)}{P(d|\Theta)},
\label{nielspost}
\end{equation}
following the approach described in \cite{OJG2014}, with the Galactic
power spectrum, the Galactic profile, and the correction factors to
the observed noise variance (indicated by $\eta_i$ in the paper by
\citealt{OJG2014}) fixed to the published values.  After fixing the
extragalactic sample, a new $\Theta$ sample is drawn from the
conditional probability
\begin{equation}
P(\Theta|\phi_{\rm e})=\frac{P(\phi_{\rm e}|\Theta)P(\Theta)}{P(\phi_{\rm e})}\propto\prod_{i=1}^{\rm N_{\mathrm{los}}} \mathcal{G}(\phi_{{\rm e}, i}, \sigma_{{\rm e}, i}^2(z_i,\Theta)) P(\Theta).
\end{equation}
Here, we drop the dependence on the data, $d$, because $\Theta$ and $d$
are conditionally independent given $\phi_{\rm e}$.  To sample from
this distribution we use a Metropolis-Hastings algorithm (\citealt{MRRT1953,H1970}). 
When direct sampling is difficult, Metropolis-Hastings algorithms can approximate a probability distribution with random samples generated from the distribution itself. At each iteration a step in the parameter space is proposed according 
to a transition kernel and then accepted according to an acceptance function. If the proposed step is not accepted, the old $\Theta$ values are kept and used
to draw a new sample of the extragalactic Faraday depths.

The convergence criteria adopted in this work are
described in Appendix~\ref{convergenceappendix}.

 \begin{table*}[!t]
        \caption{Mock catalogs. Col.\,1: scenario considered for the generation of the mock catalog. Col.\,2: identification code (ID) of the test. Col.\,3: short description of the catalog. Col.\,4: number of lines contained in the catalog. Col.\,5: Observational uncertainty in Faraday depth. Col.\,6: Figure where the results of the corresponding test are shown. }
        \centering
        \begin{tabular}{ccccccc}
          \hline
          \hline
          Scenario &  ID & Description           &   los \#    & $\sigma_{\rm noise}^{a}$ &Figure   \\
          &  &          &     & $({\rm rad}/{\rm m}^2)$ &   \\

            \hline
            Two Component       &   2C1          & Datasets used by \cite{OJR2012,OJG2014}$^b$              &   41632         & 13.0         &\ref{2comp}(a) \\
                                &   2C2          & Datasets used by \cite{OJR2012,OJG2014}$^b$              &   4003          & 13.0         &\ref{2comp}(b)\\
                                &   NPC          & LOFAR 120-160\,MHz, North Polar Cap          &   2148          & 0.05        &\ref{lofar_all}(a,b) \\
                            
                                &   GW           & LOFAR 120-160\,MHz, Great Wall               &   1036          & 0.05        &\ref{lofar_less} \\
                                &   POSSUM       & ASKAP-POSSUM 1130-1430\,MHz, South Polar Cap &  3476 & 6.0         &\ref{possum}\\
                                &   B2SPC1       & SKA Band 2, 0.65-1.67\,GHz, South Polar Cap  &  3476 & 0.8         &\ref{skamid}(a,b)\\
                                &   B2SPC2       & SKA Band 2, 0.65-1.67\,GHz, South Polar Cap  &  1129 & 0.8         &\ref{ska_even_less_loss}(b) \\
                                &   B1SPC1       & SKA Band 1, 0.35-1.05\,GHz, South Polar Cap  &  3476 & 0.3         &\ref{skalow}(a,b)\\
                                &   B1SPC2       & SKA Band 1, 0.35-1.05\,GHz, South Polar Cap  &  1129 & 0.3         &\ref{ska_even_less_loss}(a)\\
            Three Component     &   3C1       & Datasets used by \cite{OJR2012,OJG2014}$^b$                 &   41632         & 13.0        &\ref{3comp}(a)\\
                                &   3C2       & Datasets used by \cite{OJR2012,OJG2014}$^b$                 &   4003          & 13.0        &\ref{3comp}(b)\\
            Latitude Dependence &   LD1       & Datasets used by \cite{OJR2012,OJG2014}$^b$                 &   41632         & 13.0        &\ref{latdep}(a)\\
                                &   LD2       & Datasets used by \cite{OJR2012,OJG2014}$^b$                 &   4003          & 13.0        &\ref{latdep}(b)\\
            Prior 0             &   P0        & Datasets used by \cite{OJR2012,OJG2014}$^b$                 &   41632         & 13.0        &\ref{prior}(a)\\
            Prior 1             &   P1        & Datasets used by \cite{OJR2012,OJG2014}$^b$                 &   41632         & 13.0        &\ref{prior}(b)\\
            
            \hline
            \hline
        \end{tabular}
        \label{tab:A}\\
        \footnotesize $^a$ For present observations this value represents the mean value of the observed uncertainties. \normalsize
        \footnotesize $^b$ The references for the surveys and catalogs used by \cite{OJR2012} are: \cite{D1979,TI1980,SNKB1981,LD1982,RJ1983,KTIA1987,BMV1988,HOE1989,KTK1991,CCSK1992,W1993,OW1995,MS1996,CCG1998,GVMZK1998,VGKM1999,CKB2001,GDMG2001,BTJ2003,JH2003,KMGV2003,TGP2003,C2004,JHE2004,GHSS2005,MGDG2005,RRS2005,HGMDG2006,BHG2007,BOMKB2007,MGS2008,FEM2009,HBE2009,TSS2009,BFM2010,MGH2010,FCE2011,VEBS2011}
\normalsize

    \end{table*}

 \section{Results}
 \label{allresults}
 
 In the following we present tests performed with different Faraday depth catalogs. These catalogs differ in the number of components used to generate the overall extragalactic Faraday depth, the numbers of lines of sight in the sky, and the observational uncertainties that are different for the different radio surveys considered here.
In Sect.~\ref{tests} we demonstrate that the algorithm is working properly for the two-component scenario described in Sect.\,\ref{approach}. In Sect.\,\ref{pred}, we present the prospects with the surveys planned with the new generation of radio interferometers (LOFAR, ASKAP, and SKA).

We perform tests assuming an overall extragalactic Faraday depth in agreement with the values presently inferred, namely $\approx 7.0$\,rad\,m$^{-2}$ (\citealt{S2010}; \citealt{OJG2014}), and comparable intrinsic and the environmental contributions. To satisfy these two requirements, we need to use slightly different values of the $\Theta$ parameters for surveys with different frequency specifications. Indeed, the contributions depend on the frequency through the luminosity of the source, see Eq.\,(\ref{intr}). Our choice of the $\Theta$ parameters translates to a strength of magnetic fields intrinsic to the source of
\begin{equation}
  \frac{\langle B_{l 0}\rangle}{{\rm \mu G}}\sim 0.5\div 1\left(\frac{\langle n_0\rangle}{10^{-3}\,{\rm cm^{-3}}}\right)^{-1}  \left(\frac{\Lambda_{l 0}}{5\, {\rm kpc}}\right)^{-1}  \left(\frac{L}{100\, {\rm kpc}}\right)^{-1}, 
\label{Bi}
\end{equation}
where $L$ is the size of the emitting radio source, and to large scale magnetic field strengths of
\begin{equation}
  \frac{\langle B_{l 0}\rangle}{{\rm nG}}\sim 2\left(\frac{\langle n_0\rangle}{10^{-5}\,{\rm cm^{-3}}}\right)^{-1} \left(\frac{\Lambda_{l 0}}{5\,{\rm Mpc} }\right)^{-1}.
\label{Be}  
\end{equation}
In the tests for LOFAR, ASKAP, and SKA, we additionally consider an overall extragalactic Faraday rotation of
$\approx 0.7$\,rad\,m$^{-2}$ to mimic weaker fields. This corresponds to
magnetic field values weaker by a factor ten than those given in
Eq.\,(\ref{Bi}) and Eq.\,(\ref{Be}).

The two-component parameterization represents the simplest scenario. Nevertheless, the algorithm is able to successfully deal also with more complex scenarios that include a third constant or latitude-dependent component. These scenarios are discussed in Appendix\,\ref{alternative_app}. 
Moreover, to asses if different priors can have an impact on our results, we consider in Appendix\,\ref{priorappendix} a flat prior in $\sigma^2$ and a flat prior in $\ln(\sigma^2)$.
In Table~\ref{tab:A} we give a summary of all the setups we used, including those presented in the appendices.
To each of them we
assigned an identification code (ID), used in 
the paper to discriminate among the different scenarios. When the same
collection of sources is used for tests with different values of the overall
extragalactic Faraday depth, we distinguish among them by adding a
roman letter. For example, X\emph{a} and X\emph{b} indicate tests
performed using the collection of sources X and two different overall extragalactic
Faraday depth standard deviation, \emph{a} denotes $\approx 7.0$\,rad\,m$^{-2}$, while \emph{b} $\approx 0.7$\,rad\,m$^{-2}$.
In Table~\ref{tab:C} we report all the values of the $\Theta$ parameters adopted
in the different tests. 
In the main text we give both a quantitative and a visual summary of the results of all the tests we performed, while the full posteriors are shown only for the most important tests we carried out. To make the reading easier, the posteriors for all the other tests are shown in the appendices.

 \subsection{Present-instrument observations}
\label{tests}
To assess the quality of the algorithm we generate a mock
catalog for the sample of sources used by
\cite{OJR2012,OJG2014}. This mock catalog includes coordinates, redshifts, luminosities, and 
Faraday depth values.

The positions of the sources on the sky were kept the same as for the real sources.
The majority ($\approx$40000 sources) belong to the
catalog of \cite{TSS2009}, and for 4003 of them, spectroscopic
redshift measurements have been published by \cite{HRG2012}.  For most
of the sources, these catalogs give a flux density measurement that allows
the computation of the luminosity of the source.  Where available, we
use the measured redshift and flux density.
For the vast majority of the sources we generate a mock redshift, and for a few of them
a mock flux density value. Mock redshifts and flux densities are extracted independently from the
two observed distributions.  In Fig.\,\ref{reddistr}, the distribution of both
the real and mock sample of redshifts and flux densities is shown respectively
in the \emph{top} and \emph{middle} panel. In the \emph{bottom} panel
the observed flux density versus redshift distribution is presented.  
These two quantities appear to be weakly correlated.
For sake of simplicity we neglected such correlation in our mock simulation,
since it should not have any impact on our analysis. 
We assume all redshifts and luminosities to be known
with negligible uncertainty.

For all the sources in the catalog we generate a
mock Faraday depth value. 
The observed Faraday depth values consist of a Galactic, an
extragalactic, and a noise contribution.  We considered the Galactic
contribution to be given by a sample extracted from the posterior of
\cite{OJG2014}. To mimic observational uncertainties, the noise variance has been
computed for each source according to Eq.\,(37) in \cite{OJG2014},
where as observed uncertainty, $\sigma_i$, we use the uncertainties reported in the
observational catalogs and as $\eta_{i}$ we use the values recovered
by \cite{OJG2014}. The observational error of each measurement has
been extracted from a Gaussian with this standard deviation and zero-mean.
Concerning the extragalactic contribution, in this test, we consider the 2C scenario described in \S\,\ref{approach}, namely an intrinsic
and an environmental contribution,
\begin{equation}
  \sigma_{{\rm e}, i}^2(z_i,\Theta)=
  \left(\frac{L}{L_0}\right)^{\chi_{\rm lum}}\frac{\sigma^2_{{\rm int},0}}{(1+z_i)^{4}}+ \frac{D_i(z_i)}{D_0}\sigma^2_{{\rm env},0}.
\label{twocomp}
\end{equation}

\begin{figure*}[ht]
       \centering
            \begin{overpic} [width=15.5cm, angle=0]{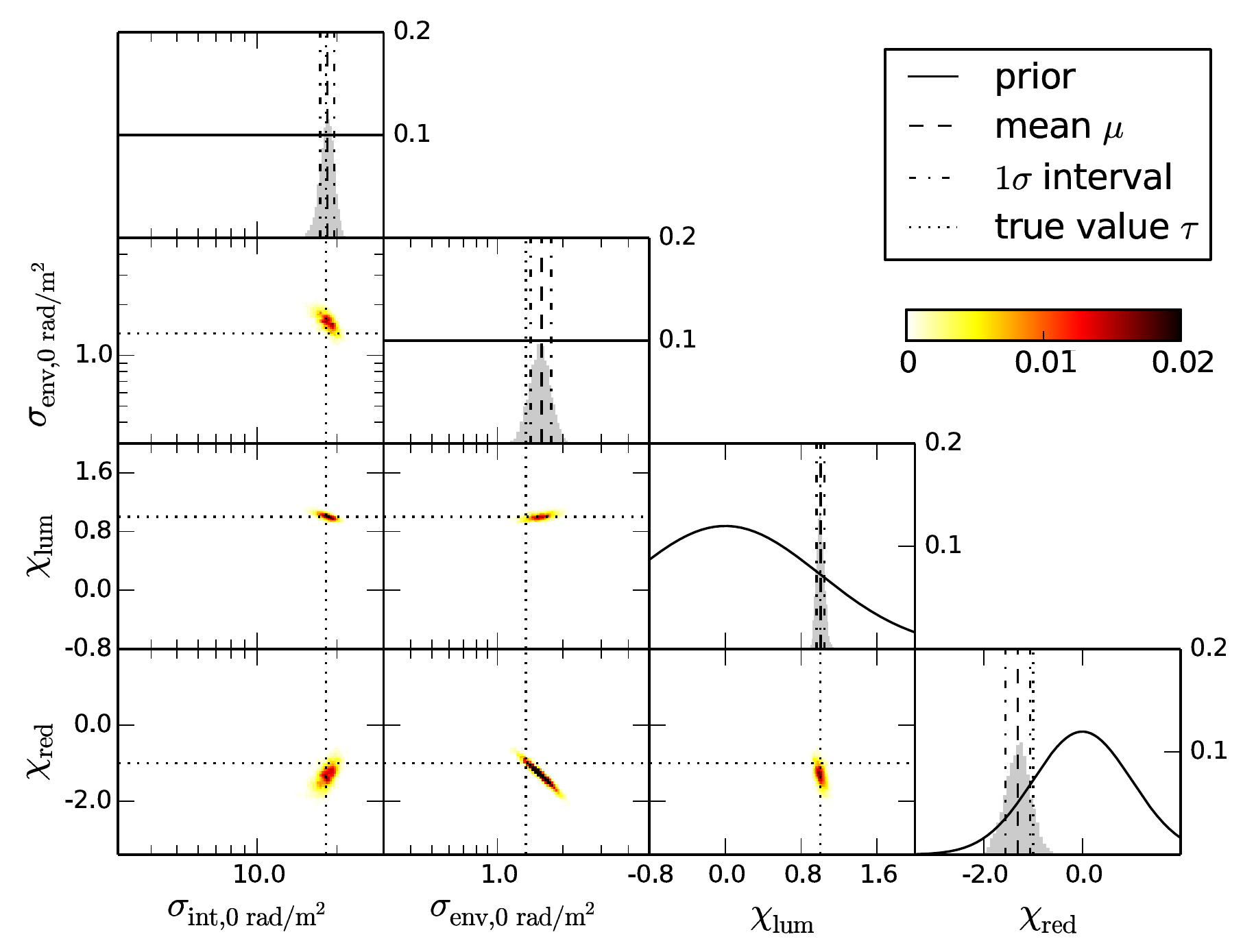} \put(-4,78){(a)} \end{overpic} \\
            \begin{overpic} [width=15.5cm, angle=0]{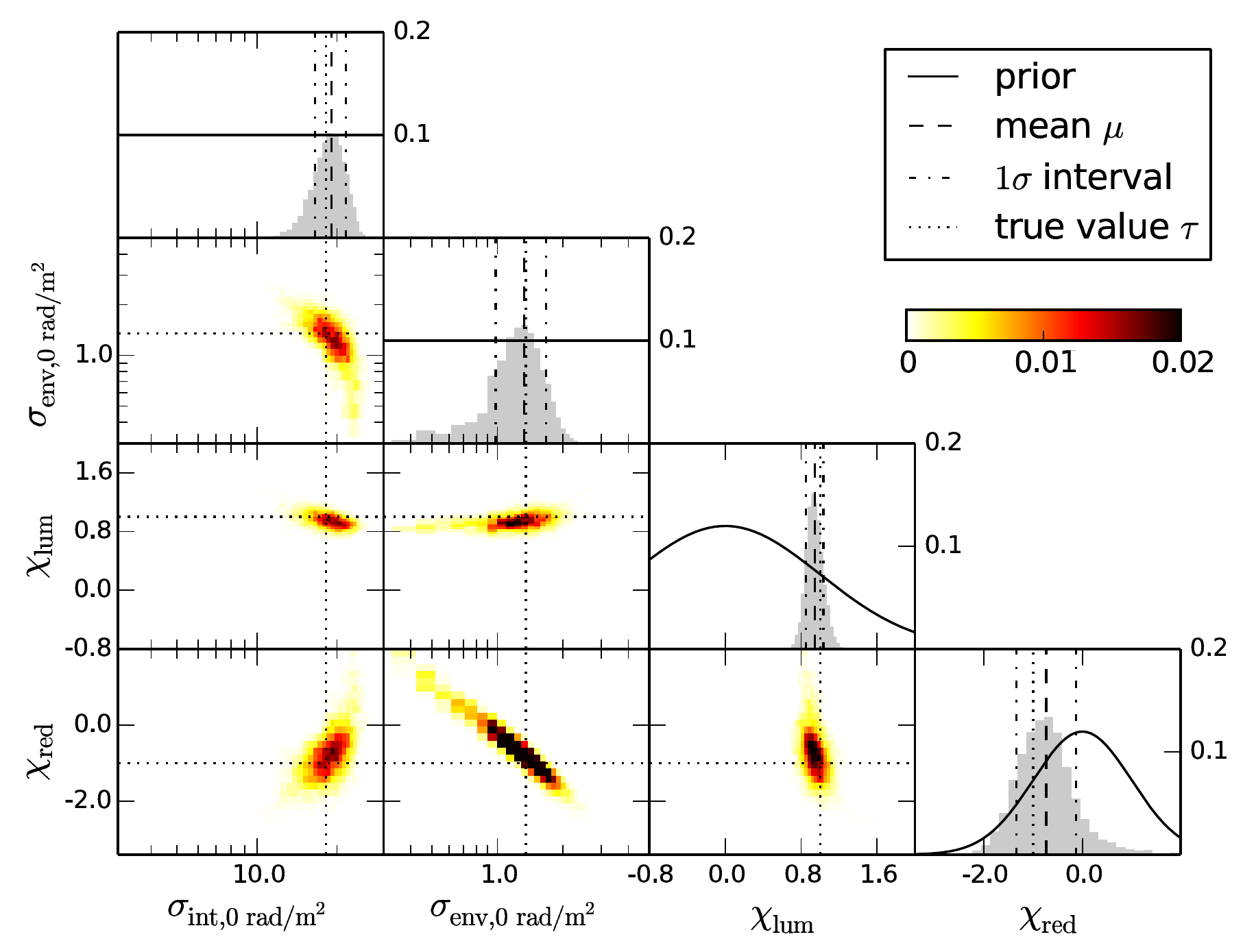} \put(-4,78){(b)}\end{overpic} \\
        \flushleft       
        \caption{
Results obtained with a two-component scenario for 41632  (2C1)
      lines of sight in panel (a) and for 4003 (2C2) in panel (b). In each panel the top plots of each column  
      show the 1-dimensional projection of the posterior as well
      as the true value (dotted line), the outcome of the analysis (dashed and dashed-dotted lines),
      the prior (continuous line). The panels in color show the 2-dimensional
      marginalized views of the posterior as sampled.}
  \label{2comp}
\end{figure*}
The variances in Eq.\,(\ref{twocomp}) depend on the redshifts and
luminosities of the sources. Therefore, each source will have a
different variance. For each source, the extragalactic contribution is
extracted from a Gaussian with this variance and zero mean.
As summarized in Table~\,\ref{tab:C}, the contributions intrinsic to the
source and  to the medium between the source and the observer are respectively, $\tau_{\sigma_{\rm int, 0}}= 18.2$\,rad\,m$^{-2}$ and $\tau_{\sigma_{\rm env, 0}}= 1.4$\,rad\,m$^{-2}$. Since the mean of the factor $L_i/L_0/(1+z_i)^4$ is $\approx 0.06$ and the mean of the factor $D_i/D_0$ is $\approx 15.5$, the standard deviation in the overall intrinsic and environmental components are  $\sigma_{\rm int}^{\rm true}\approx 4.4$\,rad\,m$^{-2}$ and $\sigma_{\rm env}^{\rm true}\approx 5.3$\,rad\,m$^{-2}$. 
For this scenario, we run two tests corresponding to a different
number of lines of sight:
\begin{itemize}
\item 41632 (2C1). This is the total number of lines of sight for
  which an estimate of the extragalactic contribution is available
  from \cite{OJG2014};
\item 4003 (2C2). This number accounts for all the sources in the catalog 2C1 for which a
  redshift measurement is available as well (\citealt{HRG2012}).
\end{itemize}

In Fig.\,\ref{2comp} we show the results for the tests 2C1 and 2C2,
meaning with a two-component scenario for 41632 lines of sight (a) and
4003 lines of sight (b).
The histograms in the top plots of each column represent the 1-dimensional projection of the posterior for the corresponding parameter.
The dotted lines mark the true value of the parameter. The dashed and dashed-dotted lines describe the posterior statistics, namely the mean and the 1$\sigma$ confidence level respectively. The continuous lines indicate the prior used in our analysis.  The panels in colors show the 2-dimensional projection of the posterior for a given couple of parameters.
These plots show that our
algorithm is able to recover the mock $\Theta$-values for this
scenario. The inferred posterior mean values agree well within
the uncertainties with the correct ones and the posterior distributions are much narrower than the prior distributions. The dispersion in the
parameters $\Theta$ increases by decreasing the number of lines of
sight, as expected.  
The plots indicate that some of the parameters are correlated,
e.g., most noticeably $\sigma_{\rm env, 0}$-$\chi_{\rm red}$
that show a strong anticorrelation.
This feature can be understood in light of Eq.\,(\ref{env}). Indeed, for a given Faraday
rotation  $\sigma_{\rm env}$ associated with the structures between the source and the observer,
larger $\sigma_{\rm env, 0}$ imply smaller $\chi_{\rm red}$ and vice versa.
We expect the correlation in the posterior to be  significant for any
reasonable parameterization that allows for the same number of degrees of
freedom.

In order to have a compact and complete visualization, in the rest of the paper we present all the tests we performed and their results as in Fig.\,\ref{2comp}.

\subsection{Future prospects}

\label{pred}
We investigate the possibility to separate the Faraday rotation
intrinsic to the emitting radio source from that due to the
extragalactic environments between the source and the observer with the specifications
of the SKA, its precursor, ASKAP, and its pathfinder, LOFAR. The mock rotation measure values for each source have been generated as described in
Sect.~\ref{tests}. 
For each catalog the noise contribution has been extracted for each line of sight from a Gaussian with zero-mean and standard deviation equal to maximum uncertainty in Faraday depth expected in the corresponding frequency range, according to \cite{SABFK2008}. Luminosities at frequencies
different from 1.4\,GHz have been spectrally adjusted\footnote{For convenience, we use a single power-law $F(\nu)\propto \nu^{-0.8}$, where $F(\nu)$ is the flux density.}.

Since our approach assumes  that all the lines of sight are independent, i.e. sufficiently separated ($\gtrsim 1^{\circ}$)
so that the cross-correlation function of their magnetic field is zero, in the following tests we compute the number of lines of
sight considering a density of sources lower than or equal to one polarized source per square degree. 
The number of sources detected by ASKAP and SKA per square degree will be at least one hundred times larger. 
Here, we  investigate if already with a small sample of lines of sight we would be able to put any constraints
on the contribution from extragalactic large-scale environments, rather than exploit the information delivered by the
full number of lines of sight and to show the full potential of ASKAP and SKA.

\subsubsection{LOFAR}
\label{lofar}
We consider the LOFAR's High Band Antennas (HBA) only, because of
their better polarization performance.
We select the HBA region of the spectrum with less
contamination from radio frequency interference (120-160\,MHz, \citealt{OBZ2013}). In this
frequency range, the uncertainty on rotation measure values is
expected to be $\leq 0.05$\,rad\,m$^{-2}$ for a signal-to-noise ratio
larger than 5 \citep{SABFK2008}.  According to \cite{MHB2014}, the
expected number of polarized extragalactic radio sources is 1 per 1.7
square degrees for 8\,h-long observations, assuming an average degree
of polarization of 1\%, a spatial resolution of 20$^{\prime\prime}$, and a detection threshold of
500\,$\mu$Jy/beam/rmsf\footnote{rmsf is the half-power width of the Faraday depth
  spread function.} that corresponds to $S/N=5$.

\begin{figure*}[ht]
  \centering
       \begin{overpic} [width=15.5cm, angle=0]{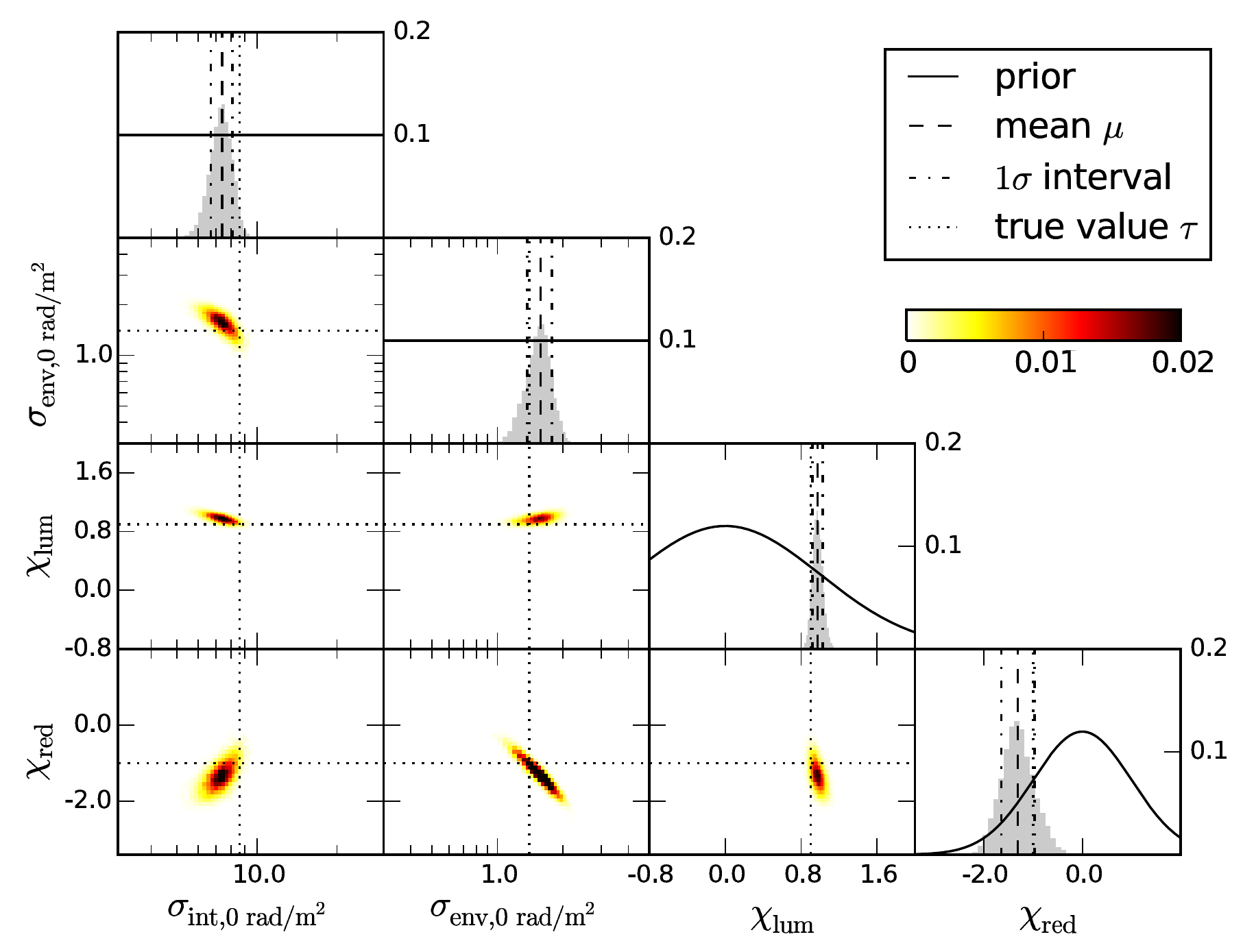} \put(-4,78){(a)}\end{overpic} \\
            \begin{overpic} [width=15.5cm, angle=0]{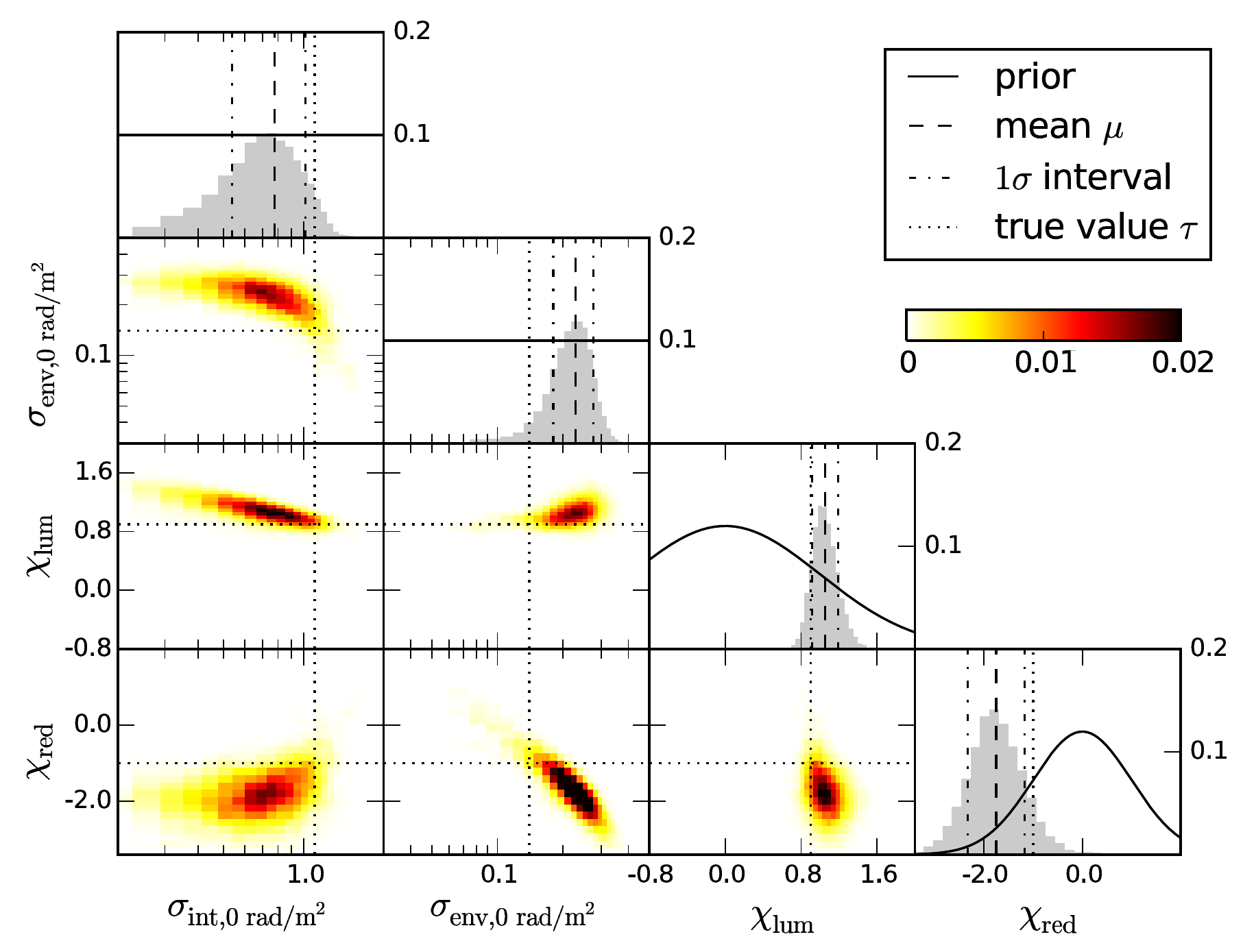} \put(-4,78){(b)} \end{overpic} \\
        \flushleft    
\caption{As Fig.\,\ref{2comp} but for results obtained with a
  two-component scenario for LOFAR HBA observations for $N_{\rm los}\approx$2200 and an overall
  extragalactic standard deviation in Faraday depth of (a)
  $\approx7$\,rad\,m$^{-2}$ (NPC\emph{a}) and (b) $\approx 0.7$\,rad\,m$^{-2}$ (NPC\emph{b}).}
\label{lofar_all}
\end{figure*}

On the basis of these assumptions we generate coordinates in the sky for a collection of sources
corresponding to a survey (8\,h per pointing) in the direction of the North Polar
Cap (NPC).  Among these sources we select those with Galactic latitude
larger than $55^{\circ}$.  This results in approximately $2200$ sources.
We derive a catalog assuming an overall
extragalactic Faraday rotation $\sigma_{\rm e}\approx$7\,rad\,m$^{-2}$  (NPC\emph{a}), 
and one assuming an overall
extragalactic Faraday rotation $\sigma_{\rm e}\approx$0.7\,rad\,m$^{-2}$  (NPC\emph{b}).
The results are shown in Fig.\,\ref{lofar_all},
respectively panel (a) and panel (b). These plots show that LOFAR can
provide good constraints for both values of the overall extragalactic
Faraday depth if a few thousand of lines of sight are used, with better
performance for larger $\sigma_{\rm e}$. The
posterior distributions are much narrower than the prior distributions
and the posterior means agree with the correct values within
1-2$\sigma$.  
Tests performed with a smaller number of lines of sight are
presented in Appendix~\ref{lofar_app}.  These plots show that the
algorithm performs well even for $N_{\rm los}\approx1000$
even if, as expected, the posterior distributions are wider.

\subsubsection{SKA}
\label{ska}
The SKA is expected to observe the entire Southern sky with a spatial resolution of
2$^{\prime\prime}$ and a sensitivity in polarization of
$\approx4\mu$Jy/beam (see, e.g., \citealt{JH2015}). The resulting sky grid of Faraday rotation
values will be 200-300 times denser than the largest catalog
available at the moment (see e.g. \citealt{H2013}).
The better resolution of 2$^{\prime\prime}$,
compared to the 45$^{\prime\prime}$ of \cite{TSS2009}, will make it
possible to identify optical counterparts uniquely and hence to assign
a redshift estimate to a larger number of sources through spectroscopic follow-up observations.
The \emph{SKA1 Re-Baseline Design 2015} 
indicates the frequency bands 1, 2, and 5 as available on SKA\_MID during SKA-Phase 1:
\begin{itemize}
\item band 1,  0.35--1.05\,GHz;
\item band 2, 0.65--1.67\,GHz;
\item band 5, 4.6--13.8\,GHz.
\end{itemize}
In the following we consider the frequency range
0.65-1.67\,GHz, since receivers in band 2 (B2) should be constructed first.

We produce mock Faraday depth values assuming a maximum standard deviation in
the noise distribution of the rotation measure of 0.8\,rad\,m$^{-2}$,
according to \cite{SABFK2008}, for this frequency range and for $S/N>5$.
We generate a catalog of coordinates in the South Polar Cap (SPC) based on the assumption
of one polarized source per square degree and we select those with Galactic latitude $b<-55^{\circ}$.  This
translates in $N_{\rm los}\approx 3500$. We will refer to this
catalog as B2SPC1. 
We produce catalogs of Faraday depth values assuming 
$\sigma_{\rm e}\approx$7.0\,rad\,m$^{-2}$
(B2SPC1\emph{a}) and $\sigma_{\rm e}\approx$0.7\,rad\,m$^{-2}$ (B2SPC1\emph{b}).

\begin{figure*}[ht]
  \centering

  \begin{overpic} [width=15.5cm, angle=0]{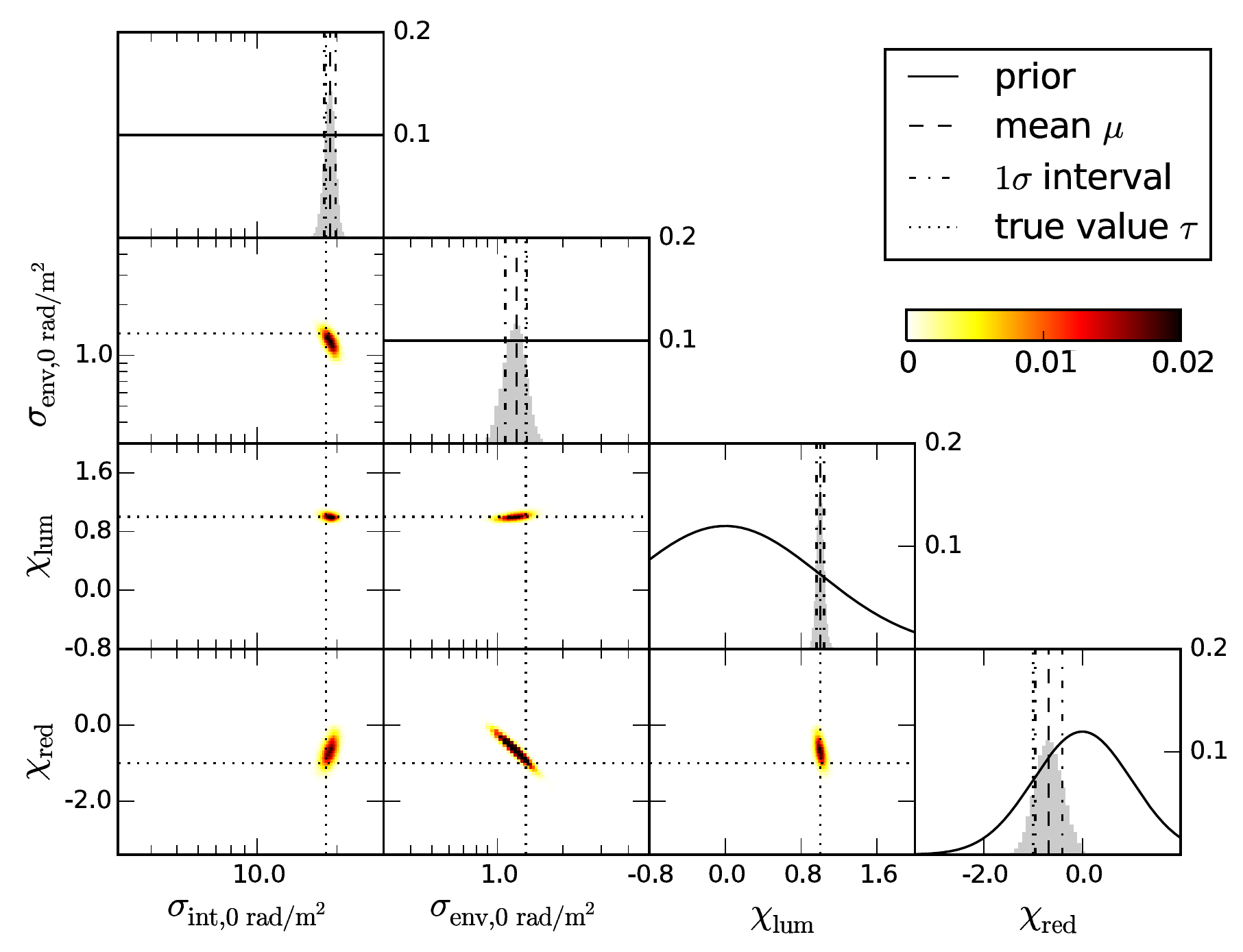} \put(-4,78){(a)} \end{overpic} \\
\begin{overpic} [width=15.5cm, angle=0]{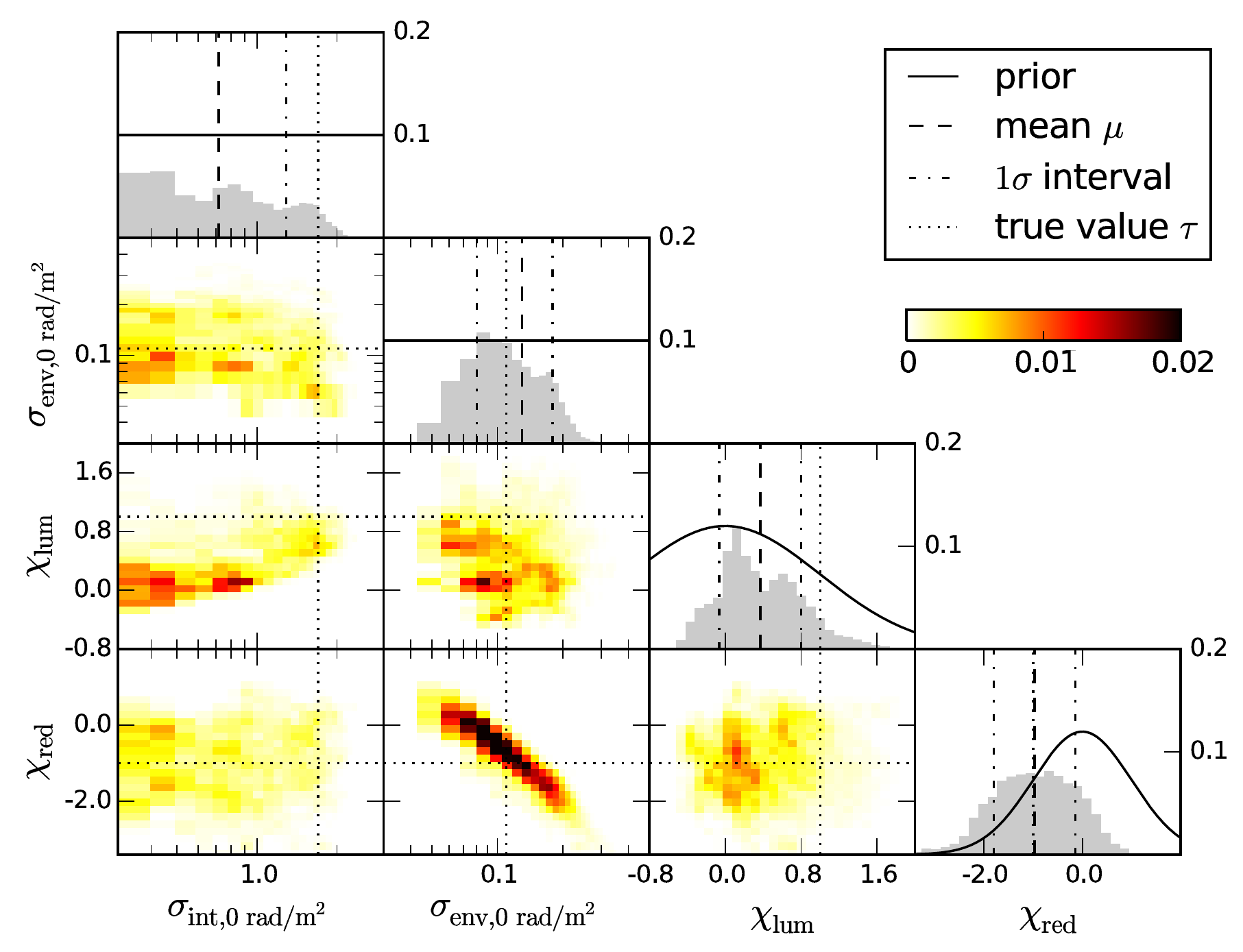} \put(-4,78){(b)}\end{overpic} \\
        \flushleft   
\caption{As Fig.\,\ref{2comp} but for results obtained with a
  two-component scenario for SKA observations in the frequency range
  650-1670\,MHz for $N_{\rm los}\approx$3500 and an overall extragalactic
  Faraday rotation of (a) $\approx 7.0$\,rad\,m$^{-2}$ (B2SPC1\emph{a}) and (b) $\approx
  0.7$\,rad\,m$^{-2}$ (B2SPC1\emph{b}).}
\label{skamid}
\end{figure*}

In Fig.\,\ref{skamid} we show the results for B2SPC1\emph{a} and B2SPC1\emph{b},
in panel (a) and (b) respectively. The plots for
$\sigma_{\rm e}\approx$7.0\,rad\,m$^{-2}$ indicate that with 
$N_{\rm los}\approx$3500 we obtain a good disentangling of the environmental and
intrinsic contributions, and their redshift and luminosity dependence, with
narrow posterior distributions and mean values
within $\approx$1$\sigma$ from the true values.
For the same amount of lines of sight but a smaller overall Faraday rotation $\sigma_{\rm e}\approx$0.7\,rad\,m$^{-2}$,
the dispersion in $\sigma_{{\rm env},0}$, $\chi_{\rm lum}$, and $\chi_{\rm red}$ becomes broader
but with mean values still within a few standard deviations from the assumed value. On the other hand,
we do not get good constraints for the parameter $\sigma_{{\rm int},0}$ that appears characterized
by a large dispersion. For a better constraint of this parameter we would need to resort to a
larger number of lines of sight.  
Since the sensitivity of the SKA will allow to detect
hundreds of sources per square degree,
narrower posterior distribution of the parameters can be obtained by increasing the number of lines of sight.

Because of the different number of lines of sight and different frequency bands we can not directly compare these
results with the results for LOFAR. For an analysis of SKA performance at lower frequencies 
see Appendix~\ref{ska_app}. 
In this appendix, results for SKA observations in band 1 as well as in band 2 but with a smaller number of lines of sight are shown.
We do not consider frequency band 5.
Even if moving to higher frequency sources have a higher degree of polarization,
the maximum uncertainty in Faraday depth is larger and the total source counts reduce.
Consequently, we expect that the parameters are
poorly constrained if the same number of lines of sight is taken.

\subsubsection{ASKAP}
\label{askap}
\begin{figure*}[ht]
  \centering
       \begin{overpic} [width=15.5cm, angle=0]{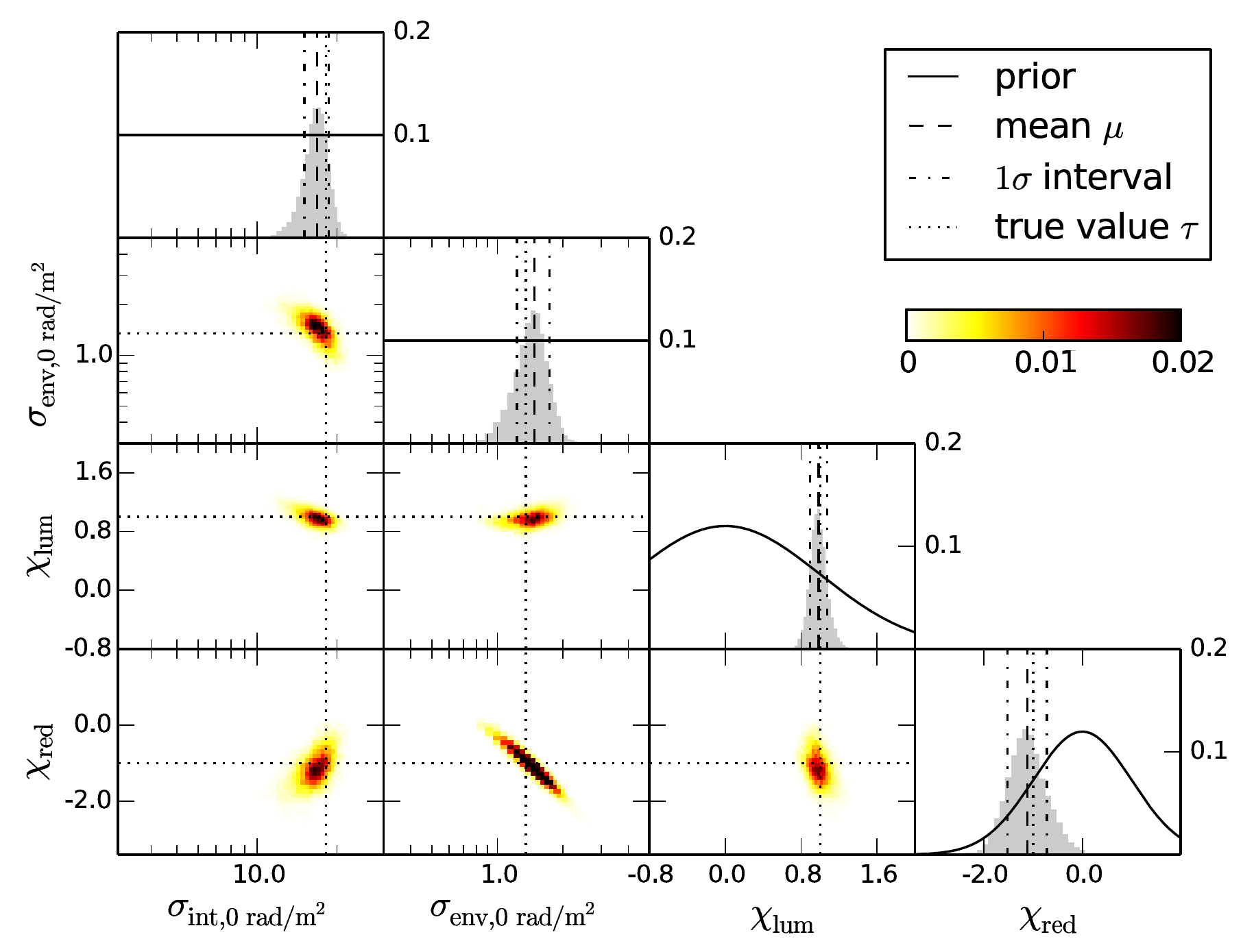} \end{overpic} \\
        \flushleft    
\caption{As Fig.\,\ref{2comp} but for results obtained with a
  two-component scenario for ASKAP observations in the frequency range
  1130-1430\,MHz for $N_{\rm los}\approx$3500 and an overall
  Faraday depth of $\approx$7.0\,rad\,m$^{-2}$ (POSSUM).}
\label{possum}
\end{figure*}

The \emph{Polarisation Sky Survey of the Universe's Magnetism}
(POSSUM, \citealt{GLT2010}) is planned between 1130 and 1430\,MHz with the Australian
Square Kilometre Array Pathfinder (ASKAP).  The survey will reach a
sensitivity in U and Q of $<$10\,$\mu$Jy/beam and a resolution
of 10$^{\prime\prime}$. This will result in a density of 70 polarized
sources per square degree \citep{HNGM2014} and a Faraday depth catalog
one hundred times denser than those currently existing
(e.g., \citealt{TSS2009}).  To test the capabilities of this survey in
constraining extragalactic magnetic fields, we use the same set-up
described in Sect.~\ref{ska}, 
but with an uncertainty drawn from a Gaussian with $\sigma_{\rm  noise}=6.$\,rad\,m$^{-2}$.
This should be a reliable approximation
of the maximum uncertainty in Faraday depth for a polarized signal
with a $S/N>5$ \citep{SABFK2008}. The results are shown in
Fig.~\ref{possum}. ASKAP observations appear already quite promising
for disentangling the contribution intrinsic to the source from that
due to the large-scale environments for 
$\sigma_{\rm e}\approx$7.0\,rad\,m$^{-2}$, already if Faraday depth values are available only for
one source per square degree.  The dispersion in the parameters is
larger than that derived in Sect.~\ref{ska} for the same collection of
sources and overall extragalactic Faraday variance
(Fig.\,\ref{skamid}a), as expected being the maximum uncertainty in
Faraday depth larger.  However, the mean values of the posterior agree
within at most about 1$\sigma$ with the true values.

\subsection{Discussion}
\label{discussion}

We developed a Bayesian algorithm to disentangle extragalactic
contributions in the Faraday depth signal. It builds upon an algorithm
to reconstruct the Galactic Faraday screen and its
uncertainty-correlation structure previously presented by \cite{OJG2014}.
We tested the algorithm by
modeling the overall extragalactic contribution as the sum of an
intrinsic and an environmental component, as described in
Sect.~\ref{tests}, but see Appendix~\ref{alternative_app} for the
inclusion of a constant or a latitude dependent term. These tests
show that the algorithm is able to discriminate among the different
components and their dependence on the luminosity and the redshift of
the source, if a suitable number of lines of sight is available.

For test purposes, we built mock catalogs according to the specifications
of the catalogs available in the literature after correcting for poorly known 
uncertainty information (see Sect.~\ref{tests}).
The results in Fig.\,\ref{2comp}(b) indicate that by applying
the algorithm to the currently available dataset we could already
infer preliminary information about extragalactic magnetic fields, if a few thousands
of sources with reliable observational uncertainty were available. For
the 4003 sources in Fig.\,\ref{2comp}(b), indeed, 
redshift and flux density values are available.  Nevertheless, since actually
we do not consider all their Faraday depth noise variances to be reliable (see Appendix~A of \citealt{OJG2014}),
and  the applied correction factors, $\eta_i$, are only statistical estimates, we would need to additionally
sample in the $\eta_{i}$-parameter space in order to use the algorithm with real
data , as described by
\cite{OJG2014}. The modern techniques of analysis, such as model fitting of
the fractional polarization components $q$ and $u$ (e.g., \citealt{FRB2011}; \citealt{ITAKR2014}),
rotation measure synthesis (e.g., \citealt{BB2005}; \citealt{AKTR2014}) and Faraday synthesis
\citep{BE2012}, and the
future radio surveys will partially overcome this problem.
The large bandwidth of the new radio interferometer will
allow to reduce the risk of $n\pi$-ambiguity, particularly strong when
the $\lambda^2$-fit approach is used, as well as to reach a
sufficiently high resolution in Faraday depth to distinguish nearby
Faraday components. We note that when the distance between two peaks
in a Faraday spectrum is smaller than the resolution, the uncertainty
in the Faraday depth may be driven by the Faraday point spread
function (e.g., \citealt{FRB2011,FGC2014,KAIKT2014}). For these
reasons, the catalogs coming from Faraday depth grids planned with LOFAR, ASKAP, and the
SKA will be more reliable both in terms of Faraday depth values and
of uncertainties.
Therefore, 
we investigated the prospects of these simpler future datasets here and
address the more complex application of our technique to present data
in a separate work.
We assumed the computation to be dominated by the
uncertainties in the Faraday depth estimates
for each source, while the coordinates, luminosities and redshifts to be
exactly known. With the advent of
highly-accurate Faraday depth catalogs, this may change and in
future work we will need to investigate to what depths and accuracy
redshifts, and luminosities are needed in order to apply the method we
propose. This is particularly important for redshifts.
Indeed, the catalog used in this work reports a spectroscopic redshift for each source \citep{HRG2012}.
Recently, optical surveys, such as
2MASS, WISE, and SuperCOSMOS, have measured the photometric redshifts for millions
of galaxies up to $z\approx0.5$ (e.g. \citealt{BPJCS2014}), and the
surveys planned with the next generation of telescopes will further
increment this number. 
Nevertheless, photometric redshifts are less accurate than spectroscopic redshifts. This will require
to evaluate the impact of their uncertainties on our results.

Our tests indicate that LOFAR, ASKAP, and the SKA will allow
us to infer information about cosmological magnetic fields already with a
few thousands of lines of sight, with better performance when lower
frequencies are used. We want to stress that the uncertainties used in
our tests only represent statistical uncertainties and any systematic
issues have been neglected.
In principle, LOFAR observations could be already used for this aim but
the development of a pipeline for the reduction of polarization data is still in progress.
The enhanced capabilities of the Jansky Very Large Array (JVLA) make the new centimeter-wavelength sky survey (VLASS, \url{https://science.nrao.edu/science/surveys/vlass}) planned with this instrument a good opportunity for the application of this algorithm and for delivering significant results in the study of cosmic magnetism already in the next years.

\begin{figure*}[ht]
       \centering
\includegraphics[width=7.5cm, angle=0]{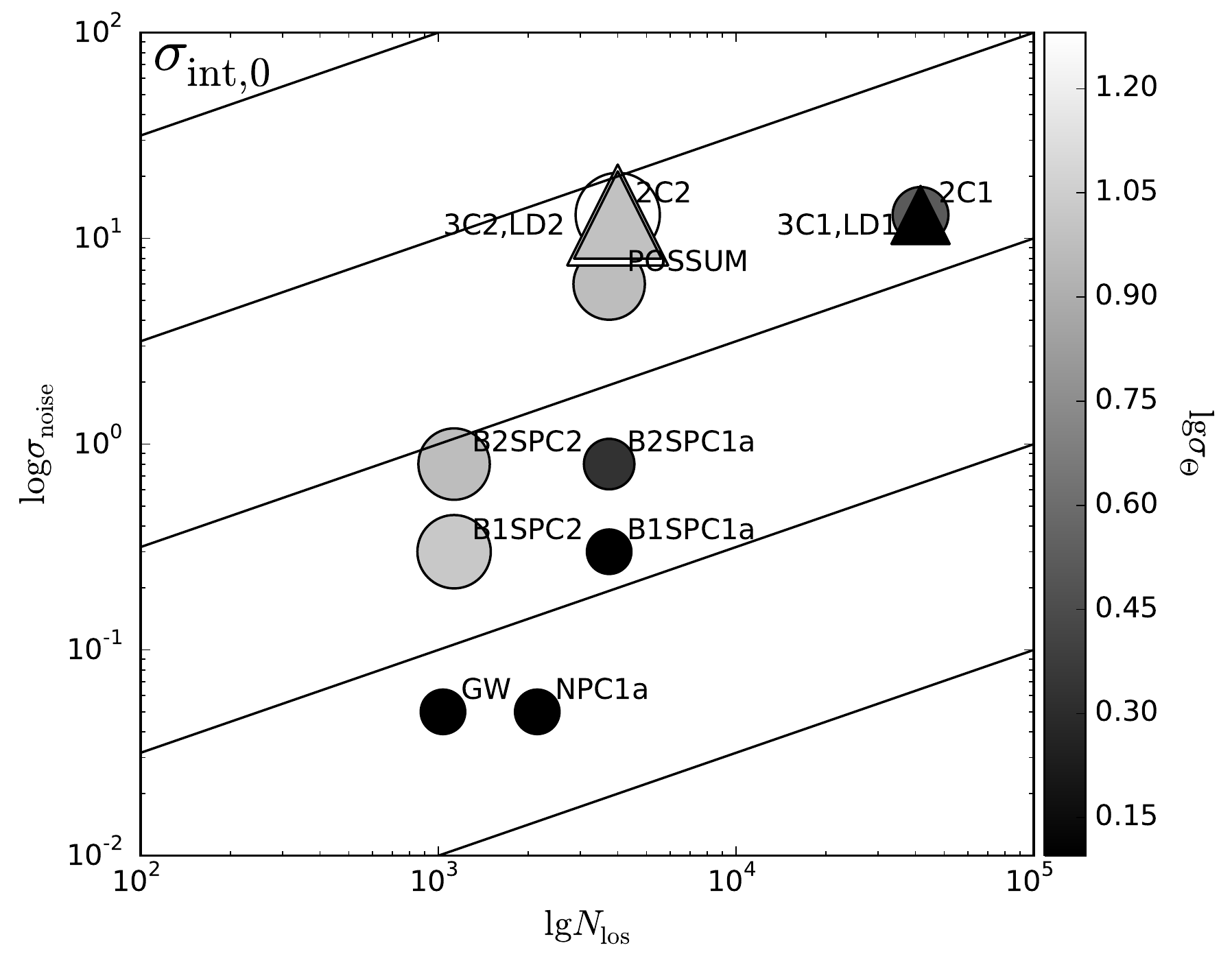} 
\includegraphics[width=7.5cm, angle=0]{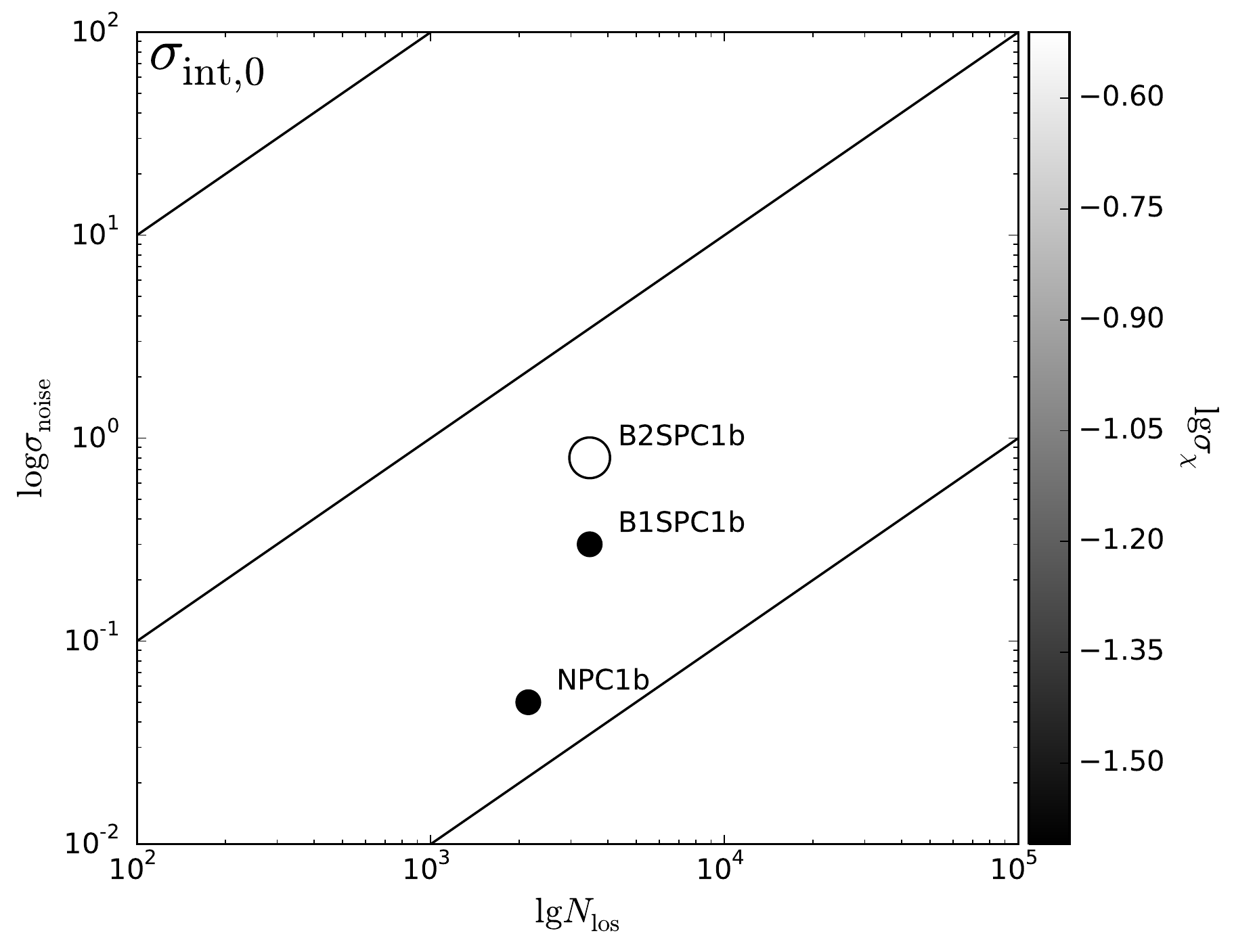}\\ 
\includegraphics[width=7.5cm, angle=0]{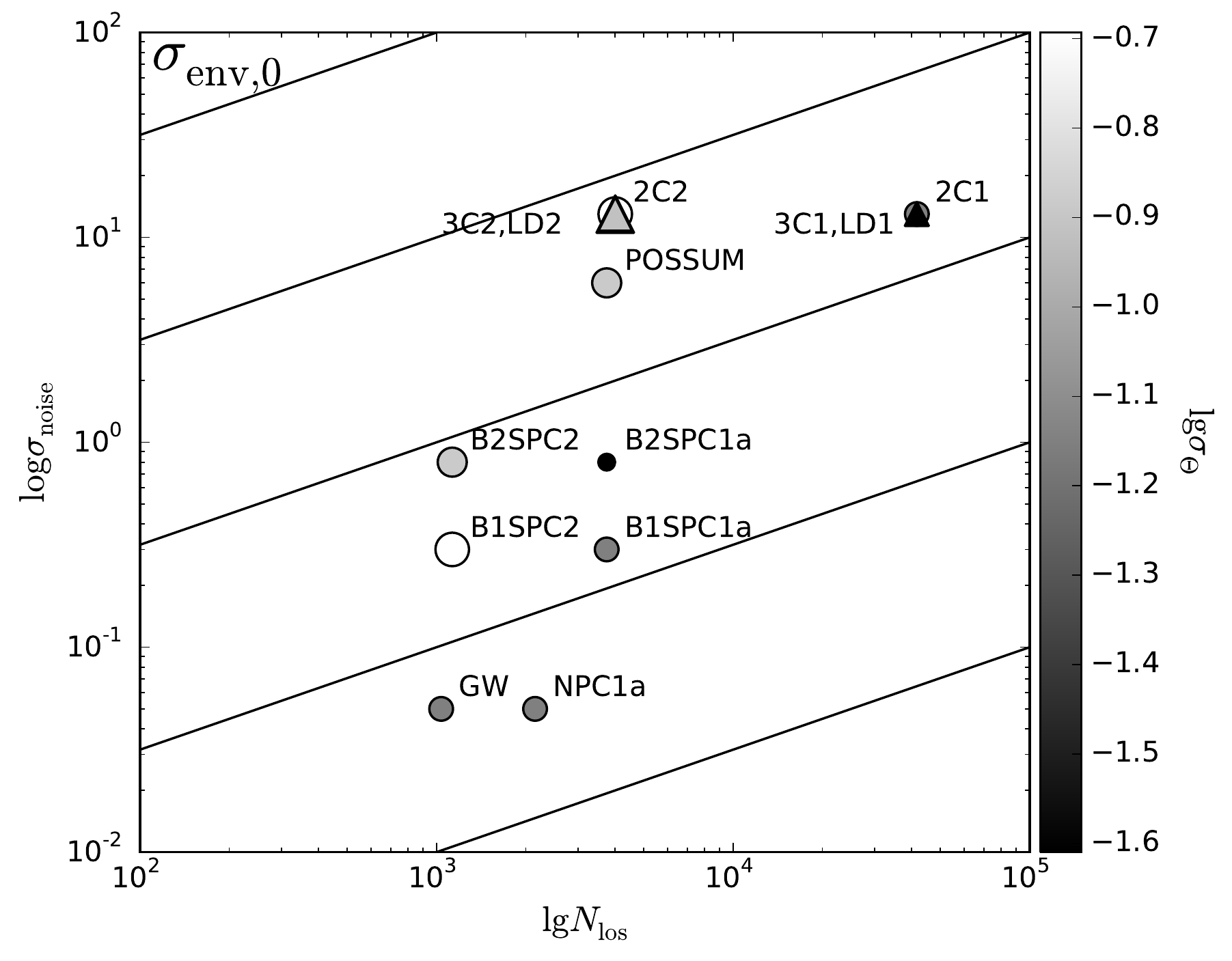} 
\includegraphics[width=7.5cm, angle=0]{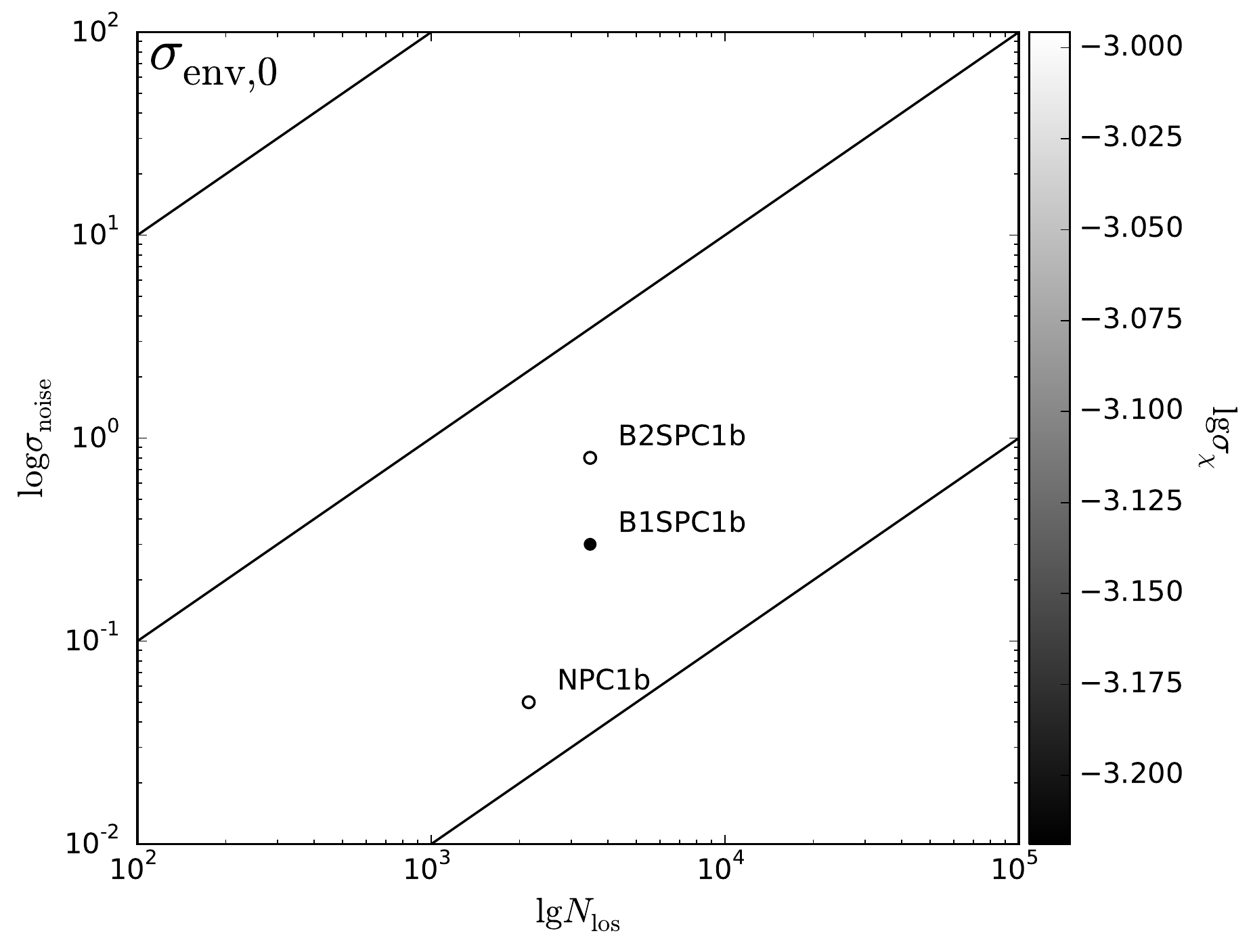}\\
\includegraphics[width=7.5cm, angle=0]{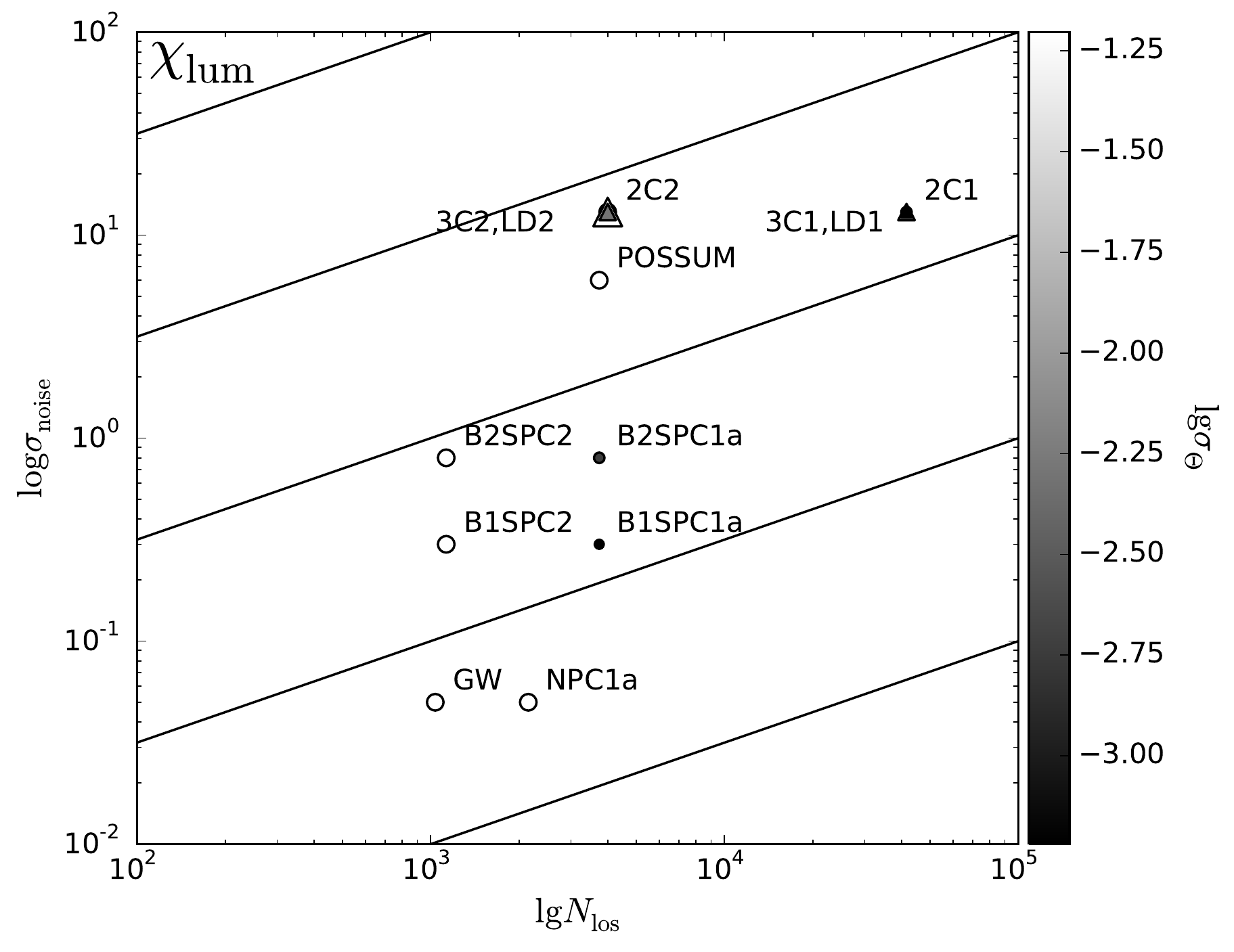} 
\includegraphics[width=7.5cm, angle=0]{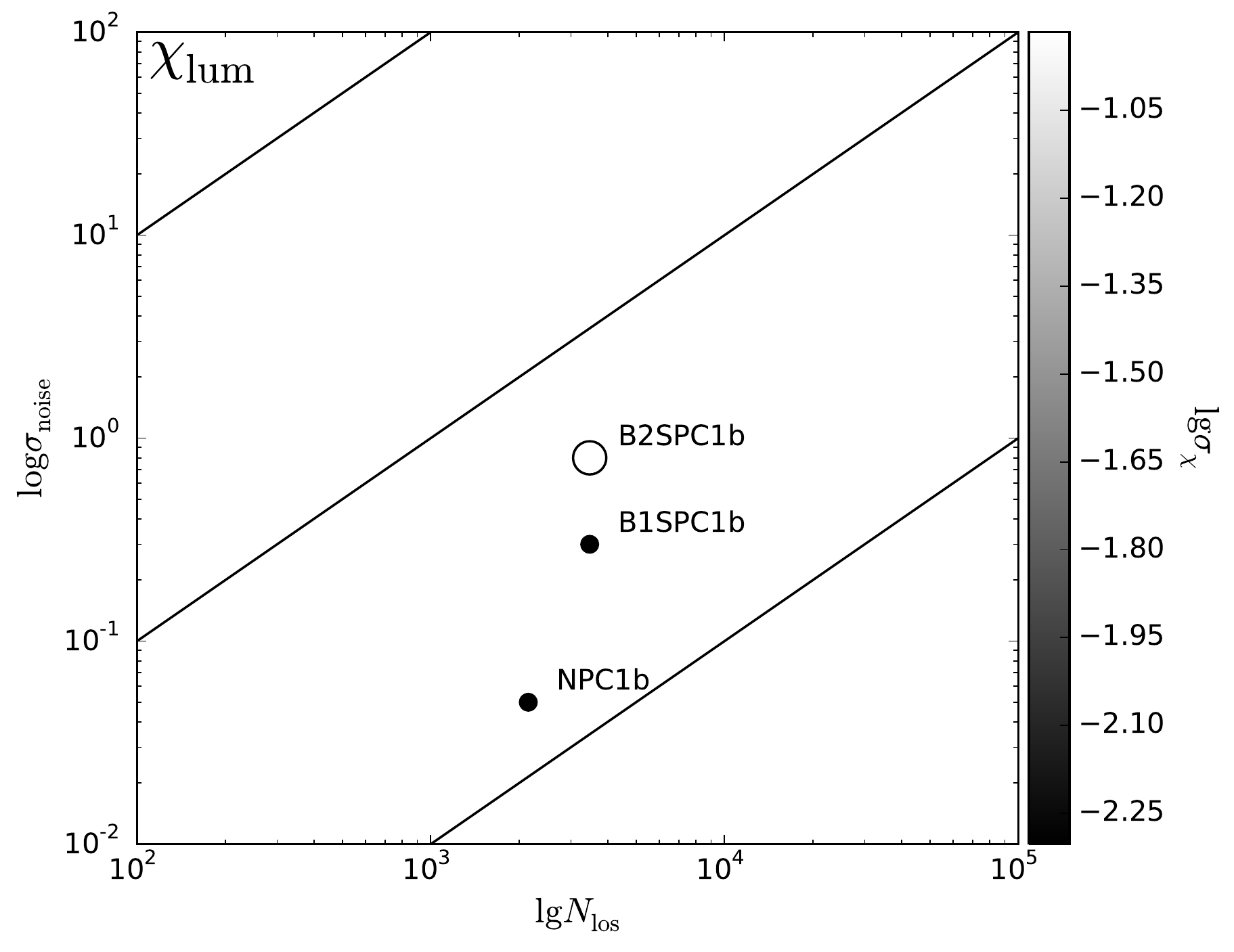}\\ 
\includegraphics[width=7.5cm, angle=0]{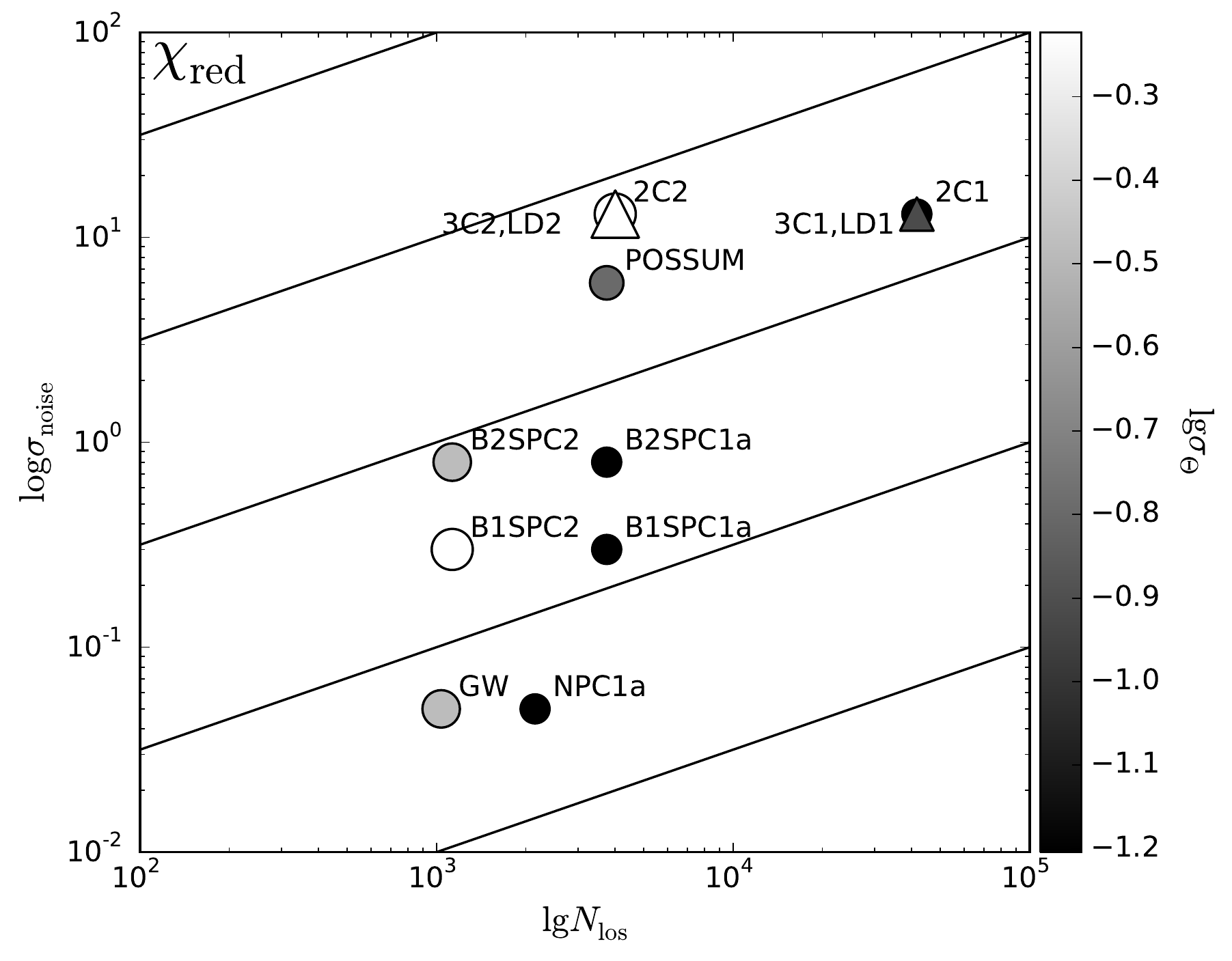} 
\includegraphics[width=7.5cm, angle=0]{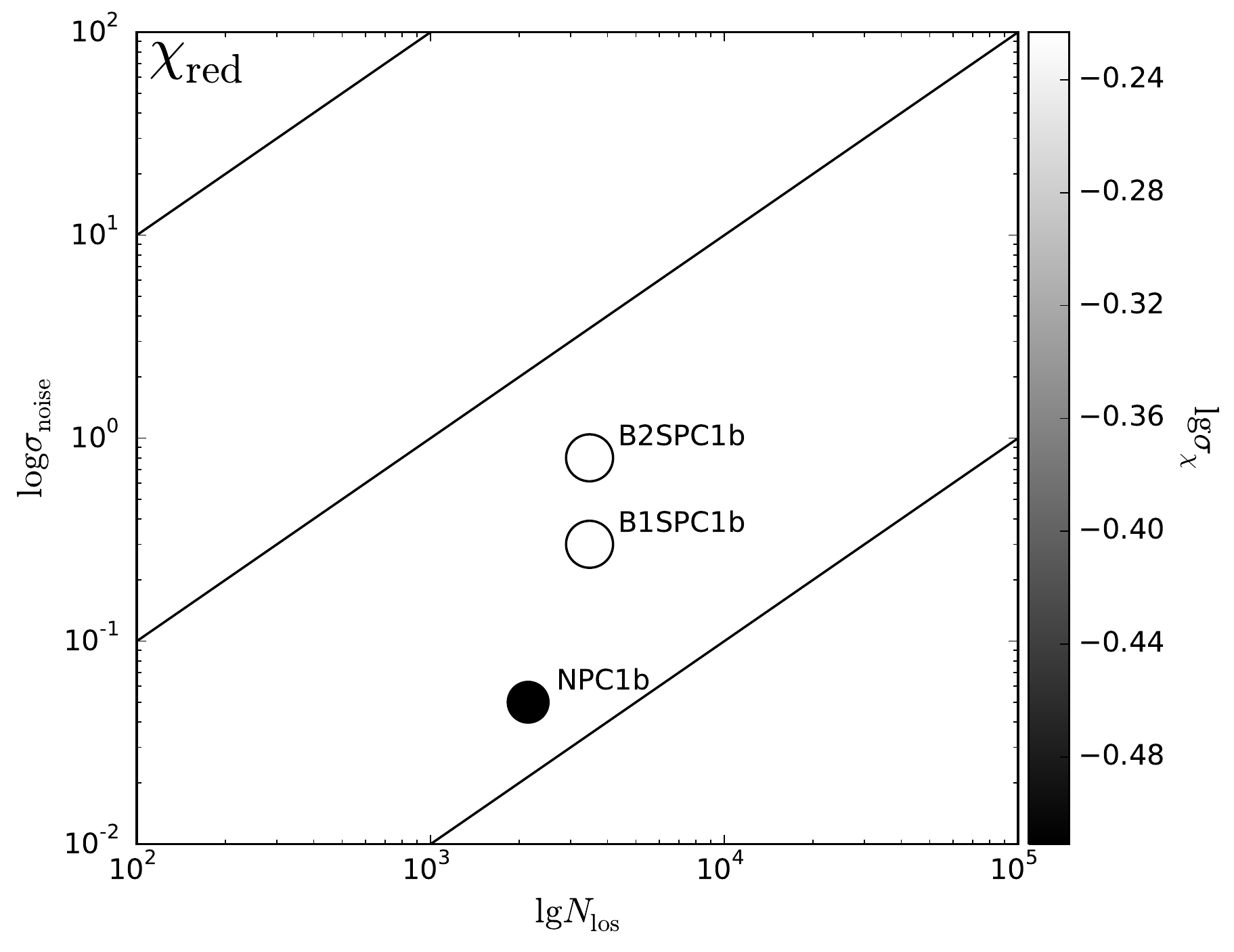}\\

        \flushleft       
        \caption{Uncertainty on $\sigma_{\rm int,0}$, $\sigma_{\rm env,0}$, $\chi_{\rm lum}$, and
          $\chi_{\rm red}$ (from the top to the bottom) as a function of the
          observational uncertainty $\sigma_{\rm noise}$ and the number of lines
          of sight $N_{\rm los}$ for an overall extragalactic Faraday depth variance of
          $\approx$7.0\,rad\,m$^{-2}$ (left panels) and $\approx$0.7\,rad\,m$^{-2}$ (right panels).
          The size of the points is proportional to the
          uncertainty on the parameter. A description of this uncertainty is
          given also by the greyscale. Circles refer to scenarios including
          two components, while triangles refer to three-component scenarios. Continuous lines
          correspond to $\sigma^2_{\rm noise}/N_{\rm los}=$const.}
  \label{summaryresults}
 \end{figure*}

In Table~\ref{tab:C} we give a quantitative summary of the results.
For each test we report the true values $\tau$ of the $\Theta$ parameters which describe the intrinsic
and environmental contribution to the Faraday depth, their mean
values $\mu$, and their uncertainties $\sigma$, as well as the
displacement of the mean from the true value in terms of the
uncertainty,
\begin{equation}
  \epsilon=\left|\frac{\mu-\tau}{\sigma}\right|.
\end{equation}
These values have been computed after discarding the burn-in samples by visual inspection.
The comparison of the results obtained with the mock catalogs created for present instruments 2C1 and 2C2 indicates that by increasing the number of sources with known redshift by a factor ten, the uncertainty in the $\Theta$ parameters can be reduced by about a factor two. This would not longer be valid if the sources without redshift information represent a different population than the observed one, since in our test we adopted mock values for these sources randomly extracted from the observed redshift distribution.
A similar result is obtained if the observational uncertainty $\sigma_{\rm noise}$ is reduced by approximately a factor ten, as shown by the comparison of results corresponding to catalog 2C2 and the SKA catalog B2SPC1\emph{a}. These two tests refer to a similar number of lines of sight and to a similar frequency range\footnote{Indeed, the 2C2-sources all belong to the catalog by \cite{TSS2009}, therefore their noise uncertainties refer to the frequency range 1365-1435\,MHz.}. 
In Fig.~\ref{summaryresults} we present a visual summary of the results
for all the two- and three-component scenarios corresponding to an overall extragalactic Faraday
rotation of approximately both $7\,$rad\,m$^{-2}$ (left panels)  and $0.7\,$rad\,m$^{-2}$ (right panels)
for the four parameters $\sigma_{{\rm int},0}$, $\sigma_{{\rm env},0}$, $\chi_{\rm lum}$, and $\chi_{\rm  red}$.
The uncertainty on the parameters $\Theta$ decreases by increasing the
number of lines of sight, $N_{\rm los}$, and by decreasing the
observational error, $\sigma_{\rm noise}$. The best results are obtained when a good
compromise between these two numbers is used, as for example shown by the
scenarios NPC\emph{a} and NPC\emph{b}. Nevertheless, the comparison of the results of tests corresponding to different instruments can be not straightforward because they refer to different frequency bands and to slightly different values of the $\Theta$ parameters.

The problem we are tackling is characterized by different complexities. 
We are looking for a very weak signal, by using the residual information left in the data after the Galactic contribution has been derived. At the same time, we expect this signal to be characterized by a possible redshift
dependence, even if very weak since not yet detected, and by cross-correlations of the extragalactic magnetic field along different lines of sight.
In this work we try to address the first two issues with a Bayesian approach able to properly combine all the available observational information and allowing for a redshift dependence in our model. On one hand, we might overestimate the
extragalactic contributions, since we are not including an uncorrelated Galactic component in the model.
This term can be easily included in our algorithm, as shown by the tests in Appendix~\ref{alternative_app}, and quantified with real data. On the other hand, we do not take into account a possible correlated component of the extragalactic magnetic field, which if present would be beneficial for our analysis.
This means
that the magnetic field values that can be inferred have to be
considered rather as a lower limit, even if we expect this error to be small (e.g., \citealt{AR2011}).
This is not a problem for surveys performed with LOFAR, since the expected density of extragalactic polarized sources is lower than the density of sources we assume. On the contrary, with ASKAP and the SKA, we will be able to detect at least one hundred sources per square degree.
An approach to deal with possible cross-correlations between lines of sight is being developed with the intention to apply this new technique to forthcoming
ultra-deep JVLA observations from the CHILES Con Pol survey (Hales et al. in preparation).
The combination of the two methods would allow
to exploit in an optimal way the information coming from the denser grids of Faraday depth measurements provided by SKA and
ASKAP surveys.

Finally, we stress that the aim of this paper is to give a proof of concept.
Indeed, the present version of the algorithm requires a high computational
burden that limits its efficiency. For example, the time required to run a test on our computer cluster   
is of the order of a couple of months per chain even when only a few thousands lines of sight are considered.
We are currently working on performance optimization to make the algorithm suitable to be applied
to the upcoming large Faraday depth catalogs.

\subsection{Future developments}
\label{generalization}
The work presented in this paper prepares a Bayesian technique to investigate of magnetic fields
in the large-scale structure, in particular in
filaments and voids.
As a next step we want to discriminate
among the amount of Faraday rotation due to each large scale structure
environment (see also \citealt{VOE2015}).  When different large-scale
environments are considered, the variance in the extragalactic Faraday rotation
can be parameterized as
\begin{align}
  \langle\phi_{{\rm e}, i}^2\rangle&\approx a_0^2\int_0^{z_{i}}\langle n_{\rm 0}^2\rangle \langle B_{l 0}^2\rangle\Lambda_{l 0}(1+z)^4\frac{c}{H(z)}\mathrm{d}z \nonumber \\
  &\approx \left[\left(\frac{L_i}{L_0}\right)^{\chi_{\rm lum}}\frac{\sigma^2_{{\rm int},0}} {(1+z_i)^{4}} +\displaystyle\sum_{j=1}^{N_{\rm env}}l_{ij}\sigma^2_{j}\right],
\label{lls}
\end{align}
where $\sigma_{1},\sigma_{2}, ..., \sigma_{N_{\rm env}}$ are the contributions from 
$N_{\rm env}$ different environments and $l_{ij}$ is the length of the line of sight,
$i$, through each environment, $j$.

With a Bayesian approach, \cite{JKLE2010} (see also \citealt{LJW2015})
reconstructed the cosmic density field. They used
optical data from the SDSS \emph{Data Release 7} \citep{AAA2009} and classified the
structures as voids, sheets, filaments, and galaxy clusters, according
to the classification scheme of \cite{HPCD2007}.  This reconstruction
enables us to compute the path covered by the radio signal through
each environment, i.e., the elements $l_{ij}$ in Eq.\,(\ref{lls}),
once the position of the radio source is identified by using the
redshift. This posterior of the large scale structure density field is
available in the form of samples.  Radio sources can belong to
different environments (e.g. galaxy clusters and filaments) and the
path covered by the signal in each environment differs for each radio
source. These facts can be statistically taken into account with the
use of a collection of sources distributed over all the sky and of
different realizations of the large scale structure. 
We plan to use the full posterior of the large scale structure in order
to statistically estimate which is the amount of variance
due to the different types of environments in the observed Faraday depths.

\section{Conclusions}
\label{conclusions}

The properties of cosmic magnetic fields constitute outstanding questions in modern cosmology. 
To get a better understanding it is essential to shed light
on the properties of magnetic fields in large-scale environments,
meaning filaments and voids,
where turbulent intracluster gas motions have not yet enhanced the magnetic field
 and, consequently, the magnetic field strength and
structure still depend on the seed field power spectrum.

Upcoming generations of radio telescopes, first LOFAR, and in the
next decades the SKA, will perform polarization sky surveys with high
sensitivity. Modern techniques based on rotation measure
synthesis and Faraday synthesis, will enable us to perform a proper
analysis of the polarization properties of extragalactic radio
sources, thus providing unprecedented, highly accurate Faraday
depth catalogs in frequency ranges from a few hundreds MHz to a few
GHz. A statistical approach is required to exploit the information
encoded in these data. For this reason we developed a Bayesian
algorithm able to combine radio observations with luminosities 
and redshifts of sources, aiming at disentangling 
contributions to the extragalactic Faraday rotation intrinsic to radio
sources and due to the large-scale structure, and in this way infer
information about large-scale magnetic fields.  Knowledge of the
redshift is essential in this approach. The present all-sky photometric
optical surveys and the surveys planned with the next
generation of telescopes will greatly enlarge the number of sources
for which this information will be available.

The work described in this paper is a proof of
concept and shows that our algorithm can be used to discriminate between 
the Faraday depth generated by the radio source itself and the contribution due
to the large-scale structures. Additionally, our algorithm is able to investigate the dependence of these terms on
the redshift and the radio luminosity of the sources.  The tests
performed with mock LOFAR, ASKAP, and SKA data suggest that this technique 
is promising for the investigation of magnetic fields
with strengths of  a few $\mu$G down to a few nG, when uncertainties
in the data are up to a few rad\,m$^{-2}$ and known with high
accuracy. We note that our modeling does not take into account any correlated component of
extragalactic magnetic fields. Consequently inferred magnetic field strengths
have to be considered as a lower limit.

The main characteristics of upcoming polarization surveys can be summarized by the number of lines of sights and by the maximum observational uncertainty in Faraday depth. 
Our tests indicate that, for a given number of lines of sight, better constraints can be obtained with observations
at lower frequencies, because of the smaller observational uncertainty. Therefore, in principle, LOFAR and, thanks to its higher sensitivity even more, SKA\_LOW (50--350\,MHz) observations would be ideal. Nevertheless, the scant number of polarized sources at these frequencies and the difficulties in the calibration of the data could make the use of these data complex. ASKAP and SKA\_MID observations respectively in the frequency range 1130-1430\,MHz and 650-1670\,MHz appear to be promising as well. We should be able to put useful constraints on large scale magnetic fields already with Faraday depth measurements for a few thousands of sources, and improve their determination by increasing the number of lines of sight. An increment in the number of lines of sight by a given factor reduces the uncertainty in the estimation of the intrinsic and environmental contribution as a reduction by the same factor in the observational uncertainty does, indicating that deeper observations of small fields could be a valuable or even better alternative to all sky surveys.      

We are aware that many aspects of our approach require improvements: e.g., computational efficiency, inclusion of a correlated extragalactic magnetic field
component, and of uncertainty in redshift,
etc. Nevertheless, we present a first step toward a Bayesian study of
magnetic fields associated with the cosmic large-scale structures.

\begin{sidewaystable*}
  \caption{In Col.~1 the Identification code (ID) of the catalog is displayed. In
    the rest of the columns for each parameter the true value $\tau$, the mean
    $\mu$, the uncertainty $\sigma$, and the displacement $\epsilon$ between mean
    and true value in terms of uncertainty is given. In the last column the
    standard deviation in the total extragalactic Faraday depth $\sigma_{\rm e}$ is reported.}
        \begin{tabular}{ccccccccccccccccc}
          \hline
           \hline
            ID ~~~~~~~~~~~~~~~~~~~~~~~~   &      &$\sigma_{\rm int,0}$ &  ~~~~~~~~~~~~~~~~~~~~~       &        &$\sigma_{{\rm env},0}$ &          ~~~~~~~~~~~~~~~~~~~~~~~      &           &$\sigma_{\rm c/lat}$&   ~~~~~~~~~~~~~~~~~~~~~~         &        &$\chi_{\rm lum}$  &     ~~~~~~~~~~~~~~~~~~~~~~~         &        &  $\chi_{\rm red}$     &   ~~~~~~~~~~~~~~~~       & $\sigma_{\rm e}$~~~~~~~  \\
\end{tabular}
            \begin{tabular}{l r D{,}{\pm}{-1} l r D{,}{\pm}{-1} l r D{,}{\pm}{-1} l r D{,}{\pm}{-1} lrD{,}{\pm}{-1} l c} 

            &$\tau$& \mu,\sigma & $\epsilon$&$\tau$  &  \mu , \sigma  & $\epsilon$       &$\tau$     & \mu,\sigma      & $\epsilon$&$\tau$  & \mu,\sigma  & $\epsilon$ &$\tau$  & \mu,\sigma       & $\epsilon$&~~~ rad\,m$^{-2}$\\     
            \hline
            2C1            &   18.2  &   18.4,1.1    & 0.2      & 1.4   &1.6,0.2    & 1.5       &           &                     &           &    1.0 &1.0,0.04    & 0.0        &$-$1.0    &-1.3,0.3         & 1.2  &~~$\approx$7.0   \\
            2C2            &   18.2  &   19.1,2.5    & 0.3      & 1.4   &1.3,0.4    & 0.1       &           &                     &           &    1.0 &0.9,0.1      & 0.7        &$-$1.0    &-0.7,0.6         & 0.4  &     \\
            3C1            &   16.5  &   14.5,1.2    & 1.6      & 1.1   &1.2,0.2    & 0.7       &3.9        &   4.2,0.3           & 1.2       &    1.0 &1.1,0.1      & 1.5        &$-$1.0    &-1.1, 0.3        & 0.2  &   \\
            3C2            &   16.5  &   11.1,3.6    & 1.5      & 1.1   &1.8,0.5    & 1.6       &3.9        &   4.0,0.7           & 0.2       &    1.0 &1.1,0.3      & 0.4        &$-$1.0    &-1.7,0.6         & 1.2  &    \\
            LD1            &   16.5  &   17.2,1.1    & 0.6      & 1.1   &1.1,0.2    & 0.2       &0.58       &   0.56,0.04         & 0.5       &    1.0 &1.0,0.04    & 0.0        &$-$1.0    &-1.1,0.4         & 0.2   &    \\
            LD2            &   16.5  &   15.6,2.7    & 0.3      & 1.1   &1.1,0.4    & 0.03      &0.58       &   0.7,0.1           & 0.9       &    1.0 &1.0,0.1      & 0.3        &$-$1.0    &-1.1,0.8         & 0.1   &    \\
            GW             &   8.6   &   9.5,0.7     & 1.2      & 1.4   &1.0,0.2    & 1.8       &           &                     &           &    0.9 &0.8,0.1      & 1.2        &$-$1.0    &-0.1,0.5         & 1.6   &   \\
            NPC\emph{a}   &   8.6   &   7.4,0.7     & 1.7      & 1.4   &1.6,0.2    & 0.9       &           &                     &           &    0.9 &1.0,0.1      & 1.4        &$-$1.0    &-1.3,0.3         & 0.9   &    \\
           POSSUM         &   18.2  &   16.9,1.8    & 0.8      & 1.4   &1.5,0.3    & 0.5       &           &                     &           &    1.0 &1.0,0.1      & 0.2        &$-$1.0    &-1.1,0.4         & 0.3   &    \\
            B2SPC1\emph{a} &   18.2  &   18.9,0.9    & 0.7      & 1.4   &1.2,0.1    & 0.9       &           &                     &           &    1.0 &1.0,0.04     & 0.0       &$-$1.0    &-0.7,0.3         & 1.1   &    \\
            B2SPC2         &   18.2  &   20.1,1.8    & 1.0      & 1.4   &1.4,0.3    & 0.3       &           &                     &           &    1.0 &1.0,0.1      & 0.0       &$-$1.0    &-1.2,0.5         & 0.3  &     \\
            B1SPC1\emph{a} &   14.9  &   15.9,0.7    & 1.5      & 1.4   &1.4,0.2    & 0.0       &           &                     &           &    1.0 &1.0,0.03     & 0.7        &$-$1.0    &-1.1,0.3         & 0.2  &     \\
            B1SPC2         &   14.9  &   14.3,1.9    & 0.3      & 1.4   &1.7,0.4    & 0.9       &           &                     &           &    1.0 &1.0,0.1      & 0.2        &$-$1.0    &-1.4,0.6         & 0.7   &    \\
            P0             &   18.2  &   18.1,1.3    & 0.1      & 1.4   &1.6,0.2    & 1.4       &           &                     &           &    1.0 &1.0,0.1     & 0.2        &$-$1.0    &-1.3,0.3         & 1.3   &    \\
            P1             &   18.2  &   18.7,1.0    & 0.5      & 1.4   &1.5,0.2    & 1.2       &           &                     &           &    1.0 &1.0,0.04     & 0.0       &$-$1.0    &-1.2,0.2         & 1.0   &    \\
            \hline
             NPC\emph{b}  &1.1      &   0.8,0.2     & 1.3      & 0.14  &0.23,0.05  & 1.8       &           &                     &           &    0.9 &1.1,0.1      & 1.1        &$-$1.0    &-1.8,0.6         & 1.3   & $~~\approx$0.7   \\
             B2SPC1\emph{b}&1.7      &   0.7,0.6     & 1.7      & 0.11  &0.13,0.05  & 0.4       &           &                     &           &    1.0 &0.4,0.4      & 1.5        &$-$1.0    &-1.0,0.8         & 0.04  &     \\
             B1SPC1\emph{b}&1.4      &   1.4,0.2     & 0.2      & 0.11  &0.11,0.04  & 0.0       &           &                     &           &    1.0 &0.9,0.1      & 0.9        &$-$1.0    &-1.0,0.8         & 0.05   &    \\
           
            \hline
           \hline

            \end{tabular}
         \label{tab:C}
\end{sidewaystable*}

\section*{Acknowledgments}
VV thanks Federica Govoni, Mark Birkinshaw, Matteo Murgia, and Gabriele Giovannini for useful discussions.
The implementation of the code makes use of the NIFTY package by
\cite{SBJ2013} and of the cosmology calculator by Ned Wright
(www.astro.ucla.edu/$\sim$wright).  This research was supported by the
DFG Forschengruppe 1254 ``Magnetisation of Interstellar and
Intergalactic Media: The Prospects of Low-Frequency Radio
Observations''. K.T. is supported by Grant-in-Aid from the Ministry of Education,
Culture, Sports, Science and Technology (MEXT) of Japan, Nos. 24340048 and 26610048. TS and HR acknowledge support from the ERC Advanced Investigator programme NewClusters 321271.

\bibliographystyle{plainnat}

\appendix
\begin{onecolumn}

\section{Extragalactic Faraday depth variance}
\label{derivation}
In this appendix we derive Eq.\,(\ref{diffenv}) from Eq.\,(5).
In Eq.\,(\ref{diffenv}) we define the extragalactic Faraday depth variance as
\begin{equation}
  \langle\phi_{{\rm e}, i}^2\rangle\approx a_0^2\int_0^{z_{i}}\frac{\mathrm{d}l}{\mathrm{d}z}\int_0^{z_{i}}\frac{\mathrm{d}l^{\prime}}{\mathrm{d}z^{\prime}}\langle n_{\rm e}(z) n_{\rm e}(z^{\prime}) B_{l}(z)B_{l}(z^{\prime})\rangle\mathrm{d}z\mathrm{d}z^{\prime}.
\end{equation}
This definition can be expressed also as a function of distance along the line of sight $l$
\begin{equation}
  \langle\phi_{{\rm e}, i}^2\rangle\approx a_0^2\int_0^{l(z_{i})}\mathrm{d}l\int_0^{l(z_{i})}\mathrm{d}l^{\prime}\langle n_{\rm e}(l) n_{\rm e}(l^{\prime}) B_{l}(l)B_{l}(l^{\prime})\rangle.
\end{equation}
If we assume that the thermal gas density is not characterized by significant fluctuations and define a new variable $r=l^{\prime}-l$, we have
\begin{equation}
 \langle n_{\rm e}(l) n_{\rm e}(l^{\prime}) B_{l}(l)B_{l}(l^{\prime})\rangle\approx \langle n_{\rm e}^2(l)\rangle\langle B_{l}(l)B_{l}(l+r)\rangle_{(B_r|n_{\rm e})}=\langle n_{\rm e}^2(l)\rangle C_B(r|n_{\rm e}),
\end{equation}
where $C_B(r|n_{\rm e})$ is the conditional magnetic field correlation function for an environment with thermal gas density $n_{\rm e}$. Indeed, we expect the magnetic field strength to be a function of the thermal gas density. 
With these new definitions, the extragalactic Faraday depth variance reads
\begin{equation}
  \langle\phi_{{\rm e}, i}^2\rangle\approx a_0^2\int_0^{l(z_{i})}\mathrm{d}l\int_{-l}^{l(z_{i})-l}\mathrm{d}r\, \langle n_{\rm e}^2(l)\rangle C_B(r|n_{\rm e}),
\end{equation}
which can be further simplified if 
we consider the limit of a infinitely far away source, 
\begin{equation}
  \langle\phi_{{\rm e}, i}^2\rangle\approx a_0^2\int_0^{z_{i}}\frac{\mathrm{d}l}{\mathrm{d}z}\mathrm{d}z\,\langle n_{\rm e}^2(l(z))\rangle\int_{-\infty}^{\infty}\mathrm{d}r C_B(r|n_{\rm e}).
\end{equation}
Recalling the definition of correlation length,
\begin{equation}
  \Lambda_{l}=\int \mathrm{d}r\frac{C_B(r|n_{\rm e})}{\langle B_l^2\rangle}=\int \mathrm{d}l^{\prime}\frac{\langle B(l)B(l^{\prime})\rangle}{\langle B(l)^2\rangle},
 \end{equation}
we obtain
\begin{equation}
  \langle\phi_{{\rm e}, i}^2\rangle= a_0^2\int_0^{z_{i}}\frac{\mathrm{d}l}{\mathrm{d}z}\mathrm{d}z\,\langle n_{\rm e}^2(l(z))\rangle\Lambda_{l}(n_{\rm e})\langle B^2\rangle_{(B|n_{\rm e})}.
\end{equation}
As described in Sec.\,\ref{approach}, in a homogeneous Universe $n_{\rm e}=n_{\rm e 0}(1+z)^3$, $\Lambda_{\rm l}=\Lambda_{\rm l 0}(1+z)^{-1}$, and $\langle B^2\rangle=\langle B_0^2\rangle(1+z)^4$. Therefore, 
\begin{equation}
  \langle\phi_{{\rm e}, i}^2\rangle= a_0^2\int_0^{z_{i}}\langle n_{\rm 0}^2\rangle\Lambda_{\rm 0 l}\langle B_{0 l}^2\rangle\frac{\mathrm{d}l}{\mathrm{d}z}(1+z)^5\mathrm{d}z,
\end{equation}
where the increment in wavelength due to the expansion of the Universe has been taken into account as well and we have assumed $z(l)\approx z(l+r)$. If we use the definition of proper displacement along a light-ray 
$\mathrm{d}l/\mathrm{d}z=c(1+z)^{-1}/H(z)$, this finally leads to Eq.\,(\ref{diffenv}),
\begin{equation}
  \langle\phi_{{\rm e}, i}^2\rangle= a_0^2\int_0^{z_{i}}\langle n_{\rm 0}^2\rangle \Lambda_{\rm 0 l}\langle B_{0 l}^2\rangle\frac{c}{H(z)}(1+z)^4\mathrm{d}z.
\end{equation}

In an inhomogeneous Universe with different environments, the differential variance in each environment is
\begin{equation}
\frac{\mathrm{d}\sigma_{RM}^2(n_{\rm e})}{\mathrm{d}x}=\langle n_{\rm e}^2(z)\rangle\langle
B^2\rangle_{(B|n_{\rm e})}.
\end{equation}
Therefore, it follows 
\begin{equation}
  \langle\phi_{{\rm e}, i}^2\rangle= a_0^2\int_0^{z_i}\frac{\mathrm{d}x}{\mathrm{d}z}\frac{\mathrm{d}z}{(1+z)^4}\int \mathrm{d}n_{\rm e}P(n_{\rm  e}|z)\frac{\mathrm{d}\sigma_{RM}^2(n_{\rm e})}{\mathrm{d}x},
\end{equation}
where the integral can be replaced by a discrete sum over typical environments (see also Eq.\,(\ref{lls})).

\section{Convergence}
\label{convergenceappendix}

We start the algorithm from two random positions in the $\Theta$-space
and then explore the space\footnote{We tuned the variance
  of the Gaussian step proposal to ensure an acceptance rate of the
  Metropolis-Hastings algorithm of approximately $15$-$30$\%} until
convergence. Each of these two sequences of steps in the $\Theta$-space is called Gibbs chain.
To assess convergence of
each Gibbs chain, we require the following
conditions to be satisfied:
\begin{itemize}
\item the number of steps taken from each chain to be at least
  about $10l_{c}$ for each parameter, where $l_{c}$ is the number of
  steps at which the correlation coefficient drops to 10\%;
\item the Gelman and Rubin test \citep{GR1992,BG1997}. We
  evaluate the intra-chain variance
  \begin{equation}
  I=\frac{1}{m}\Sigma_{j=1}^{m}s_j^2
  \end{equation}
  and the  inter-chain variance 
  \begin{equation}
    B=\frac{n}{m-1}\Sigma_{j=1}^{m}(\theta_j-\theta)^2
    \end{equation}
  for our $\Theta$ parameters and use them to compute the potential scale
  reduction factor $R$ defined as
\begin{equation}
R=\sqrt{1-\frac{1}{n}+\frac{B}{nI}},
\end{equation}
where $m$ is the number of chains, $n$ is the half-length of each chain, $\theta_j$ and $s_j$ are respectively the mean and the standard deviation of the $j$th chain, while $\theta_j$ is the mean of the chain $j$ and $\theta$ is the variance of the chain means. We require $R=1$ within a
few percents for each parameter.
\end{itemize}
We consider the chains to be converged when both these conditions are satisfied.
\begin{figure*}[ht]
       \centering
            \includegraphics[width=7cm, angle=0]{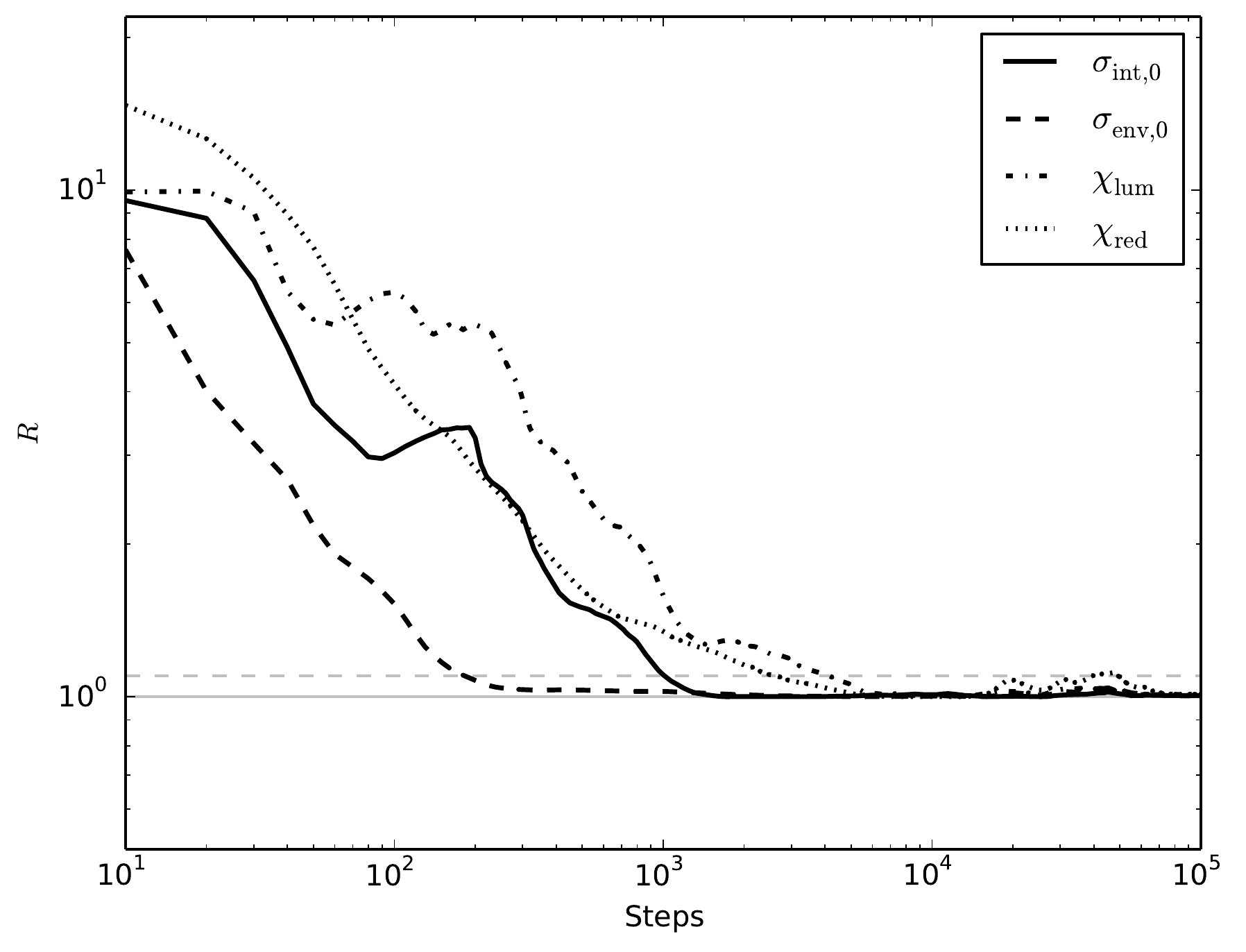} 
            
        \flushleft       
    \caption{Potential scale reduction factor $R$ evaluated every
      10 samples for the parameters $\sigma_{\rm int,0}$, $\sigma_{\rm env,0}$, $\chi_{\rm lum}$, and $\chi_{\rm red}$,
       as a function of the number of steps in the MCMC. The continuous and the dashed lines represent a potential scale
      reduction factor equal to 1.0 and 1.1, respectively. It can be seen that the Gelman and Rubin test indicates convergence after typically a few thousand steps.}
  \label{scale_factor}
\end{figure*}

As an example, in Fig.\,\ref{scale_factor} we show the plot of
the potential scale reduction factor $R$ versus the number of steps
for the $\Theta$ parameters in the scenario 2C1. 
The potential scale reduction factor
has been evaluated every 10 samples.  After about 5000 steps we obtain
$R=1$ within a few percent. In Table~\ref{tab:D} the potential scale
reduction factor $R$ for each parameter of each test is reported.
For present-instrument (scenario 2C1), LOFAR (scenario NPC\emph{a}) and
 SKA (scenario B2SPC1a) observations, in Fig.\,\ref{corr_coeff} we show the correlation
coefficient $\rho_j$
\begin{equation}
\rho_j(k)=\frac{\sum_{s=1}^{2n-k}(x_s-\theta_j)(x_{s+k}-\theta_j)}{\sum_{s=1}^{2n}(x_{s}-\theta_j)},
  \end{equation}
as a function of the number of steps for each parameter
$\sigma_{{\rm int},0}$, $\sigma_{{\rm env},0}$, $\chi_{\rm lum}$, and $\chi_{\rm red}$.
Here $x_s$ is the value of one of these parameters for a given step. For all
the parameters in the scenario 2C1, $l_{\rm c}$ turns out to be
$\approx$3000.  while for the scenarios NPC\emph{a} and B2SPC1\emph{a}, the correlation
length is about 500-1000.  In these plots the steps in the burn-in phase
have been discarded by visual inspection.

\begin{figure*}[ht]
       \centering
            \includegraphics[width=7cm, angle=0]{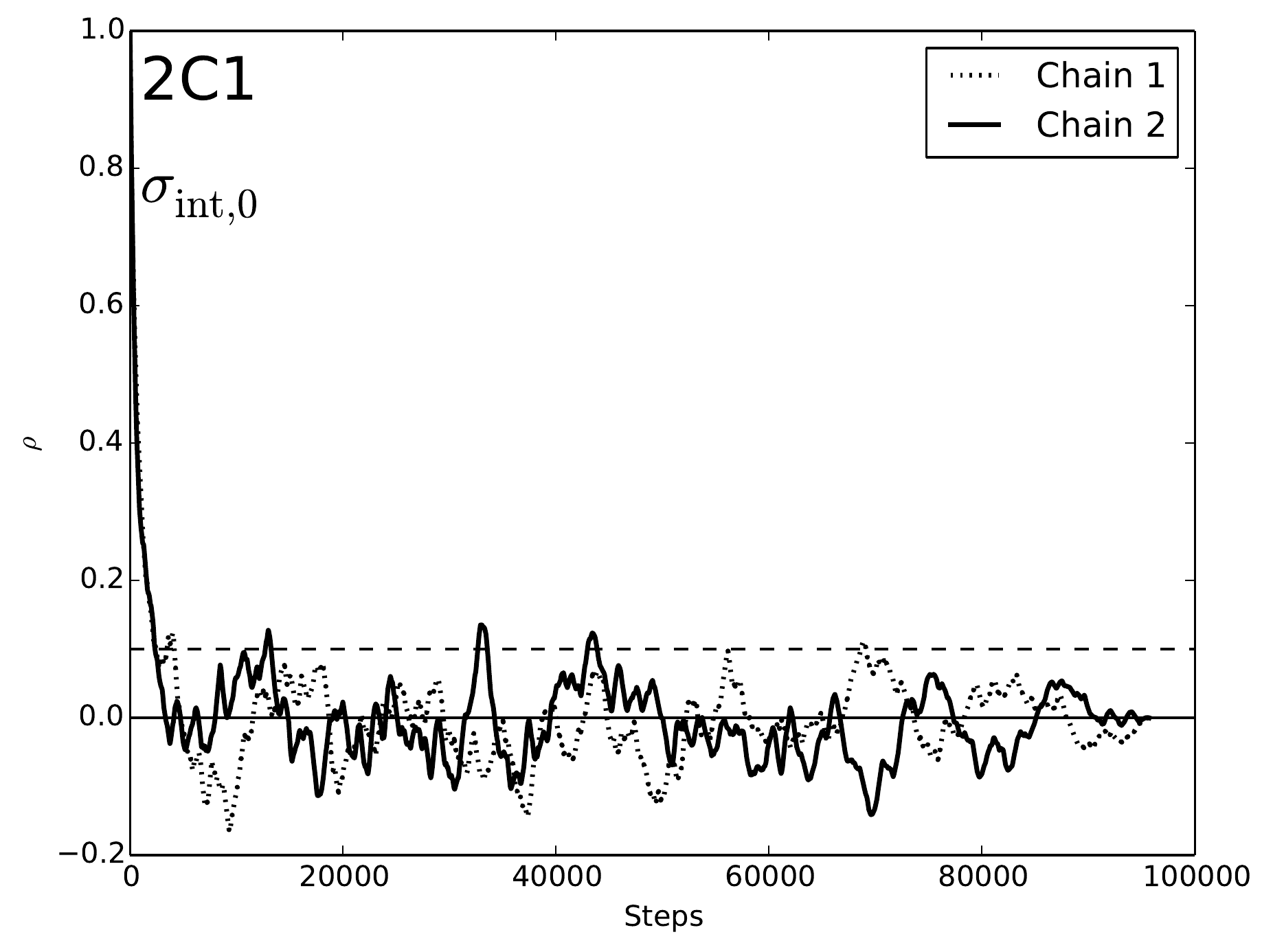} 
            \includegraphics[width=7cm, angle=0]{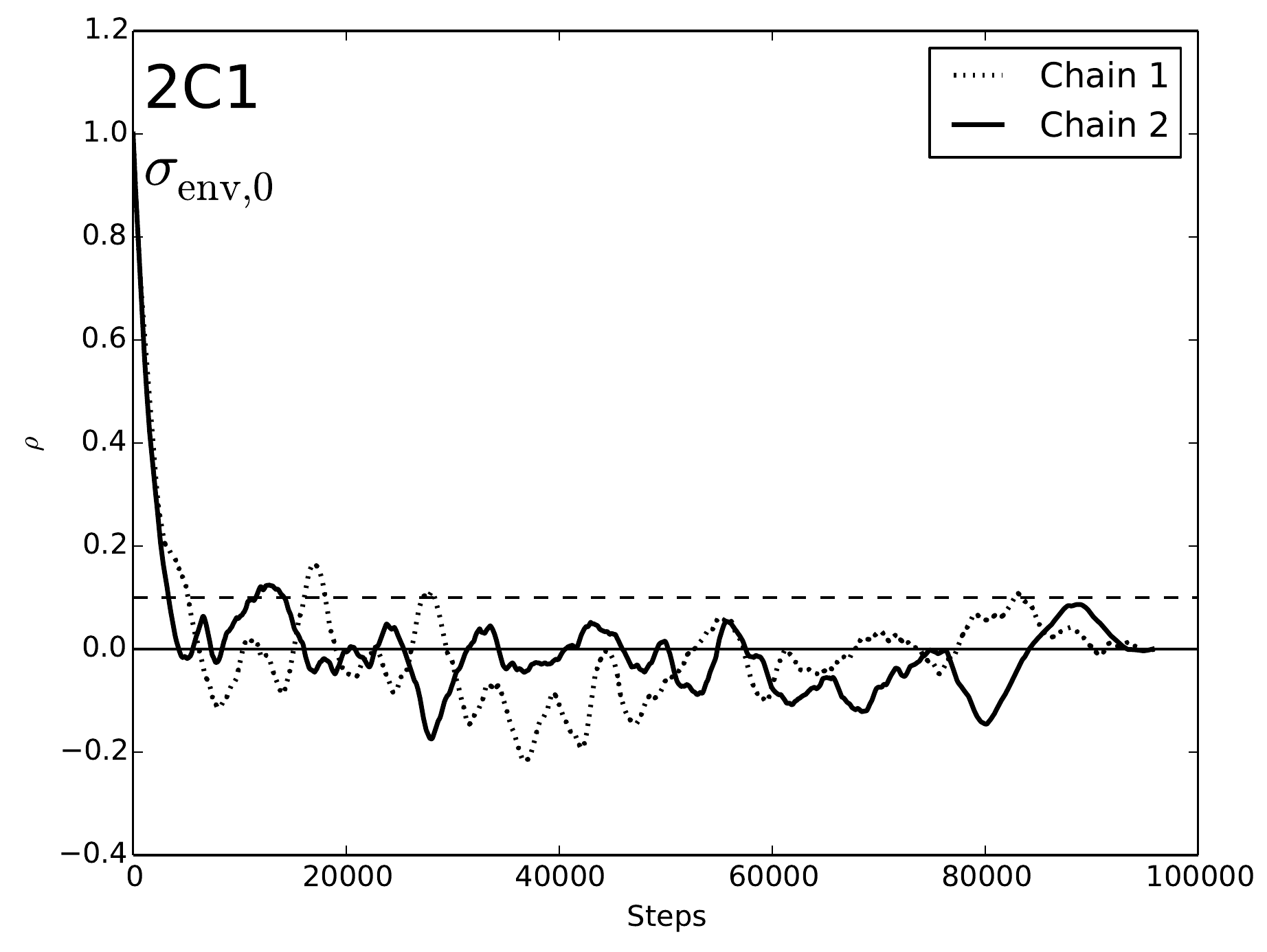}\\ 
            \includegraphics[width=7cm, angle=0]{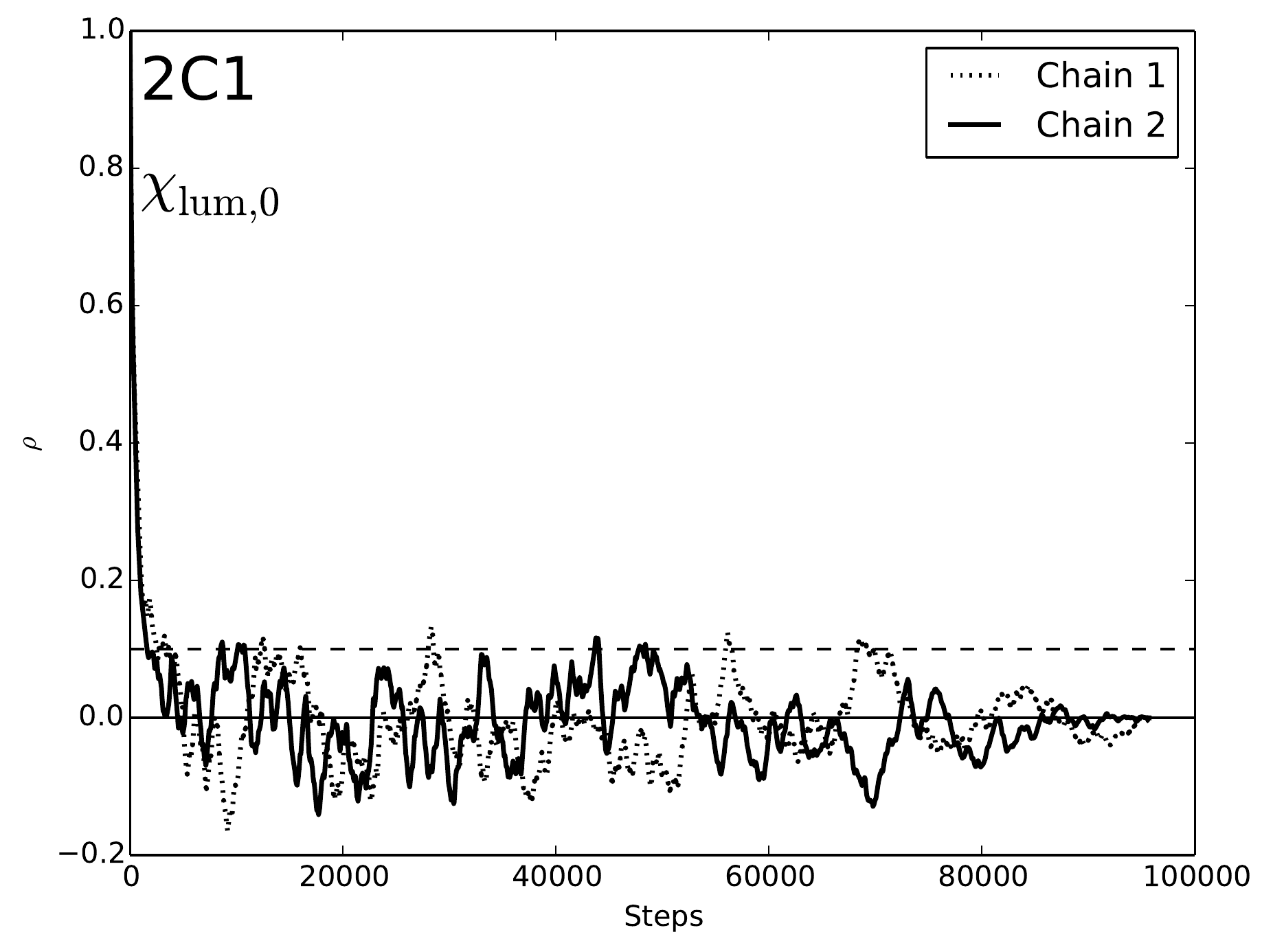} 
            \includegraphics[width=7cm, angle=0]{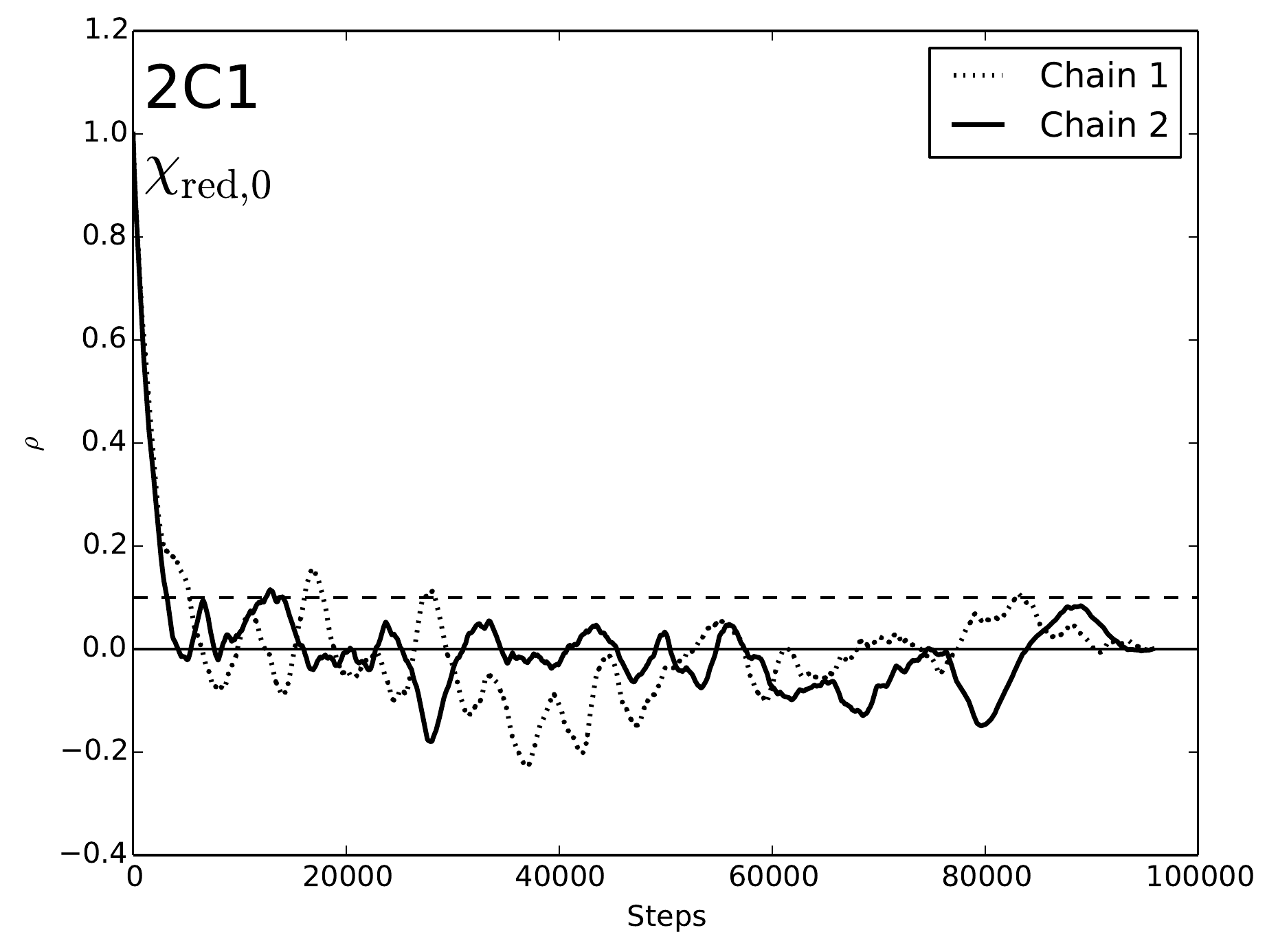}\\
            \includegraphics[width=7cm, angle=0]{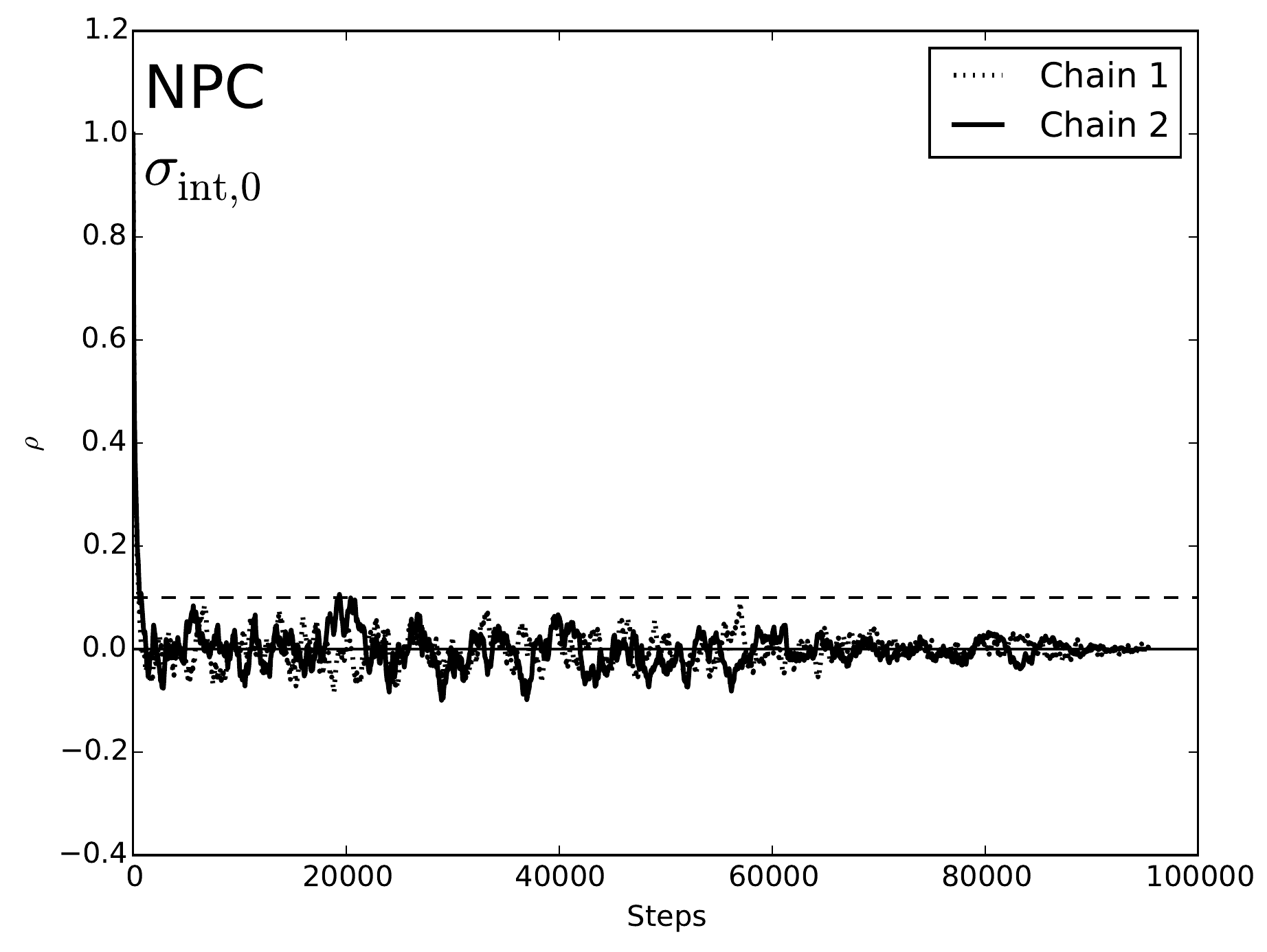} 
            \includegraphics[width=7cm, angle=0]{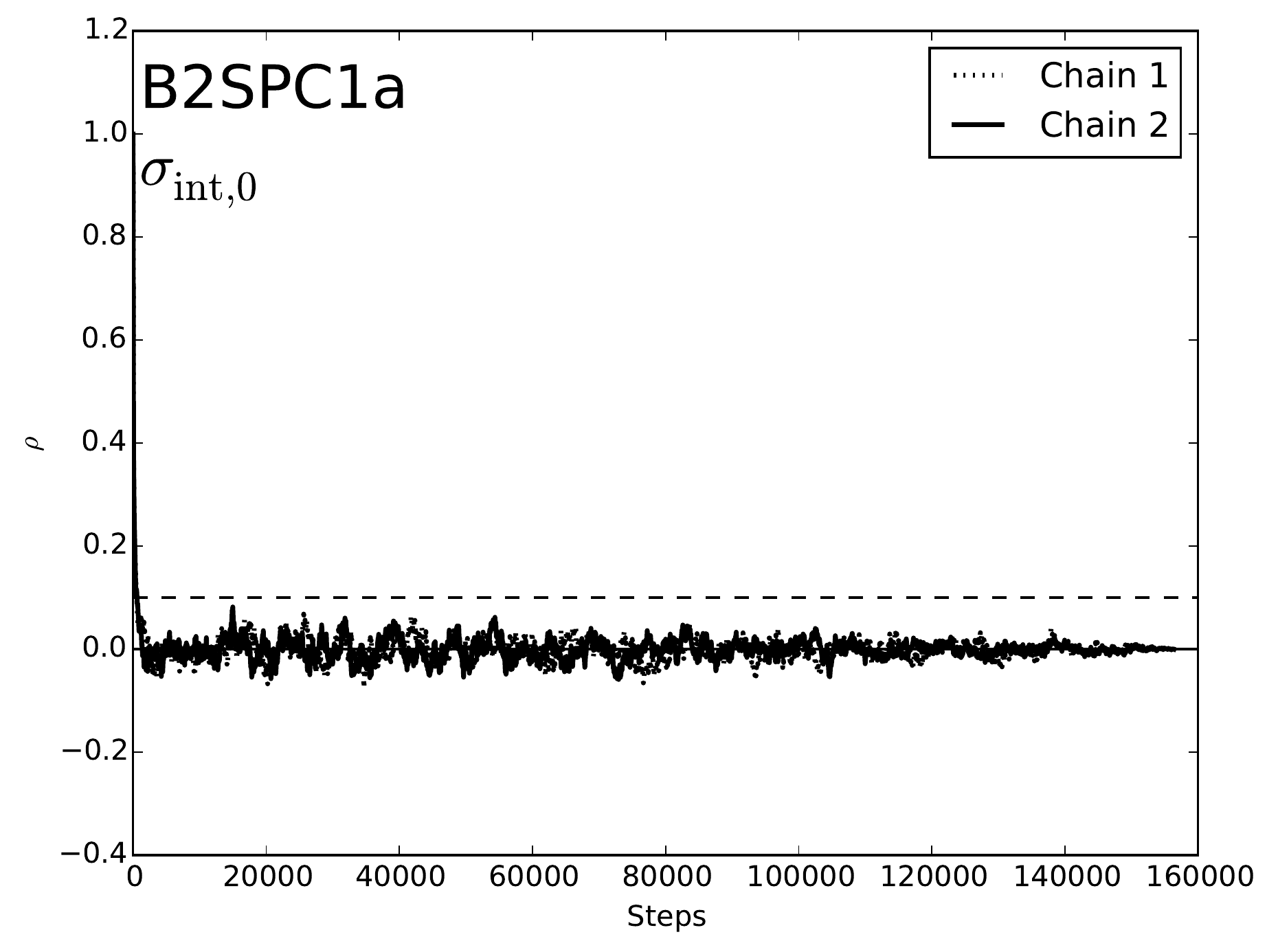}\\ 
        \flushleft       
    \caption{Correlation coefficient $\rho$ as a function of the number of
      steps in the MCMC for the four $\Theta$ parameters in the 2C1 test (top and middle panels) and for the
      $\sigma_{\rm int,0}$ parameter in the LOFAR NPC and the SKA B2SPC1\emph{a} tests (bottom panels). The steps
      in the burn-in phase have been discarded. The dashed line
      indicates a correlation coefficient of 0.1, while the continuous line the zero-level.}
  \label{corr_coeff}
\end{figure*}

\begin{table}[!t]
  \caption{Potential scale reduction factor $R$ from the Gelman and Rubin test.}
  \small
        \centering
        \begin{tabular}{cccccc}
          \hline
           \hline
            ID &$\sigma_{{\rm int},0}$ &$\sigma_{{\rm env},0}$ &$\chi_{\rm c/lat}$&$\chi_{\rm lum}$  &  $\chi_{\rm red}$   \\
            \hline
            2C1           &  1.00  & 1.01      &      & 1.00   & 1.01\\
            2C2           &  1.00  & 1.01      &      & 1.00   & 1.01\\
            3C1           &  1.01  & 1.12      & 1.07 & 1.00   & 1.09\\
            3C2           &  1.00  & 1.00      & 1.00 & 1.01   & 1.00\\
            LD1           &  1.03  & 1.05      & 1.00 & 1.02   & 1.04\\
            LD2           &  1.00  & 1.01      & 1.01 & 1.00   & 1.01\\
            GW            &  1.00  & 1.00      &      & 1.00   & 1.00\\
            NPC\emph{a}   &  1.00  & 1.00      &      & 1.00   & 1.00\\
            POSSUM        &  1.02  & 1.03      &      & 1.01   & 1.03\\
            B2SPC1\emph{a}   &  1.00  & 1.01      &      & 1.00   & 1.01\\
            B2SPC2           &  1.00  & 1.00      &      & 1.00   & 1.00\\
            B1SPC1\emph{a}   &  1.00  & 1.00      &      & 1.00   & 1.00\\
            B1SPC2           &  1.00  & 1.00      &      & 1.00   & 1.00\\ 
            P0            &  1.02  & 1.03      &      & 1.01   & 1.03\\
            P1            &  1.01  & 1.02      &      & 1.00   & 1.02\\
            \hline
            NPC\emph{b}   &  1.00  & 1.00      &      & 1.00   & 1.00\\
            B2SPC1\emph{b}   &  1.45  & 1.08      &      & 1.01   & 1.03\\
            B1SPC1\emph{b}   &  1.04  & 1.02      &      & 1.01   & 1.01\\
            \hline
           \hline
        \end{tabular}
        \begin{flushleft}
          \normalsize
            \end{flushleft}
          \label{tab:D}
        
    \end{table}

  \section{Alternative scenarios}
  \label{alternative_app}

In this appendix we present the application to scenarios including three components, representing:
\begin{itemize}
\item an intrinsic, an environmental, and a constant contribution (scenario 3C),
\begin{equation}
  \sigma_{{\rm e},i}^2(z_i,\Theta)= \left(\frac{L}{L_0}\right)^{\chi_{\rm lum}}\frac{\sigma^2_{\rm int,0}}{(1+z_i)^{4}}+ \frac{D_i(z_i,\chi_{\rm red})}{D_0}\sigma^2_{\rm env, 0}+\sigma^2_{\rm c}
\label{constant}
\end{equation}
The constant contribution $\sigma^2_{\rm c}$ takes into account terms that are not
described by the parameterization of the other two but that could
nevertheless be present in our data (e.g. the ionosphere, under the
assumption that this does not show any direction-dependence);
\item an intrinsic, an environmental, and a latitude-dependent contribution  (scenario LD),
\begin{equation}
  \sigma_{{\rm e},i}^2(z_i,\Theta)= \left(\frac{L}{L_0}\right)^{\chi_{\rm lum}}\frac{\sigma^2_{\rm int,0}}{(1+z_i)^{4}}+ \frac{D_i(z_i,\chi_{\rm red})}{D_0}\sigma^2_{\rm env, 0}+p(b)\sigma^2_{\rm lat}
\label{latitude}
\end{equation}
where $p(b)$ is the Galactic profile from \cite{OJG2014}. The latitude-dependent contribution $p(b)\sigma^2_{\rm lat}$ may explain a residual latitude
dependence not taken into account in the modeling of \cite{OJG2014}, e.g. an 
uncorrelated Galactic signal not captured by their analysis. 

\end{itemize}
As for the simplest 2C-scenario, for each of these scenarios we run
two tests corresponding to 41632 and 4003 lines of sight.
Fig.\,\ref{3comp} and Fig.\,\ref{latdep} show the results respectively when a third
constant component (tests 3C1 and 3C2) and
latitude-dependent component (tests LD1 and LD2) are included. In both figures, we show 
the plots for 41632 lines of sight in panel (a) and for
4003 lines of sight in panel (b).

These plots indicate that the algorithm performs well also when three
components are considered. As expected, the values of the $\Theta$
parameters recovered are less accurate when a lower number of lines of
sight is used. The additional parameters tend to lead to a slight increase in the posterior uncertainty
for the other parameters when comparing with the results of 2C1 and 2C2.

\begin{figure*}[ht]
  \centering
            \begin{overpic} [width=15.5cm, angle=0]{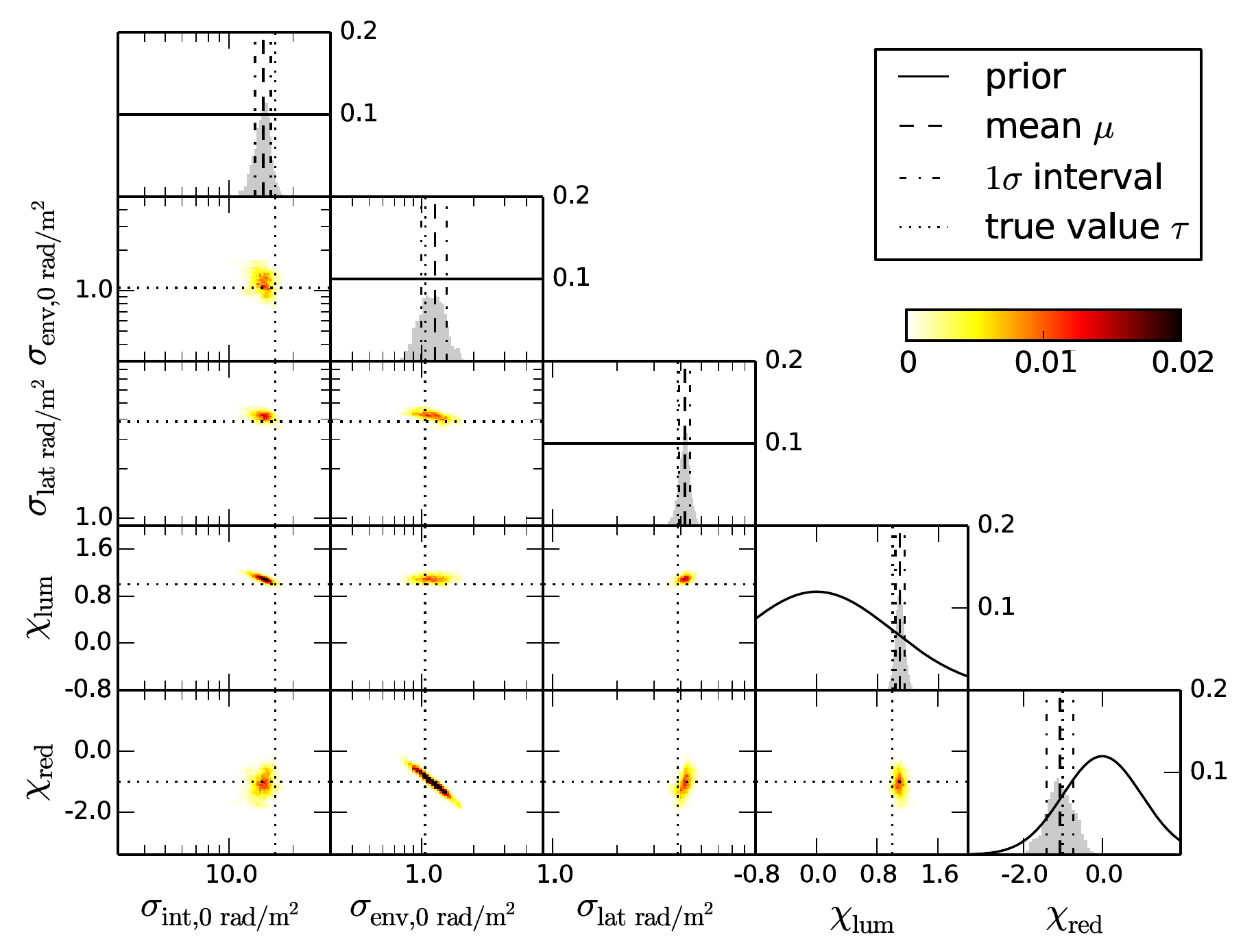} \put(-4,78){(a)} \end{overpic} \\
            \begin{overpic} [width=15.5cm, angle=0]{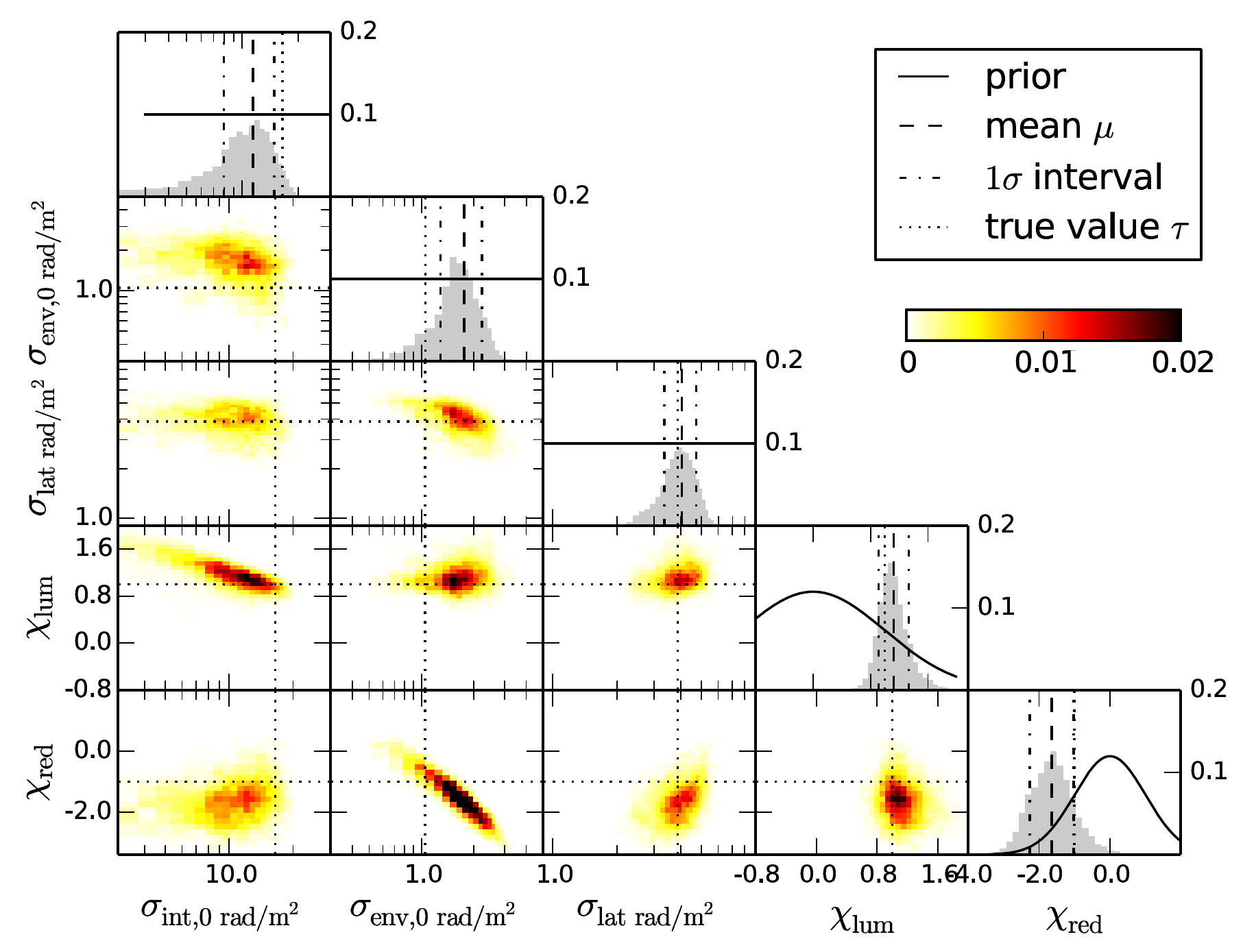} \put(-4,78){(b)} \end{overpic} \\
        \flushleft       
 
    \caption{Results obtained with a three-component scenario, including
      a constant contribution, for
      41632 (3C1) lines of sight (a) and 4003 (3C2) lines of sight (b).}
  \label{3comp}
\end{figure*}
\begin{figure*}[ht]
  \centering
           \begin{overpic} [width=15.5cm, angle=0]{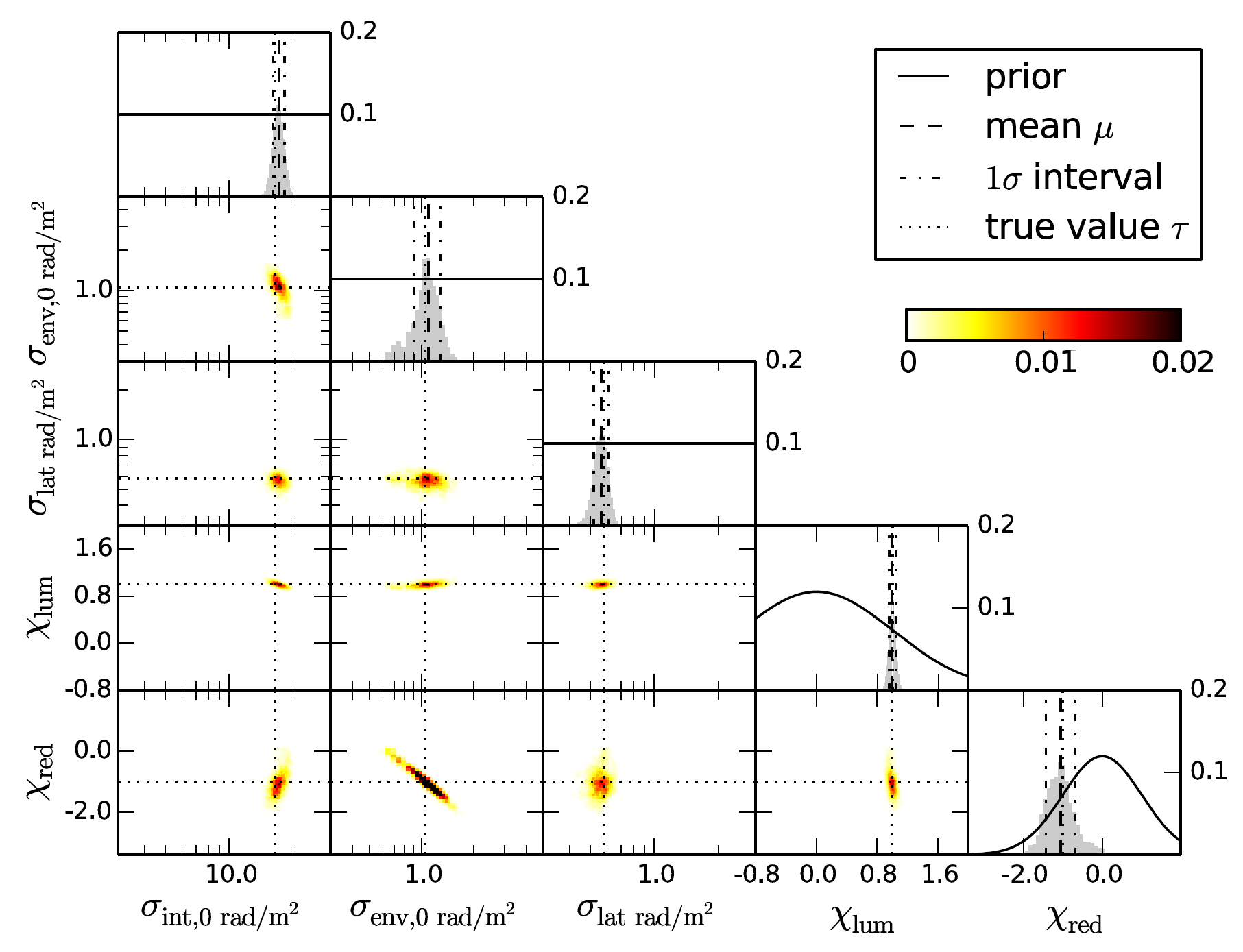} \put(-4,78){(a)}\end{overpic} \\
            \begin{overpic} [width=15.5cm, angle=0]{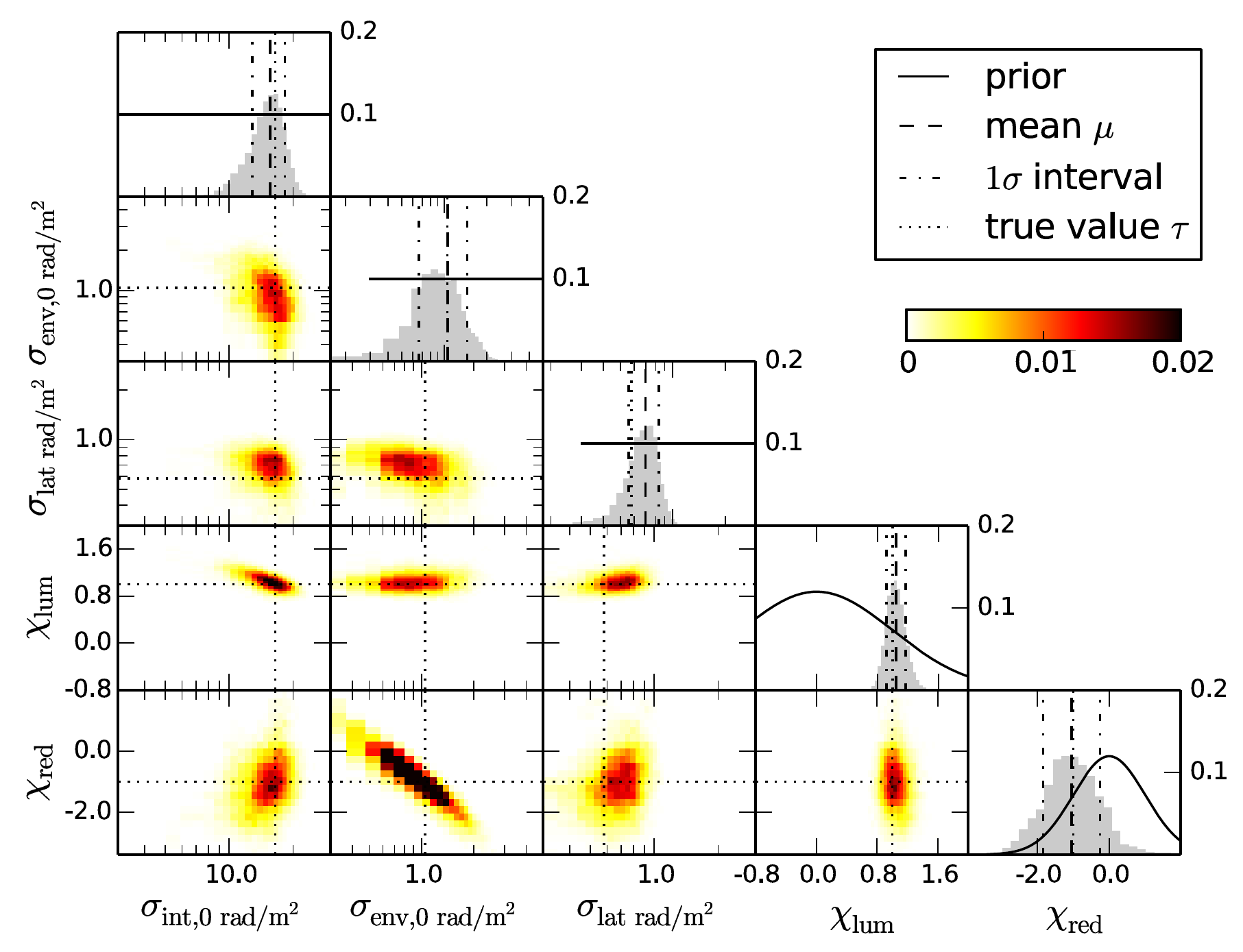} \put(-4,78){(b)} \end{overpic} \\
        \flushleft       
     \caption{Results obtained with a three-component scenario, including
       a latitude-dependent contribution, for
       41632 (LD1) lines of sight (a) and 4003 (LD2) lines of sight (b).}
  \label{latdep}
\end{figure*}
\section{Priors}
\label{priorappendix}
In order to get a data-driven solution and to keep our assumptions as
general as possible, an uninformative prior should be adopted for the
$\Theta$ parameters.  Since we included in our model all the main
redshift and luminosity dependencies, the Gaussian prior in Eq.\,(\ref{gaussprior}) is suitable for $\chi_{\rm lum}$ and $\chi_{\rm red}$. Concerning
$\sigma_{{\rm int},0}$ and $\sigma_{{\rm env},0}$, we may ask if different priors
can have an impact on our results. In Sect.\,\ref{tests}, we adopted a
flat-prior,
\begin{equation}
P(\sigma)= {\rm const}.
\end{equation}
In this appendix we present two tests corresponding to 
extreme choices of these priors. Indeed, we considered a flat
prior in $\sigma^2$ (scenario P1)
\begin{equation}
P(\sigma^2)= {\rm const},
\end{equation}
and a flat prior in $\ln(\sigma^2)$ (scenario P0)
  \begin{equation}
P(\ln(\sigma^2))= {\rm const}.
  \end{equation}

In Fig.\,\ref{prior} we show the results for the two priors: flat in
$\sigma^2$ in panel (a) and flat in $\ln(\sigma^2)$ in panel (b).

The first choice is an optimistic prior, since it implies a
suppression of $\sigma$ values $<<1$, pushing for the recovery of 
larger, possibly $\sim 1$ values of $\sigma$. The second choice is a
pessimistic prior since it would weight all small and large
$\sigma$ values in the same way, favoring negligible extragalactic
contributions, easily compatible with the data due to the shape of the
likelihood.
We stress that the final results are not affected by the choice of the
prior. However, this choice has an impact on the convergence-time,
since the starting point of each chain is randomly extracted from the
prior. For example, for a flat prior in $\ln(\sigma^2)$, the chances to
extract a very small $\sigma$ value ($<< 1$) are larger than
for the prior used in Sect.\,\ref{tests}, possibly making the convergence time very long. 
For theses tests, the number of lines of sights and the assumed noise properties are the same as in the 2C1 test.  

\begin{figure*}[ht]
  \centering
  \begin{overpic} [width=15.5cm, angle=0]{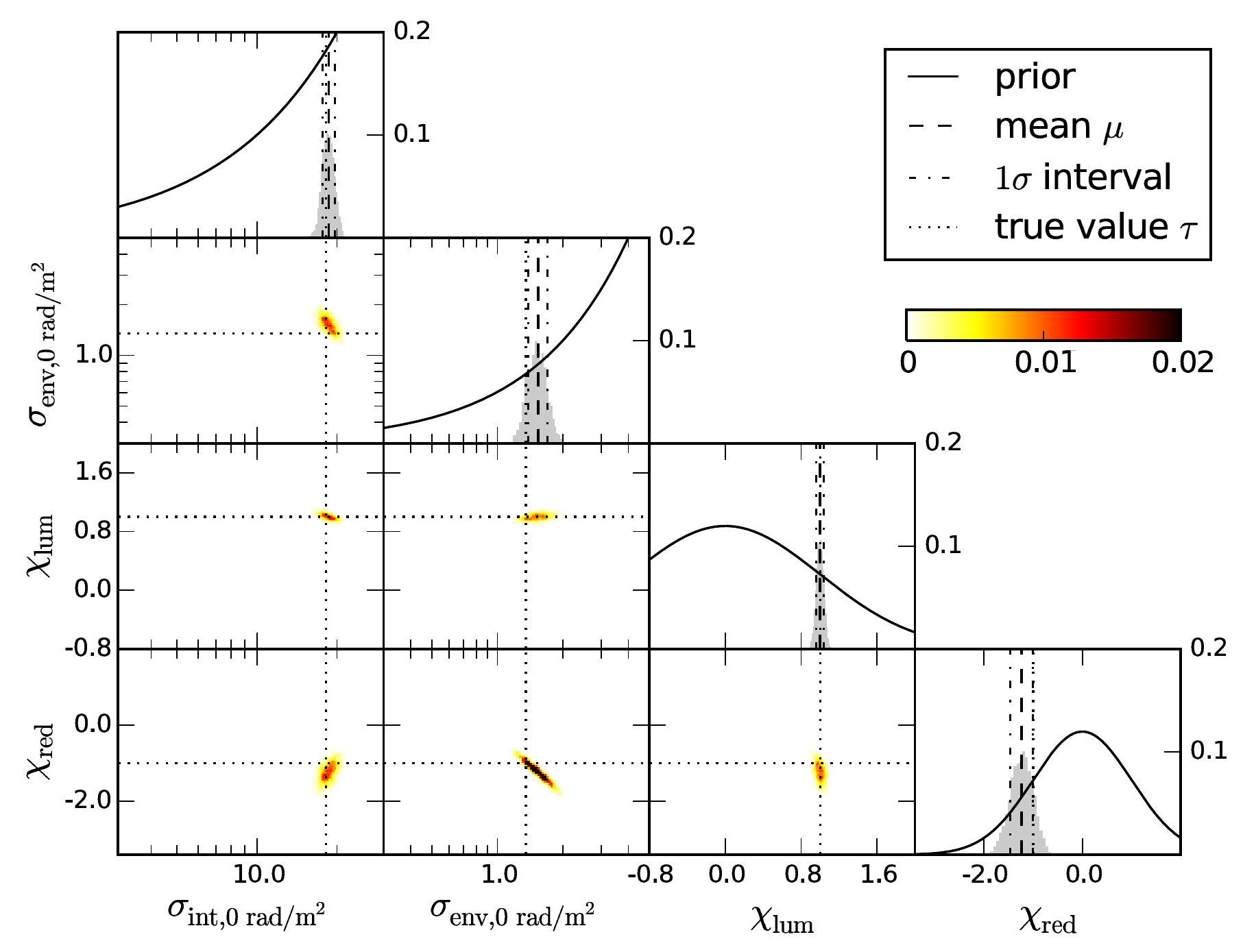} \put(-4,78){(a)}\end{overpic} \\
            \begin{overpic} [width=15.5cm, angle=0]{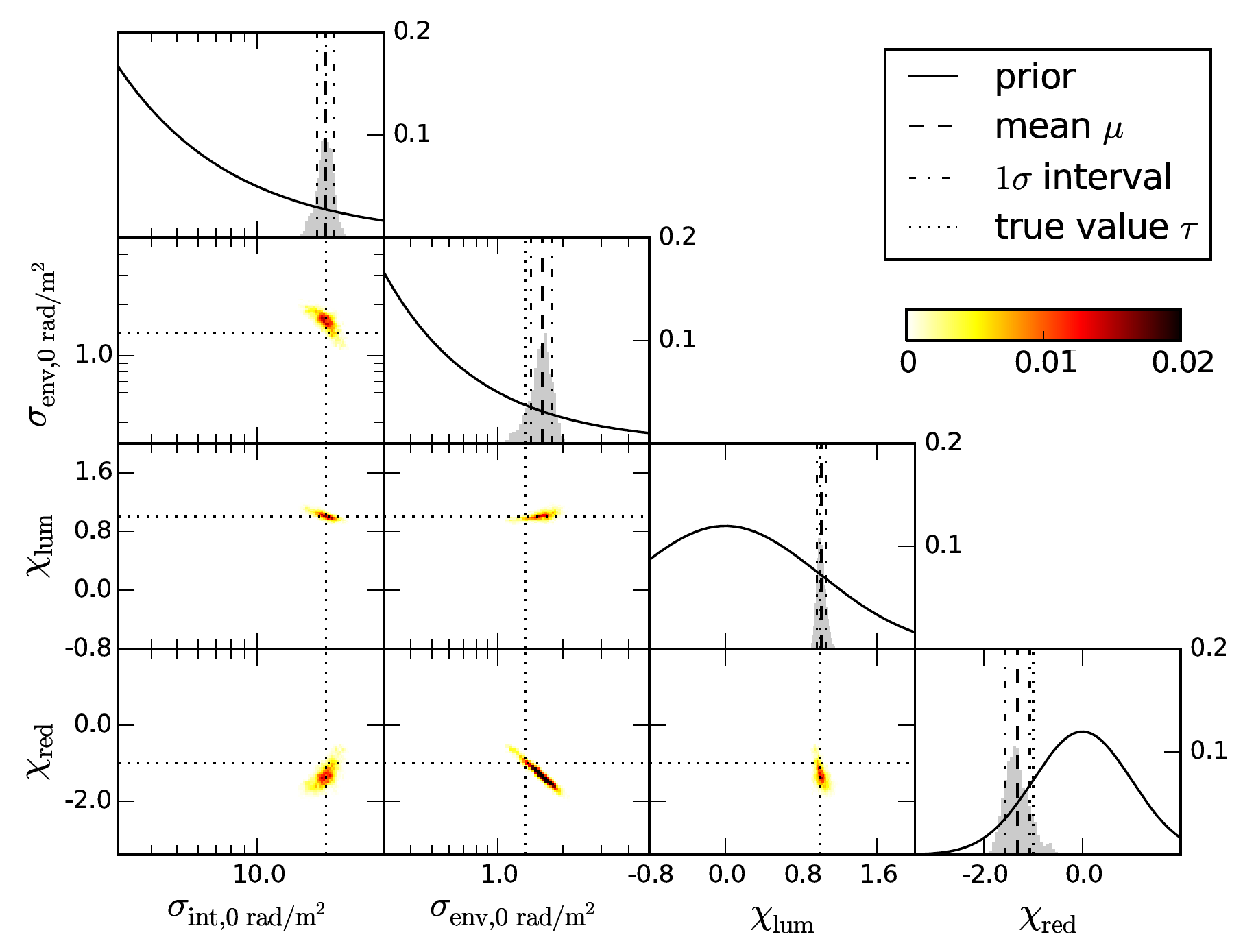} \put(-4,78){(b)} \end{overpic} \\
        \flushleft    
      \caption{Results obtained with a two-component scenario and an overall extragalactic Faraday 7\,rad\,m$^{-2}$ for (a) a
        flat prior in $\sigma^2$ (P1) and for (b) a flat prior in $\ln(\sigma^2)$
        (P0) for 41632 lines of sight.}
  \label{prior}
\end{figure*}

  \section{LOFAR}
  \label{lofar_app}

  An interesting region of the sky is represented by the Great Wall, where one of the
  largest filaments of optical galaxies has been observed (\citealt{G2005}). 
  We generate a mock collection of source coordinates 
  in a region of the sky as large as this region
 ($7.5^{\rm h}<$RA$<17.5^{\rm h}$ and
  $25^{\circ}<$Dec$<65^{\circ}$), considering a density of one polarized sources per 1.7 square degrees
  (survey of 8\,h per pointing) and Galactic latitude  $b>55^{\circ}$. 
  This results in $N_{\rm los} \approx 1000$ and a maximum uncertainty in Faraday depth $\sigma_{\rm noise}=0.05$\,rad\,m$^{-2}$. We will refer to
  this mock catalog in the following as GW.
  \begin{figure*}[ht]
  \centering
  
  \begin{overpic} [width=15.5cm, angle=0]{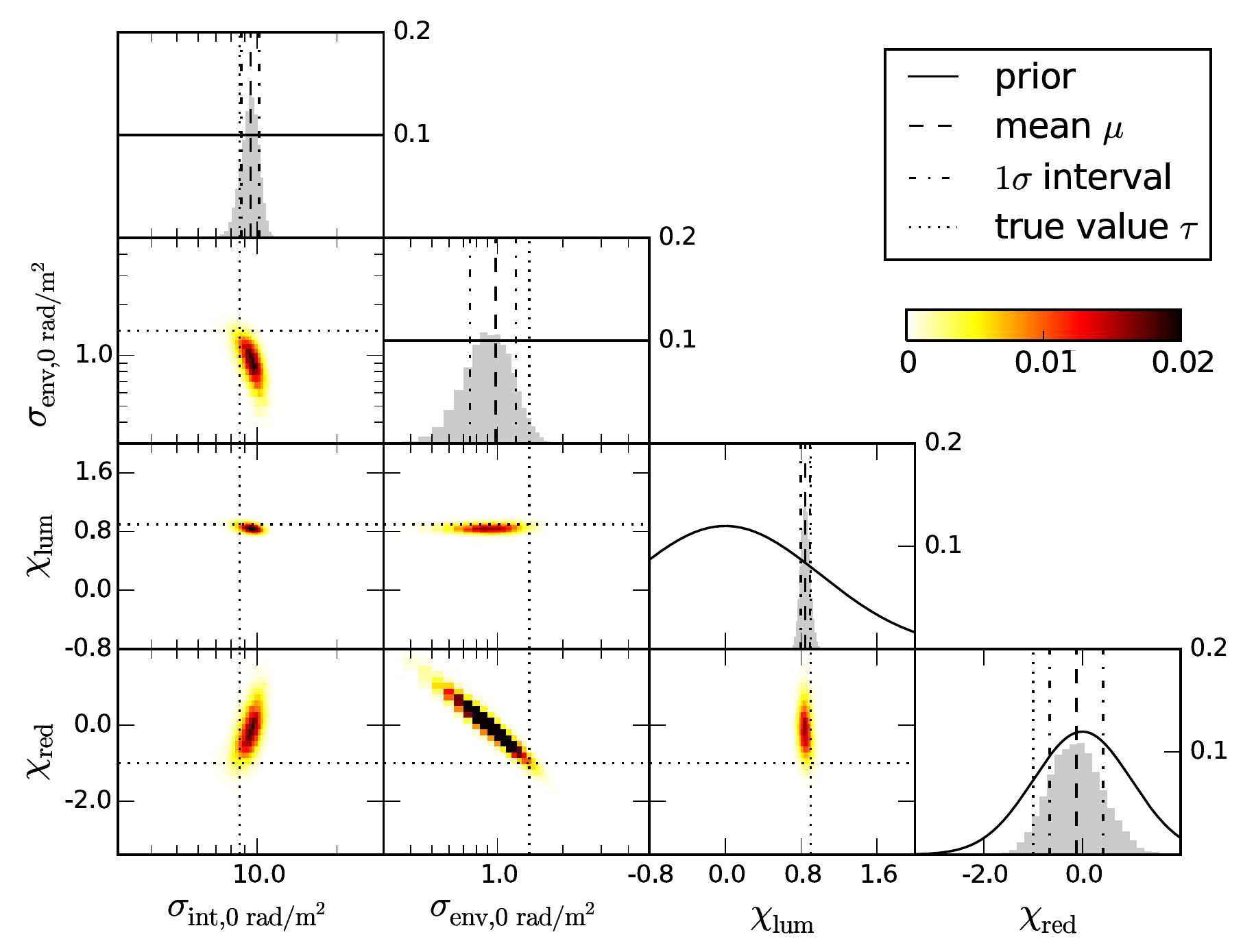}\end{overpic} \\

        \flushleft    
\caption{Results obtained with a two-component scenario for LOFAR HBA
  observations for an overall Faraday depth of $\approx$7.0\,rad\,m$^{-2}$ and $N_{\rm los}\approx$1000 (GW).}
\label{lofar_less}
\end{figure*}
The results for an overall extragalactic Faraday rotation of $\approx 7.0$\,rad\,m$^{-2}$ are shown in Fig.\,\ref{lofar_less} and indicate that it is possible to disentangle the
intrinsic and environmental contributions already with a number of lines of sight of about one thousand.

  \section{SKA}
  \label{ska_app}

  In this appendix we
  investigate the performance in the frequency band 1 (B1), meaning
  350-1050\,MHz,
  with the same collection of sources (approximately 3500) described in Sect.\,\ref{ska}.
  According to \cite{SABFK2008},
  this frequency range corresponds to an expected maximum uncertainty 
  in Faraday depth of 0.3\,rad\,m$^{-2}$, for a
  $S/N>5$. We will refer to this catalog as B1SPC1. We generate mock rotation measure catalogs 
  considering an overall extragalactic Faraday rotation of
  $\approx 7.0$\,rad\,m$^{-2}$ (B1SPC1\emph{a}) as well as $\sigma_{\rm e}\approx 0.7$\,rad\,m$^{-2}$ (B1SPC1\emph{b}).
  In Fig.\,\ref{skalow} we show the results. These tests indicate that a catalog in band 1 allows derivation of better constraints then a catalog in band 2. This becomes particularly evident for an overall extragalactic Faraday rotation of
$\approx 0.7$\,rad\,m$^{-2}$.

  Additionally, we examine the case when a smaller number of lines of sight is available.
  Therefore, we generate a catalog corresponding to a density of one polarized source per
  three square degrees, including
  all the sources with Galactic latitude $b<-55^{\circ}$. This translates in
  $N_{\rm los}\approx 1000$. We will refer to these catalogs as
  B1SPC2 and B2SPC2, respectively for the frequency ranges corresponding to band 1 and band 2.
  We generate these catalogs assuming
  an overall extragalactic Faraday rotation of $\approx$7\,rad\,m$^{-2}$.
  In Fig.\,\ref{ska_even_less_loss} we show the results for both frequency bands.

When data in the frequency range from 350 to 1670\,MHz are used,
already a number of lines of sight of the order of one thousand is enough
to disentangle the contribution intrinsic to the source from the one
due to the environment between the source and the observer for $\sigma_{\rm e}\approx 7.0$\,rad/m$^2$. 
Our results indicate that the posterior distributions narrow moving to lower frequencies
with the best performance obtained in the frequency band 350-1050\,MHz.
At higher frequencies the posteriors are broader but with mean values still in agreement with the real values.

\begin{figure*}[ht]
  \centering
  \begin{overpic} [width=15.5cm, angle=0]{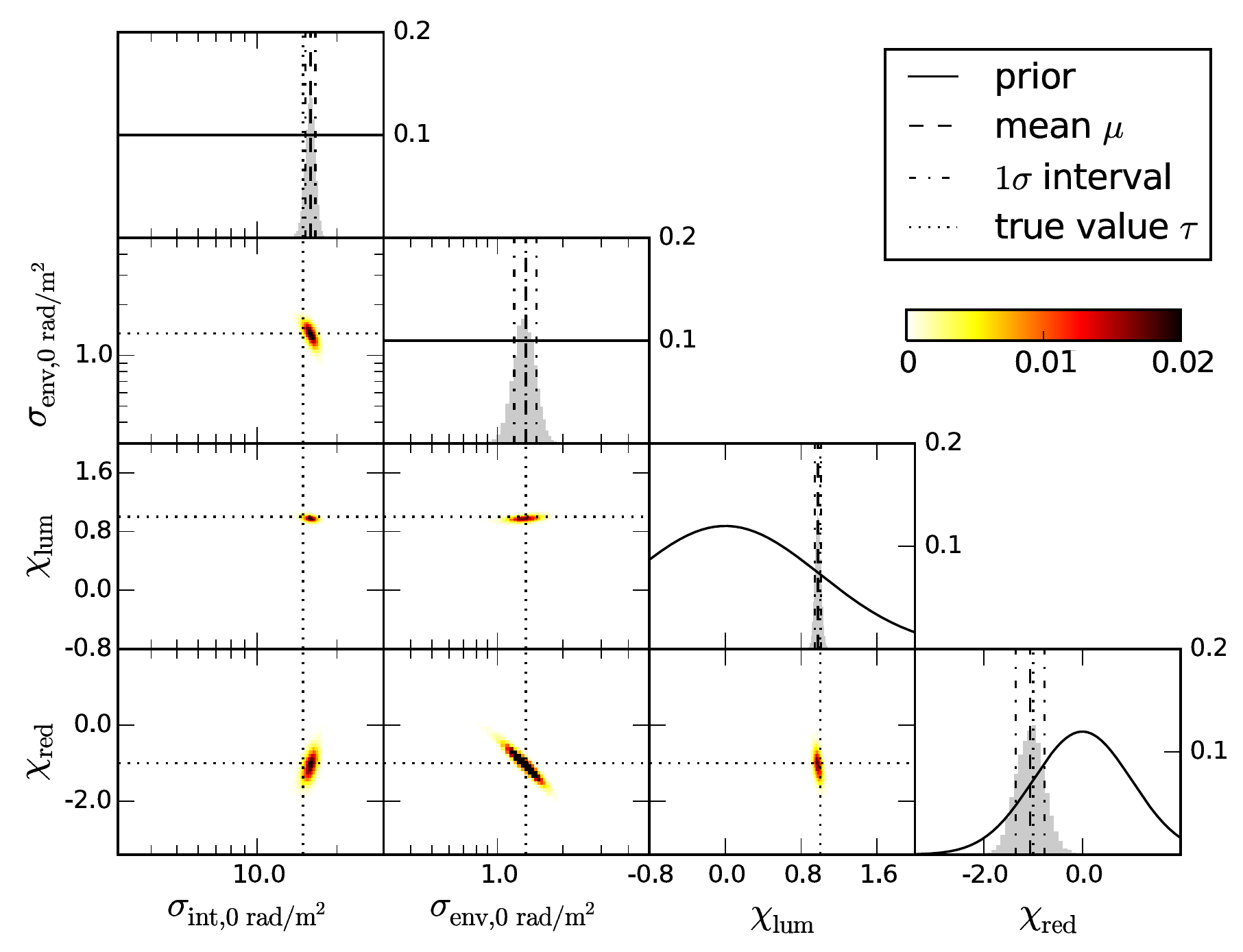} \put(-4,78){(a)}\end{overpic} \\
            \begin{overpic} [width=15.5cm, angle=0]{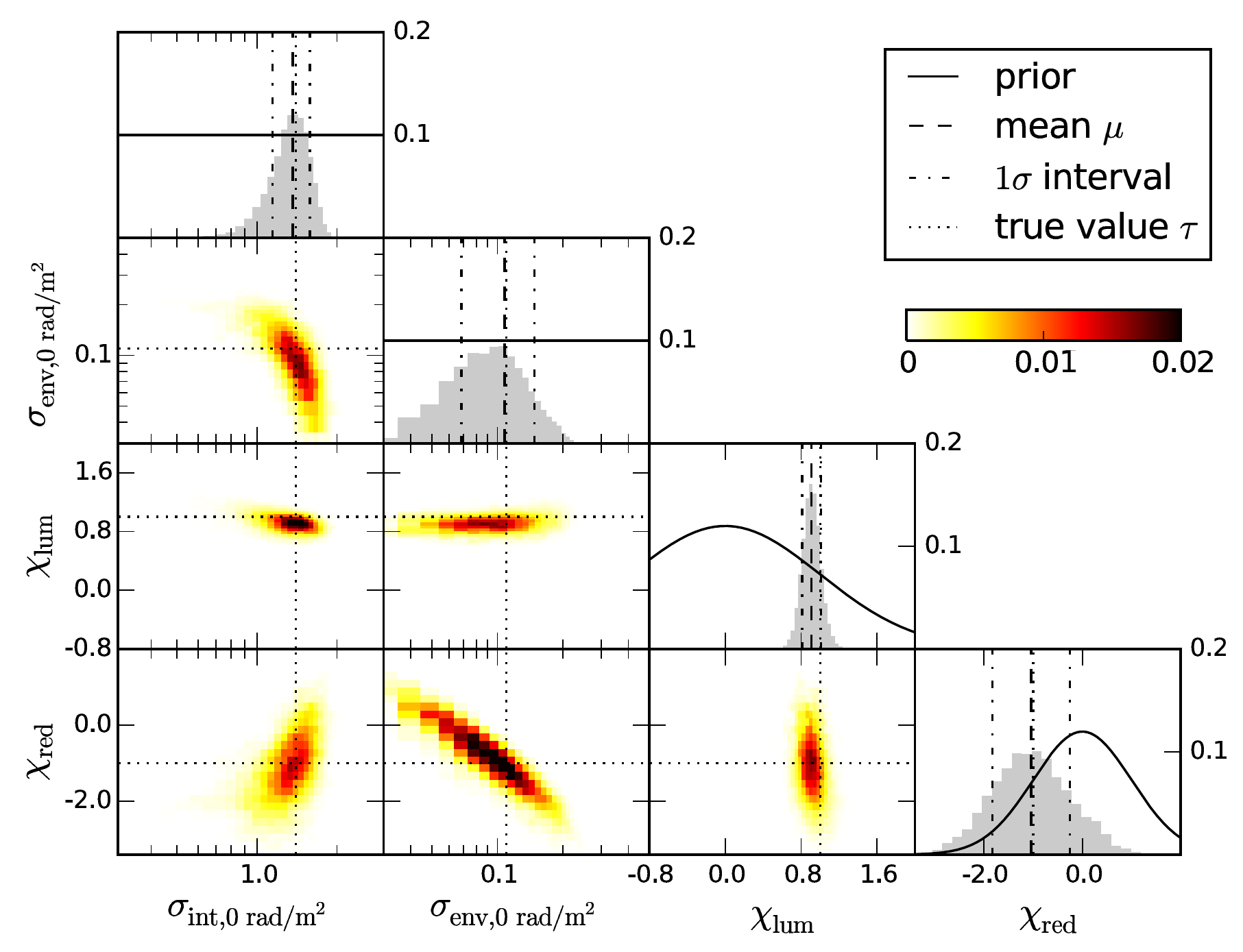} \put(-4,78){(b)} \end{overpic} \\
        \flushleft    
 \caption{Results obtained with a two-component scenario for SKA
   observations in the frequency range 350-900\,MHz for $N_{\rm los}\approx$3500 and an overall Faraday
  depth of (a) $\approx$7.0\,rad\,m$^{-2}$ (B1SPC1\emph{a}) and (b)  $\approx$0.7\,rad\,m$^{-2}$ (B1SPC1\emph{b}).}
\label{skalow}
\end{figure*}

\begin{figure*}[ht]
  \centering
           \begin{overpic} [width=15.5cm, angle=0]{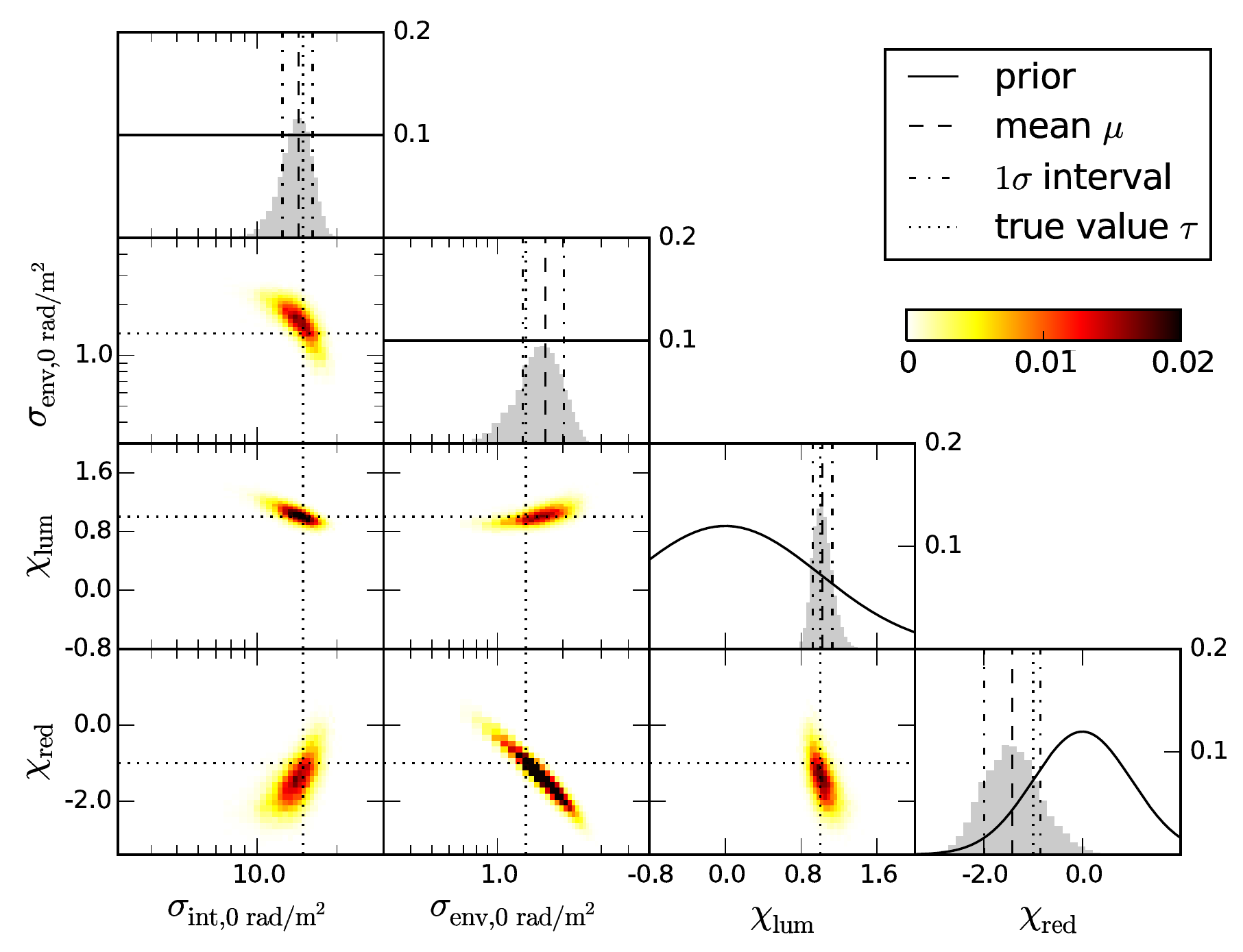}\put(-4,73){(a)}\end{overpic}
                    \begin{overpic} [width=15.5cm, angle=0]{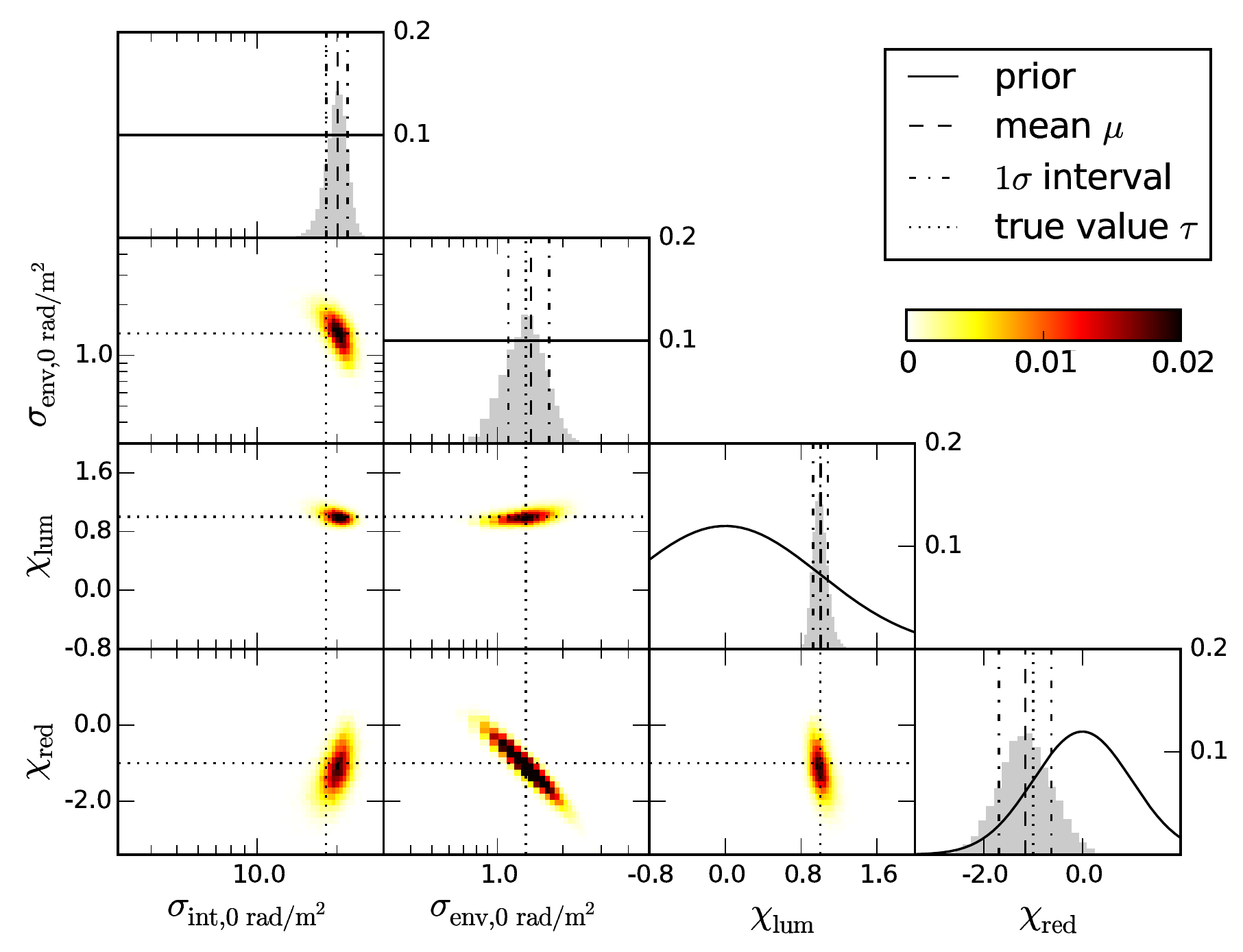}\put(-4,73){(b)}\end{overpic}
        \flushleft   
\caption{Results obtained with a two-component scenario for SKA
  observations in the frequency range (a) 350-900\,MHz (B1SPC2) and (b) 650-1670\,MHz (B2SPC2) for $N_{\rm los}\approx$1000
  and an overall Faraday depth of $\approx$7.0\,rad\,m$^{-2}$.}
\label{ska_even_less_loss}
\end{figure*}

\end{onecolumn}

\end{document}